\documentclass{article}

 \usepackage[final]{nips_2016}

\usepackage[utf8]{inputenc} 
\usepackage[T1]{fontenc}    
\usepackage{hyperref}       
\usepackage{url}            
\usepackage{booktabs}       
\usepackage{amsfonts}       
\usepackage{nicefrac}       
\usepackage{microtype}      
\usepackage{times}
\usepackage{pifont}
\usepackage{bm}
\usepackage{amsmath}
\usepackage{amssymb}
\usepackage{hyperref}
\usepackage{url}
\usepackage{algorithm}
\usepackage{algorithmic}
\usepackage{xfrac}
\usepackage{graphicx}
\usepackage{tensor}
\usepackage{float}
\usepackage{wrapfig}
\usepackage{color}
\usepackage{framed}
\usepackage{eurosym}
\usepackage{supertabular}
\usepackage{colortbl}
\usepackage{sidecap}
\newenvironment{claim}[1]{\par\noindent\underline{Claim:}\space#1}{}

\usepackage{times}
\usepackage{graphicx} 
\usepackage{subfigure} 
\usepackage{amsthm}

\usepackage{blkarray}
\usepackage{algorithm}
\usepackage{algorithmic}

\usepackage{hyperref}

\newtheorem{theorem}{Theorem}
\newtheorem{assumption}{Assumption}
\newtheorem{lemma}{Lemma}
\theoremstyle{definition}
\newtheorem{definition}{Definition}

\title{A distributed Newton Method for Large Scale Consensus Optimization}

\author{
Rasul Tutunov \\
  Department of Computer and Information Science\\
University of Pennsylvania\\
  \texttt{tutunov@seas.upenn.edu} \\
\And
  Haitham Bou Ammar \\
Operations Research \& Financial Engineering Department \\
Princeton University \\
\texttt{hammar@princeton.edu} \\
\AND
  Ali Jadbabaie \\
  Department of Computer and Information Science\\
  University of Pennsylvania\\
  \texttt{jadbabai@seas.upenn.edu} \\
}

\begin{document}

\maketitle

\begin{abstract}
In this paper, we propose a distributed Newton method for consensus optimization. Our approach outperforms state-of-the-art methods, including ADMM. The key idea is to exploit the sparsity of the dual Hessian and recast the computation of the Newton step as one of efficiently solving symmetric diagonally dominant linear equations. We validate our algorithm both theoretically and empirically. On the theory side, we demonstrate that our algorithm exhibits superlinear convergence within a neighborhood of optimality. Empirically, we show the superiority of this new method on a variety of machine learning problems. The proposed approach is scalable to very large problems and has a low communication overhead. 
\end{abstract}
\section{Introduction}
Data analysis through machine and statistical learning has become an important tool in a variety of fields including artificial intelligence, biology, medicine, finance, and marketing. Though arising in diverse applications, these problems share key characteristics, such as an extremely large number (in the order of tens of millions) of training examples typically residing in high-dimensional spaces. With this unprecedented growth in data, the need for \emph{distributed computation} across multiple processing units is ever-pressing. This direction holds the promise for algorithms that are both rich-enough to capture the complexity of modern data, and scalable-enough to handle ``Big Data'' efficiently.

In the distributed setting, central problems are split across multiple processors each having access to local objectives. The goal is then to minimize a sum of local costs while ensuring consensus (agreement) across all processors. To clarify, consider the example of linear regression in which the goal is to find a latent model for a given dataset. Rather than searching for a centralized solution, one can distribute the optimization across multiple processors each having access to local costs defined over random subsets of the full dataset. In such a case, each processor learns a separate ``chunk'' of the latent model, which is then unified by incorporating consensus constraints.

Generally, there are two popular classes of algorithms for distributed optimization. The first is sub-gradient based, while the second relies on a decomposition-coordination procedure. Sub-gradient algorithms proceed by taking a gradient related step then followed by an averaging with neighbors at each iteration. The computation of each step is relatively cheap and can be implemented in a distributed fashion~\cite{NedicO09}. Though cheap to compute, the best known convergence rate of sub-gradient methods is relatively slow given by $\mathcal{O}\left(\frac{1}{\sqrt{t}}\right)$ with $t$ being the total number of iterations~\cite{WeiO12,Goffin}. The second class of algorithms solve constrained problems by relying on dual methods. One of the well-know methods (state-of-the-art) from this class is the Alternating Direction Method of Multipliers (ADMM)~\cite{Boyd}. ADMM decomposes the original problem to two subproblems which are then solved sequentially leading to updates of dual variables. In~\cite{WeiO12}, the authors show that ADMM can be fully distributed over a network leading to improved convergence rates in the order of $\mathcal{O}\left(\frac{1}{t}\right)$.  

Apart from accuracy problems inherent to ADMM-based methods~\cite{Kadkhodaie}, much rate improvements can be gained from adopting second-order (Newton) methods. Though a variety of techniques have been proposed~\cite{Wei11b,GurbuzbalabanOP15,ScheinbergTang}, less progress has been made at leveraging ADMM's accuracy and convergence rate issues. In a recent attempt~\cite{Aryan,Aryan2}, the authors propose a distributed second-order method for general consensus by using the approach in~\cite{zargham2014accelerated} to compute the Newton direction. As detailed in Section~\ref{Sec:Experiments}, this method suffers from two problems. First, it fails to outperform ADMM and second, faces storage and computational defficiencies for large data sets, thus ADMM retains state-of-the-art status\footnote{The approach involves power of matrices of sizes $np \times np$ with $n$ being the total number of nodes and $p$ the number of features.}. 

\textbf{Contributions:} In this paper, we contribute to the above problems and propose a distributed Newton method for general consensus with the following characteristics:  \textit{i)} approximating the \emph{exact newton direction} up-to any arbitrary $\epsilon > 0$, \textit{ii)} exhibiting super-linear convergence within a neighborhood of the optimal solution similar to exact Newton, and \textit{iii)} outperforming ADMM and others in terms of iteration count, running times, and total message complexity on a set of benchmark datasets, including one on a real-world application of fMRI imaging. One can argue that our improvements arrive at increased communication costs compared to other techniques. In a set of experiments, we show that such an increase is relatively small for low accuracy requirements and demonstrate a growth proportional to the condition number of the processors' graph as accuracy demands improve. Of course, as shown in our results (see Section~\ref{Sec:Experiments}), this increase is slower compared to other methods which can be exponential. 

\section{SDD Linear Systems}
Symmetric Diagonally Dominant Matrices (SDD) play a vital role in the computation of the Newton direction in a distributed fashion. In this section, we briefly review SDD systems and present a summary of efficient methods for solving them. SDD systems are linear equations of the form:
\begin{equation}
\label{Eq:SDDM}
\bm{M} \bm{x}_{0} = \bm{b}_{0},
\end{equation}
with $\bm{M}$ being a Symmetric Diagonally Dominant Matrix (SDD). Namely, $\bm{M}$  is symmetric positive semi-definite with non-positive off-diagonal elements, such that for all $i = 1, \dots, n$: $\left[\bm{M}\right]_{ii} \geq - \sum_{j=1, i \neq j}^{n} \left[\bm{M}\right]_{ij}$. The goal is to determine an $\epsilon$-approximate solution, $\tilde{\bm{x}}$, to the exact solution $\bm{x}^{\star}$ of Equation~\ref{Eq:SDDM} (bounded under the $\bm{M}$ norm), which is defined as: 
\begin{definition}\cite{PengS14}
Let $\bm{x}^{\star}$ be the solution to $\bm{M} \bm{x}_{0} = \bm{b}_{0}$. A vector $\tilde{\bm{x}}$ is called an $\epsilon$-approximate solution to $\bm{x}^{\star}$, if: 
\begin{equation*}
\left|\left| \bm{x}^{\star} - \tilde{\bm{x}}\right|\right|_{\bm{M}} \leq \epsilon \left|\left|\bm{x}^{\star}\right|\right|_{\bm{M}}, \ \ \ \text{with $||\bm{u}||_{\bm{M}}^{2} = \bm{u}^{\mathsf{T}} \bm{M} \bm{u}$.}
\end{equation*}
\end{definition} 

\paragraph{Standard Splittings \& Approximations:} The story of computing an approximation to the exact inverse, $\bm{M}^{-1}$, starts from standard splittings of symmetric matrices. Here, $\bm{M}$ is decomposed as: 
\begin{equation*}
\bm{M} = \bm{D}_{0} - \bm{A}_{0},
\end{equation*}
where $\bm{D}_{0}$ is a diagonal matrix consisting of diagonal elements in $\bm{M}$, while $\bm{A}_{0}$ is a non-negative symmetric matrix collecting all off-diagonal components, i.e., $[\bm{A}_{0}]_{ij}= - [\bm{M}]_{ij}$ if $i \neq j$ and $0$ otherwise. Based on this splitting, the authors in~\cite{PengS14} recognize that the inverse of $\bm{M}_{0}$ can be written as: 
\begin{align}
\label{Eq:StandardSplitting}
\left(\bm{D}_{0} - \bm{A}_{0}\right)^{-1} &= \frac{1}{2} \Bigg[ \bm{D}^{-1}_{0} + (\bm{I} + \bm{D}_{0}^{-1}\bm{A}_{0}) (\bm{D}_{0} - \bm{A}_{0}\bm{D}_{0}^{-1}\bm{A}_{0})^{-1} \left(\bm{I}+ \bm{A}_{0}\bm{D}_{0}^{-1}\right)
\Bigg].
\end{align}
Since $\bm{D}_{0} - \bm{A}_{0}\bm{D}_{0}^{-1}\bm{A}_{0}$ is also symmetric diagonally dominant, Spielman and Peng recurse the above for a length of $d=\mathcal{O}(\log n)$ to arrive at the inverse approximated chain, $\mathcal{C} = \{\bm{D}_{i}, \bm{A}_{i}\}_{i=1}^{d}$ with: 
\begin{equation*}
\bm{D}_{i} = \bm{D}_{0}, \ \ \ \ \ \ \  \ \text{and}  \ \ \ \ \ \ \  \ \bm{A}_{i} =\bm{D}_{0}(\bm{D}_{0}^{-1}\bm{A}_{0})^{2^{i}}.
\end{equation*} 

Rewriting Equation~\ref{Eq:StandardSplitting} in terms of the approximated chain, an $\epsilon$-close solution to $\bm{x}^{\star}$ can be determined using a two step procedure. In the first a ``crude'' solution to $\bm{x}^{\star}$ is returned  (Algorithm~\ref{Algo:AlgorithmCrude}). The procedure operates in two loops, both running to an order $d=\mathcal{O}(\log n)$. In the forward loop, intermediate vectors are constructed which are then used in the backward loop to determine the solution $\bm{x}_{0}=\bm{Z}_{0}\bm{b}_{0}$, where $\bm{Z}_{0} \approx_{\epsilon_{d}} \bm{M}^{-1}$.
\begin{algorithm} 
\label{Algo:AlgorithmCrude}
\caption{``Crude'' SDD Solver~\cite{PengS14}}
\begin{algorithmic}
\STATE \textbf{Inputs:} Inverse approximated chain $\mathcal{C}$, vector $\bm{b}_{0}$
\FOR {$i=1 \dots d$}
\STATE Compute $\bm{b}_{i} = \left(\bm{I} - \bm{A}_{i-1}\bm{D}_{i-1}^{-1}\right)\bm{b}_{i-1}$
\ENDFOR
\STATE Determine $\bm{x}_{d} = \bm{D}_{d}^{-1}\bm{b}_{d}$
\FOR {$i=d-1, \dots, 0$}
\STATE Compute $\bm{x}_{i} = \frac{1}{2} \left[\bm{D}_{i}^{-1}\bm{b}_{i} + (I + \bm{D}_{i}^{-1}\bm{A}_{i})\bm{x}_{i+1}\right]$
\ENDFOR
\STATE \textbf{Return:} ``Crude'' approximation, $\bm{x}_{0}$, to $\bm{x}^{\star}$. 
\end{algorithmic}
\end{algorithm}
 Since $\bm{Z}_{0}$ incurs an $\epsilon_{d}$ error (a constant error) to the real inverse $\bm{M}^{-1}$,   Spielman and Peng introduce a Richardson pre-conditioning scheme, detailed in Algorithm~\ref{Algo:Richard}, to arrive at any arbitrary precision.
\begin{algorithm} 
\label{Algo:Richard}
\caption{``Exact'' SDD Solver~\cite{PengS14}}
\begin{algorithmic}
\STATE \textbf{Inputs:} Inverse approximated chain $\mathcal{C}$, length $q =\mathcal{O}(\frac{1}{\epsilon})$, ``crude'' solution $\bm{x}_{0}$
\FOR {$i=1 \dots q$}
\STATE $\bm{y}_{k} = \bm{y}_{k-1} + \bm{u}_{k} + \chi$, where $\bm{M}\bm{u}_{k} = \bm{M} \bm{y}_{k-1}, \ \ \ \text{and} \ \ \ \chi = \bm{x}_{0}$.
\ENDFOR
\STATE \textbf{Return:} $\epsilon$-close solution $\tilde{\bm{x}} = \bm{y}_{q}$.
\end{algorithmic}
\end{algorithm}

To be used in determining the Newton direction, a distributed version of the above SDD solver has to be developed. In recent work, the authors in~\cite{Tutunov} distribute the parallel SDD solver across multiple processors. To do so, the system in Equation~\ref{Eq:SDDM} is interpreted as represented by an undirected weighted graph, $\mathcal{G}$, with $\bm{M}$ being its Laplacian. The authors then introduce an inverse approximate chain which can be computed in a distributed fashion. Consequently, both the ``crude'' and exact solutions can be determined using only local communication between nodes on the undirected graph $\mathcal{G}$. In the consensus problem, we adapt this solver for distributing the computation of the Newton direction (see Section~\ref{Sec:DistNewton}). 

\section{Distributed Global Consensus}
Much of distributed machine learning can be interpreted within the framework of global consensus. 
Global consensus considers a network of $n$ agents represented by a connected undirected graph $\mathcal{G} = \left( \mathcal{V}, \mathcal{E}\right)$ with $|\mathcal{V}| =n$ and $|\mathcal{E}| =m$. Each agent, $i$, corresponding to a node, can exchange information among its first-hop neighborhood denoted by $\mathcal{N}(i)= \left\{j \in \mathcal{V}: (i,j) \in \mathcal{E} \right\}$. The size of such $\mathcal{N}(i)$ is referred to as the degree of node $i$, i.e., $d(i)= |\mathcal{N}(i)|$. In the general form, the goal is for each agent to determine an unknown vector $\bm{x}_{i} \in \mathbb{R}^{p}$ which minimizes a sum of multivariate cost functions $\left\{f_{i}\right\}_{i=1}^{|\mathcal{V}|}$ with $f_{i} : \mathbb{R}^{p} \rightarrow \mathbb{R}$ distributed over the network while abiding by consensus constraints: 
\begin{equation}
\label{Eq:GlobalCons}
\min_{\bm{x}_{1}:\bm{x}_{n}} f\left(\bm{x}_{1}, \dots, \bm{x}_{n}\right) = \min_{\bm{x}_{1}:\bm{x}_{n}}\sum_{i=1}^{n} f_{i}(\bm{x}_{i}) \ \ \text{s.t.} \ \ \ \ \bm{x}_{1} = \bm{x}_{2} = \dots = \bm{x}_{n}.
\end{equation}

\subsection{Distributed Global Consensus}
Though multiple attempts have been made at distributing the global consensus problem, the majority of these works suffer from the following drawbacks. The first line of work is that introduced in~\cite{WeiO12}. This work mostly focuses on the univariate and separable settings and fails to generalize to the multivariate case. The second, on the other hand, is that of~\cite{Boyd}, where the focus is mainly on a parallelized setting and not a distributed one. Parallelized methods assume shared memory that can become restrictive for problems with large data sets. In this work, we focus on the ``true'' distributed setting where each processor abides by its own memory constraints and the framework does not invoke any central node. We start by introducing a set of vectors $\bm{y}_{1}, \dots, \bm{y}_{p}$, each in $\mathbb{R}^{n}$. Each vector $\bm{y}_{j}$ acts as a collector for every dimension of the solution across all nodes. In other words, each vector $\bm{y}_{i}$ contains the $i^{th}$ components of $\bm{x}_{1}, \dots, \bm{x}_{n}$:
\begin{align*}
\bm{y}_{1} = \left[
x_{1}(1), x_{2}(1), \dots x_{n}(1)
\right]^{\mathsf{T}}, \hspace{.1em} \bm{y}_{2} = \left[ x_{1}(2), \dots, x_{n}(2) \right]^{\mathsf{T}},
\hspace{.1em} \dots \hspace{.1em}, \bm{y}_{p} = \left[ x_{1}(p), x_{2}(p), \dots, x_{n}(p)
\right]^{\mathsf{T}}.
\end{align*}

Clearly, the collection of $\bm{y}_{1}, \dots, \bm{y}_{p}$ is locally distributed among the nodes of graph $\mathcal{G}$, since each node $i \in \mathcal{V}$ need only to have access to the $i^{th}$ components of such vectors. Consequently, we can rewrite the problem of Equation~\ref{Eq:GlobalCons} in an equivalent distributed form: 
\begin{equation}
\label{Eq:Two_1}
\min_{\bm{y}_{1}:\bm{y}_{p}} f\left(\bm{y}_{1}, \dots, \bm{y}_{p}\right) = \sum_{i=1}^{n}f_{i}\left(y_{1}(i), \dots, y_{p}(i)\right) \ \  \text{s.t.} \ \ \mathcal{L}\bm{y}_{1} = \bm{0}, \ \ \ \ \ \ \mathcal{L}\bm{y}_{2} = \bm{0}, \ \ \ \ \dots \ \ \ \ \mathcal{L} \bm{y}_{p} =\bm{0}.
\end{equation}
with $\mathcal{L} \in \mathbb{R}^{n \times n}$ being the unweighted graph Laplacian of $\mathcal{G}$ defined as follows: $\mathcal{L}_{i,j} = d(i)$ if $i=j$, -1 if $e=(i,j) \in \mathcal{E}$, and $0$ otherwise. To finalize the definition, we write the problem in Equation~\ref{Eq:Two_1} in a vectorized format as: 
\begin{equation}
\label{Eq:Two}
\min_{\bm{y}} f\left(\bm{y}\right) = \sum_{i=1}^{n}f_{i}\left(y_{1}(i), \dots, y_{p}(i)\right), \ \ \  \text{s.t.} \underbrace{\bm{I}_{p \times p} \otimes \mathcal{L}}_{ \mathbb{R}^{np \times np}} \underbrace{\bm{y}}_{\bm{y} \in \mathbb{R}^{np}} = \bm{0},
\end{equation}
where $\bm{M} = \bm{I}_{p \times p} \otimes \mathcal{L}$ is a block-diagonal matrix with Laplacian diagonal entries, and $\bm{y}$ is a vector concatenating $\bm{y}_{1}:\bm{y}_{p}$. At this stage, our aim is to solve the problem in Equation~\ref{Eq:Two} using dual techniques. Before presenting properties of the dual problem, we next introduce a  standard assumption~\cite{Boyd,zargham2014accelerated} on the associated functions $f_{i}$'s: 
\begin{assumption}\label{Ass:Two}
The cost functions $\left\{f_{i}\right\}_{i=1}^{n}$ are convex and:
\begin{itemize}
\setlength\itemsep{.1em}
\item twice continuously differentiable with: 
$\gamma \preceq \nabla^{2} f_{i} \preceq \Gamma$
\item Lipschitz-Hessian invertible: $\left|\left| \left(\nabla^{2} f_{i}\left(\hat{\bm{x}}_{i}\right)\right)^{-1}-  \left(\nabla^{2} f_{i}\left({\bm{x}}_{i}\right)\right)^{-1}\right|\right|_{2} \leq \delta \left|\left| \hat{\bm{x}}_{i} - \bm{x}_{i} \right|\right|_{2}$,  for some $\delta$ and all $\hat{\bm{x}}_{i}, \bm{x}_{i} \in \mathbb{R}^{p}$. 
\end{itemize}
\end{assumption}
\subsection{Dual Problem}
To acquire the dual formulation of distributed general consensus, we first introduce a vector of dual variables $\bm{\lambda} = \left[\bm{\lambda}_{1}^{\mathsf{T}}, \dots, \bm{\lambda}_{p}^{\mathsf{T}}\right]^{\mathsf{T}} \in \mathbb{R}^{np}$, where each $\bm{\lambda}_{i} \in \mathbb{R}^{n}$ are Lagrange multipliers, one for each dimension of the unknown vector. For distributed computations, we assume that each node need only to store its corresponding components $\lambda_{1}(i), \dots, \lambda_{p}(i)$. Consequently, the Lagrangian of Equation~\ref{Eq:Two}  can be written as follows:
\begin{align*}
\bm{l}\left(\bm{y}_{1}, \dots, \bm{y}_{p}, \bm{\lambda}_{1}, \dots, \bm{\lambda}_{p}\right)= \sum_{i=1}^{n}\Bigg(f_{i}\left(y_{1}(i), \dots, y_{p}(i)\right)  + y_{1}(i)
(\mathcal{L}\bm{\lambda}_{1})_{i} + \dots + y_{p}(i)
(\mathcal{L}\bm{\lambda}_{p})_{i}\Bigg).
\end{align*}
Hence, the dual has the following form: 
\begin{align*}
q(\bm{\lambda})  =  \sum_{i=1}^{n} \inf{_{y_{1}(i):y_{p}(i)}}\Bigg(f_{i}\left(y_{1}(i), \dots, y_{p}(i)\right)  +  y_{1}(i)
(\mathcal{L}\bm{\lambda}_{1})_{i} + \dots + y_{p}(i)
(\mathcal{L}\bm{\lambda}_{p})_{i}\Bigg).
\end{align*}
Having determined the dual variables, we still require a procedure which allows us to infer about the primal. Using the above, the primal variables are determined as the solution to the following system of differential equations: 
\begin{equation}
\label{Eq:System}
     \frac{\partial f_{i}(\cdot)}{\partial y_{1}(i)} = - (\mathcal{L}\bm{\lambda}_{1})_{i},  \ \ \  \dots \ \ \ 
       \frac{\partial f_{i}(\cdot)}{\partial y_{p}(i)} = - (\mathcal{L}\bm{\lambda}_{p})_{i}. 
\end{equation}
Clearly, Equation~\ref{Eq:System} is locally defined for each node $i \in \mathcal{V}$, where for each $r=1,\dots, p$: 
$$- (\mathcal{L} \bm{\lambda}_{r})_{i} = \sum_{j \in \mathcal{N}(i)} \lambda_{r}(j) - d(i) \lambda_{r}(i).$$ 
Hence, each node $i$ can construct its own system of equations by collecting $\{\lambda_{1}(j), \dots, \lambda_{p}(j)\}$ from its neighbors $j \in \mathcal{N}(i)$ without the need for full communication. Denoting the solution of the PDE as: 
$y_{1}(i)= \phi_{1}^{(i)}\left((\mathcal{L}\bm{\lambda}_{1})_{i},\dots, (\mathcal{L}\bm{\lambda}_{p})_{i}\right), \ \ \  \dots, \ \ \  y_{p}(i)= \phi_{p}^{(i)}\left((\mathcal{L}\bm{\lambda}_{1})_{i},\dots, (\mathcal{L}\bm{\lambda}_{p})_{i}\right)$, we can show the following essential theoretical guarantee on the partial derivatives: 
\begin{lemma}
Let $z_{1} = (\mathcal{L}\bm{\lambda}_{1})_{i}$, $z_{2} = (\mathcal{L}\bm{\lambda}_{2})_{i}$, \dots, $z_{p} = (\mathcal{L}\bm{\lambda}_{p})_{i}$. Under Assumption~\ref{Ass:Two}, the functions $\phi_{1}^{(i)}$, \dots, $\phi^{(i)}_{p}$ exhibit bounded partial derivatives with respect to $z_{1}$, \dots, $z_{p}$. In other words, for any $r=1,\dots, p$: $\left|\frac{\partial \phi_{r}^{(i)}}{\partial z_{1}}\right| \leq \frac{\sqrt{p}}{\gamma} \dots \left|\frac{\partial \phi_{r}^{(i)}}{\partial z_{p}}\right| \leq \frac{\sqrt{p}}{\gamma}$, 
for any $(z_{1},\dots, z_{p}) \in \mathbb{R}^{p}$. 
\end{lemma}
The above result is crucial in our analysis, as an obvious corollary is that each function, $\phi_{r}^{(i)}$,  is Lipschitz continuous, i.e., for any two vectors $\tilde{\bm{z}}= (\tilde{z}_{1}, \dots, \tilde{z}_{p})$ and $\bm{z} = (z_{1}, \dots, z_{p})$: 
\begin{align*}
\left|\phi_{r}^{(i)}(\tilde{\bm{z}}) - \phi_{r}^{(i)}(\bm{z})\right| \leq \frac{\sqrt{p}}{\gamma} \left|\left|\tilde{\bm{z}} - \bm{z}\right|\right|_{2}, \ \ \ \ \ \text{for $r=1,\dots, p$}. 
\end{align*}
\paragraph{Dual Function Properties:} Our method for computing the Newton direction relies on the fact that the Hessian of the dual problem is an SDD matrix. We prove this property in the following lemma:
\begin{lemma}\label{Lemma:Props}
The dual function $q(\bm{\lambda})= q(\bm{\lambda}_{1}, \dots, \bm{\lambda}_{p})$ shares the following characteristics: 
\begin{itemize}
\item The dual Hessian $\bm{H}(\bm{\lambda})$ and gradient $\nabla q(\bm{\lambda})$ are given by:
\begin{equation*}
\bm{H}(\bm{\lambda})= - \bm{M} \left(\nabla^{2} f(\bm{y}(\bm{\lambda})), \right)^{-1}\bm{M}, \ \ \ \ \ \  \nabla q(\bm{\lambda}) = \bm{M} \bm{y}(\bm{\lambda}). 
\end{equation*}
\item The dual Hessian is Lipschitz continuous with respect to $\bm{M}$-weighted norm, where for any $\tilde{\bm{\lambda}}$ and $\bm{\lambda}$: $\left|\left|\bm{H}(\tilde{\bm{\lambda}})- \bm{H}\left({\bm{\lambda}}\right)\right|\right|_{\bm{M}} \leq B \left|\left|\tilde{\bm{\lambda}} - \bm{\lambda}\right|\right|_{\bm{M}}$, with $B=\frac{\delta {p}}{\gamma} \mu_{n}^{2}(\mathcal{L})\sqrt{\mu_{n}(\mathcal{L})}$, where $\mu_{n}(\mathcal{L})$ is the largest eigenvalue of $\mathcal{L}$ and the constants $\gamma$ and $\delta$ are these given in~\ref{Ass:Two}. 
\end{itemize}
\end{lemma}
\vspace{-1em}
\section{Distributed Newton  for General Consensus}\label{Sec:DistNewton}
We solve the consensus problem using Newton-like techniques, where our method follows the \emph{approximate}  Newton direction in the dual: $\bm{\lambda}^{[k+1]}= \bm{\lambda}^{[k]} + \alpha^{[k]} \tilde{\bm{d}}^{[k]}$, where $k$ is the iteration number, and $\alpha^{[k]}$ the step-size. $\tilde{\bm{d}}^{[k]}$ is the $\epsilon$-approximation to the exact Newton direction at iteration $k$. For efficient operation, the main goal is to \emph{accurately} approximate the Newton direction in a fully distributed fashion. This can be achieved with the help of the SDD properties of the dual hessian proved earlier. Recalling that exact Newton computes \footnote{Above we used the following notation: $\bm{y}^{[k]}= \bm{y}\left(\bm{\lambda}^{[k]}\right)$, $\bm{H}^{[k]} = - \bm{M} \left(\nabla^{2} f \left(\bm{y}\left(\bm{\lambda}^{[k]}\right)\right)\right)^{-1}\bm{M}$, and $\bm{g}^{[k]}= \nabla q\left(\bm{\lambda}^{[k]}\right)= \bm{M} \bm{y}^{[k]}$ to denote the primal variables, dual Hessian and gradient, respectively.}:
\begin{align}
\label{Eq:NewtonSystem}
\bm{H}^{[k]} \bm{d}^{[k]} = - \bm{g}^{[k]},  \ \ \ \ \bm{M}\left(\nabla^{2} f \left(\bm{y}^{[k]}\right)\right)^{-1}\bm{M} \bm{d}^{[k]} &= \bm{M} \bm{y}^{[k]},
\end{align}
we notice that Equation~\ref{Eq:NewtonSystem} can be split to two SDD linear systems of the form: 
\begin{equation}
\label{Eq:LinearSystems}
\bm{M}\bm{z}^{[k]} =  \bm{M} \bm{y}^{[k]}, \ \ \bm{M} \bm{d}^{[k]} = \nabla^{2} f\left(\bm{y}^{[k]}\right)\bm{z}^{[k]}. 
\end{equation}
The first equation is by itself SDD which can be solved in a distributed fashion using the approach in~\cite{Tutunov}. Having attained that solution, we map the second system to $p$-SDD systems by introducing $\bm{d}^{[k]}=\left(\left(\bm{d}_{1}^{[k]}\right)^{\mathsf{T}}, \dots, \left(\bm{d}_{p}^{[k]}\right)^{\mathsf{T}}\right)^{\mathsf{T}}$ with each $\bm{d}_{r}^{[k]} \in \mathbb{R}^{n}$. It is easy to see that this can be split to the following collection of $p$ linear systems for $r=1,\dots, n$: 
\begin{equation}
\label{Eq:Seventeen}
\mathcal{L}\bm{d}_{1}^{[k]} = \bm{b}_{1}^{[k]},  \ \ \ \ \ \dots \ \ \ \ \ , \mathcal{L}\bm{d}_{p}^{[k]} = \bm{b}_{p}^{[k]},  
\end{equation}
where $\bm{b}_{1}^{[k]}, \dots, \bm{b}_{p}^{[k]} \in \mathbb{R}^{n}$ defined as: 
\begin{align*}
b_{1}^{[k]} (r) = \sum_{l=1}^{p} \frac{\partial^{2} f_{r}(\cdot)}{\partial y_{1}(r)\partial y_{l} (r)} z^{[k]}_{l}(r), \ \ \ \dots \ \ \ \ 
b_{p}^{[k]} (r) = \sum_{l=1}^{p} \frac{\partial^{2} f_{r}(\cdot)}{\partial y_{p}(r)\partial y_{l} (r)} z^{[k]}_{l}(r) 
\end{align*}
for $r = 1,\dots, n$. Interestingly, the above computations can be performed completely locally by noting that each node $r \in \mathcal{V}$ can compute the $r^{th}$ component of each vector $\bm{b}_{1}^{[k]}, \dots, \bm{b}_{p}^{[k]}$. This is true as such a node stores $f_{r}$ as well as the variables $z^{[k]}_{1}(r), \dots, z^{[k]}_{p}(r)$. Before commencing to the convergence analysis, the final step needed is to establish the connection between the approximate solutions:
\begin{lemma}
Let $\tilde{\bm{d}}_{1}^{[k]}, \dots, \tilde{\bm{d}}_{p}^{[k]}$ be the $\epsilon_{0}$-approximate solution to Equation~\ref{Eq:Seventeen}, then $\tilde{\bm{d}}^{[k]}$ is an $\epsilon$-approximate solution to~\ref{Eq:NewtonSystem} with $\epsilon = \epsilon_{0} \sqrt{\frac{\Gamma}{\gamma} \frac{\mu_{n}(\mathcal{L})}{\mu_{2}(\mathcal{L})}}\left[1 + \epsilon_{0} \frac{\mu_{n}(\mathcal{L})}{\mu_{2}(\mathcal{L})}\sqrt{\frac{\Gamma}{\gamma}} + \sqrt{\frac{\mu_{n}(\mathcal{L})}{\mu_{2}(\mathcal{L})}}\right]$.
\end{lemma} 
\vspace{-1em}
\section{Convergence Guarantees}
We next analyze the convergence properties of our approximate Newton method showing similar three convergence phases to approximate Newton methods. We start with the following lemma: 
\begin{lemma}
Let $\bm{g}^{[k]} = \nabla q\left(\bm{\lambda}^{[k]}\right)$ be the dual gradient at the $k^{th}$ iteration. Then: 
\begin{align*}
\left|\left|\bm{g}^{[k+1]}\right|\right|_{\bm{M}} &\leq \left[1 -\alpha_{k} + \epsilon \alpha_{k} \sqrt{\frac{\Gamma}{\gamma} \frac{\mu_{n}^{3}(\mathcal{L})}{\mu_{2}^{3}(\mathcal{L})}}\right]\left|\left|\bm{g}^{[k]}\right|\right|_{\bm{M}} +\frac{B(\alpha_{k} (1+\epsilon))^{2}}{2\mu_{2}^{4}(\mathcal{L})} \left|\left|\bm{g}^{[k]}\right|\right|_{\bm{M}}.
\end{align*}
\end{lemma} 
Now, we are ready to provide the theorem summarizing the three convergence phases: 
\begin{theorem}
Let $\Gamma$, $\gamma$, be the constants defined in Assumption~\ref{Ass:Two}, $\mu_{n}(\mathcal{L})$, $\mu_{2}(\mathcal{L})$ be the largest and the second smallest eigenvalues of $\mathcal{L}$, respectively, and $\epsilon \in \left(0, \frac{\mu_{2}(\mathcal{L})}{\mu_{n}(\mathcal{L})}\sqrt{\frac{\Gamma}{\gamma}\frac{\mu_{2}(\mathcal{L})}{\mu_{n}(\mathcal{L})}}\right]$. Consider the following iteration scheme: $\bm{\lambda}^{[k+1]}=\bm{\lambda}^{[k]} + \alpha^{\star} \tilde{\bm{d}}^{[k]}$, where $\alpha^{\star}= \left(\frac{\gamma}{\Gamma}\right)^{2}\left(\frac{\mu_{2}(\mathcal{L})}{\mu_{n}(\mathcal{L})}\right)^{4}\frac{1-\epsilon}{(1+\epsilon)^{2}}$. Then the distributed Newton algorithm exhibits the following convergence phases: 
\begin{itemize}
\item \textbf{Strict Decrease Phase:} while $\left|\left|\bm{g}^{[k]}\right|\right|_{\bm{M}} \geq \eta_{1}$: $q\left(\bm{\lambda}^{[k+1]}\right)-q\left(\bm{\lambda}^{[k]}\right) \leq -\frac{\gamma^{3}}{\Gamma^{2}}\left(\frac{1-\epsilon}{1+\epsilon}\right)^{2}\frac{\mu_{2}^{4}(\mathcal{L})}{\mu_{n}^{7}(\mathcal{L})} \eta_{1}^{2}$,
\item \textbf{Quadratic Decrease Phase:} while $\eta_{0} \leq \left|\left|\bm{g}^{[k]}\right|\right|_{\bm{M}} \leq \eta_{1}$: $
\left|\left|\bm{g}^{[k+1]}\right|\right|_{\bm{M}} \leq \frac{1}{\eta_{1}}\left|\left|\bm{g}^{[k]}\right|\right|_{\bm{M}}^{2}$,
\item \textbf{Terminal Phase:} while $\left|\left|\bm{g}^{[k]}\right|\right|_{\bm{M}} \leq \eta_{0}$: $
\left|\left|\bm{g}^{[k+1]}\right|\right|_{\bm{M}} \leq \zeta \left|\left|\bm{g}^{[k]}\right|\right|_{\bm{M}}$,
where $\eta_{0} = \frac{\zeta (1 - \zeta)}{\xi}$, $\eta_{1} = \frac{1-\zeta}{\xi}$, and 
$\zeta = \sqrt{\left[1-\alpha_{k} + \epsilon \alpha_{k} \sqrt{\frac{\Gamma}{\gamma}\frac{\mu_{n}^{3}(\mathcal{L})}{\mu_{2}^{3}(\mathcal{L})}}\right]}, \ \ \ \xi= \frac{B(\alpha_{k}\Gamma(1+\epsilon))^{2}}{2\mu_{2}^{4}(\mathcal{L})}$. 
\end{itemize}
\end{theorem}



\vspace{-1em}
\section{Evaluation}\label{Sec:Experiments}
We evaluate our method against five other approaches: 1) distributed Newton ADD, an adaptation of ADD~\cite{zargham2014accelerated} that we introduce to compute the Newton direction of general consensus, 2) distributed ADMM~\cite{WeiO12}, 3) distributed averaging~\cite{Olshevsky1}, an algorithm solving general consensus using local averaging, 4) network newton 1 and 2~\cite{Aryan,Aryan2}, and 5) distributed gradients~\cite{NedicO09}. We are chiefly interested in the convergence speeds of both the objective value and the consensus error. The comparison against ADMM allows us to understand whether we surpass state-of-the-art, while comparisons against ADD and network newton sheds-the-light on the accuracy of our Newton's direction approximation. \\
\textbf{Real-World Distributed Implementation:} To simulate a real-world distributed environment, we used the Matlab parallel pool running on an 8 core server. After  generating the processors' graph structure with random edge assignment (see below for specifics on the node-edge configuration), we split nodes equally across the 8 cores. Hence, each processor was assigned a collection of nodes for performing computations. Communication between these nodes was handled using the Matlab Message Passing Interface (MatlabMPI) of~\cite{kepner2004matlabmpi}, which allows for efficient scripting. As for bandwidth, it has been shown in~\cite{kepner2001parallel} that MatlabMPI can match C-MPI for large messages, and can maintain high-bandwidth even when multiple processors are communicating. 
\subsection{Benchmark Data Sets}
We performed three sets of experiments on standard machine learning problems: 1) linear regression, 2) logistic regression, and 3) reinforcement learning. We transformed centralized problems to fit within the distributed consensus framework. This can be easily achieved by factoring the summation running over all the available training examples to partial summations across multiple processors while introducing consensus (see Appendix H). Due to space constraints, we report two of these in this section and leave the others to the supplementary material (see Appendix G). We considered both synthetic as well as real-world data sets:\\
\textbf{Synthetic Regression Task:} We created a dataset for regression with $10^8$ data points each being an $80$ dimensional vector. The task parameter vector $\bm{\theta}$ was generated as a linear combination of these features. The training data set $\bm{X}$ was generated from a standard normal distribution in $80$ dimensions. The training labels were given as $\bm{y} = \bm{X} \bm{\theta} + \bm{\zeta}$, where each element in $\bm{\zeta}$ was an independent univariate Gaussian noise. \\
\textbf{MNIST Data} The MNIST data set is a large database of handwritten digits which has been used as a benchmark for classification algorithms~\cite{mnistlecun}. The goal is to classify among 10 different digits amounting from 1 to 10. After reading each image, we perform dimensionality reduction to reduce the number of features of each instance image to $150$ features using principle component analysis and follow a one-versus-all classification scheme.\\  

\begin{figure*}[tb!]
\centering

\subfigure[Obj. Synthetic]{
	\label{fig:ObjSynth}
\includegraphics[trim = 32mm 20mm 35mm 22mm, clip,height=0.2\textwidth,width=0.32\textwidth]{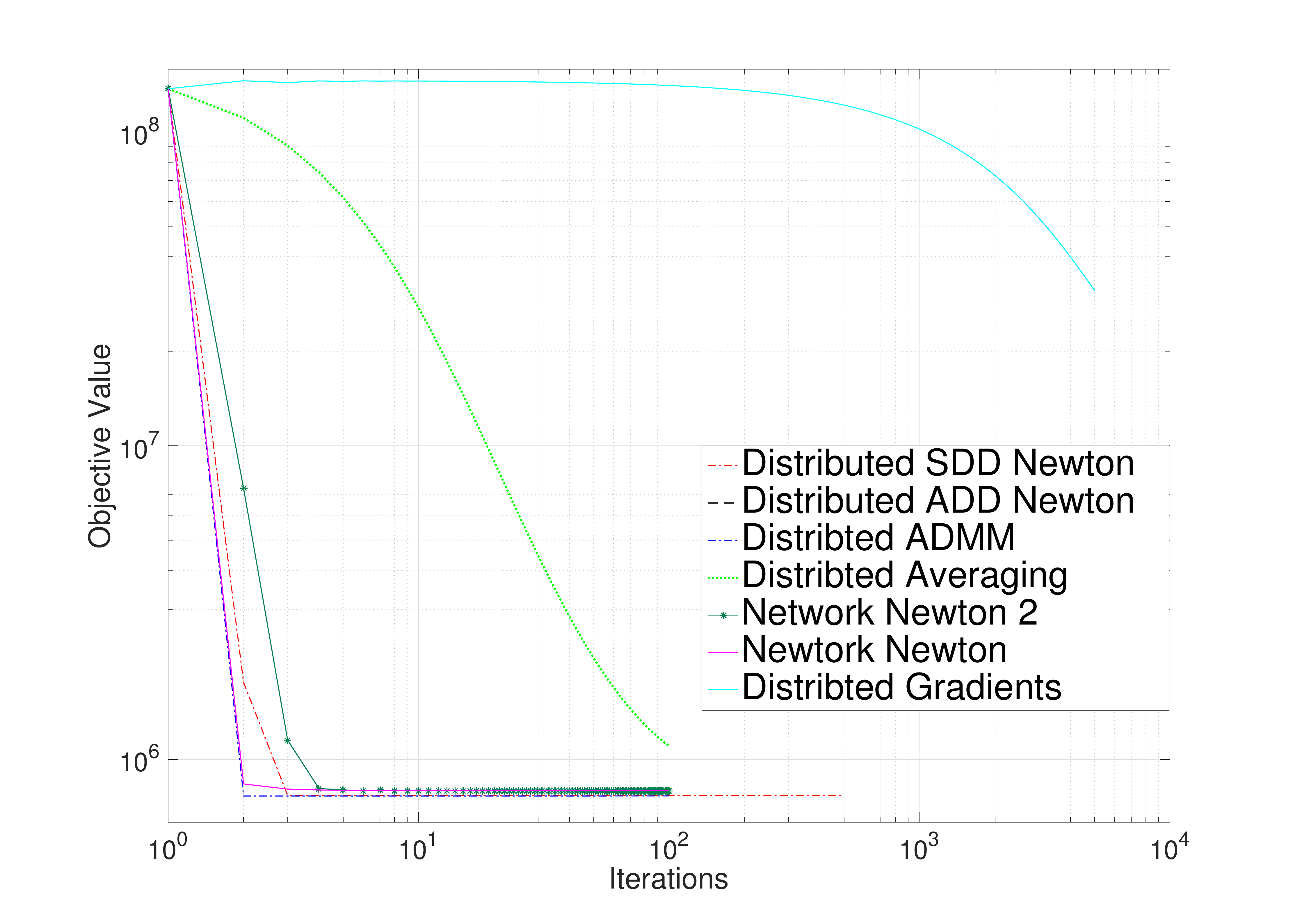}
}
\hfill\hspace{-1.1em}\hfill
\subfigure[Con. Synthetic]{
	\label{fig:ConsSynth}
\includegraphics[trim = 17mm 20mm 25mm 20mm, clip, height=0.2\textwidth,width=0.32\textwidth]{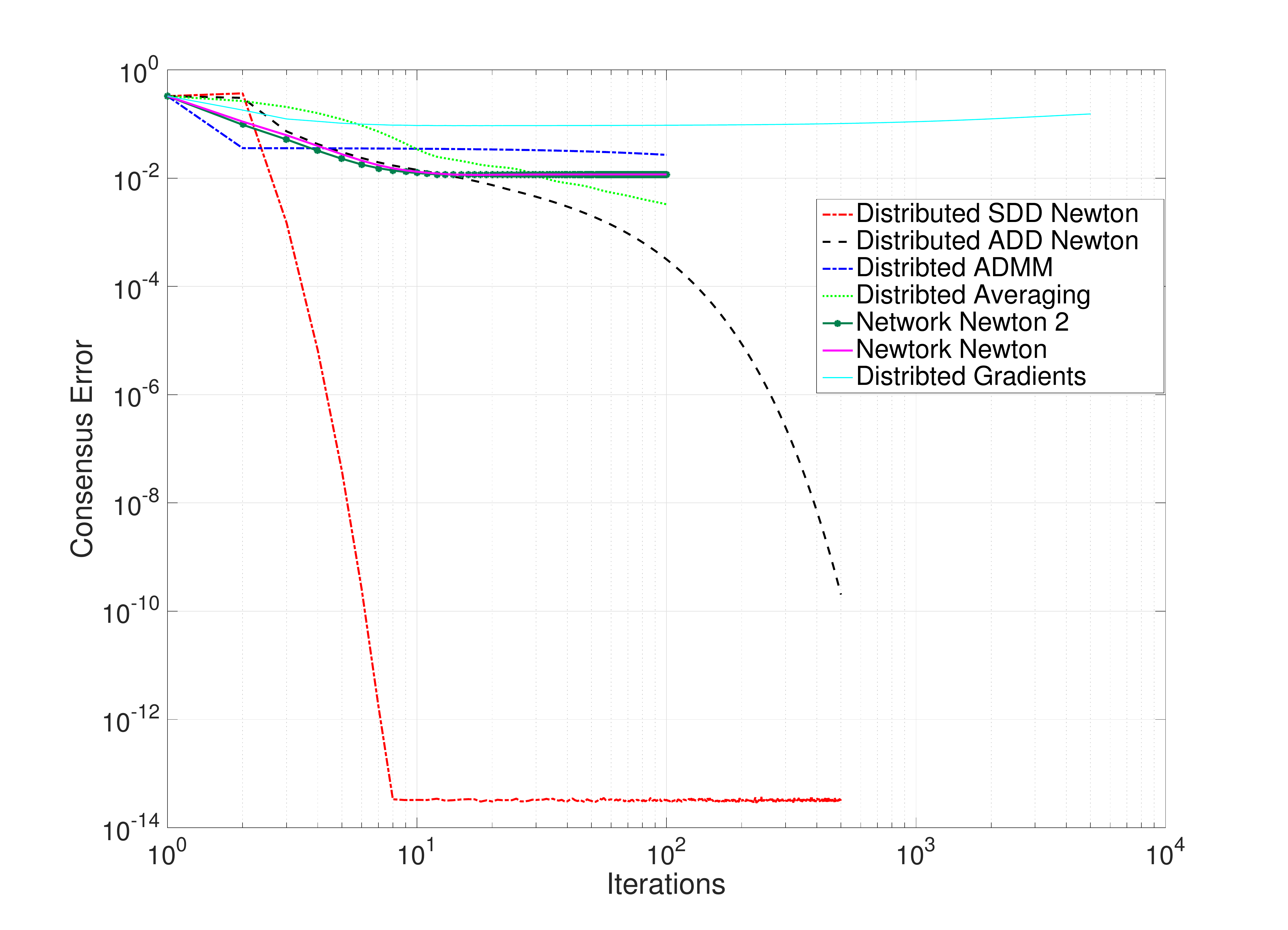}
}
\hfill
\subfigure[Obj. MNIST L$_{2}$]{
	\label{fig:ObjClassNormal}
\includegraphics[trim = 18mm 20mm 25mm 25mm, clip,height=0.20\textwidth,width=0.32\textwidth]{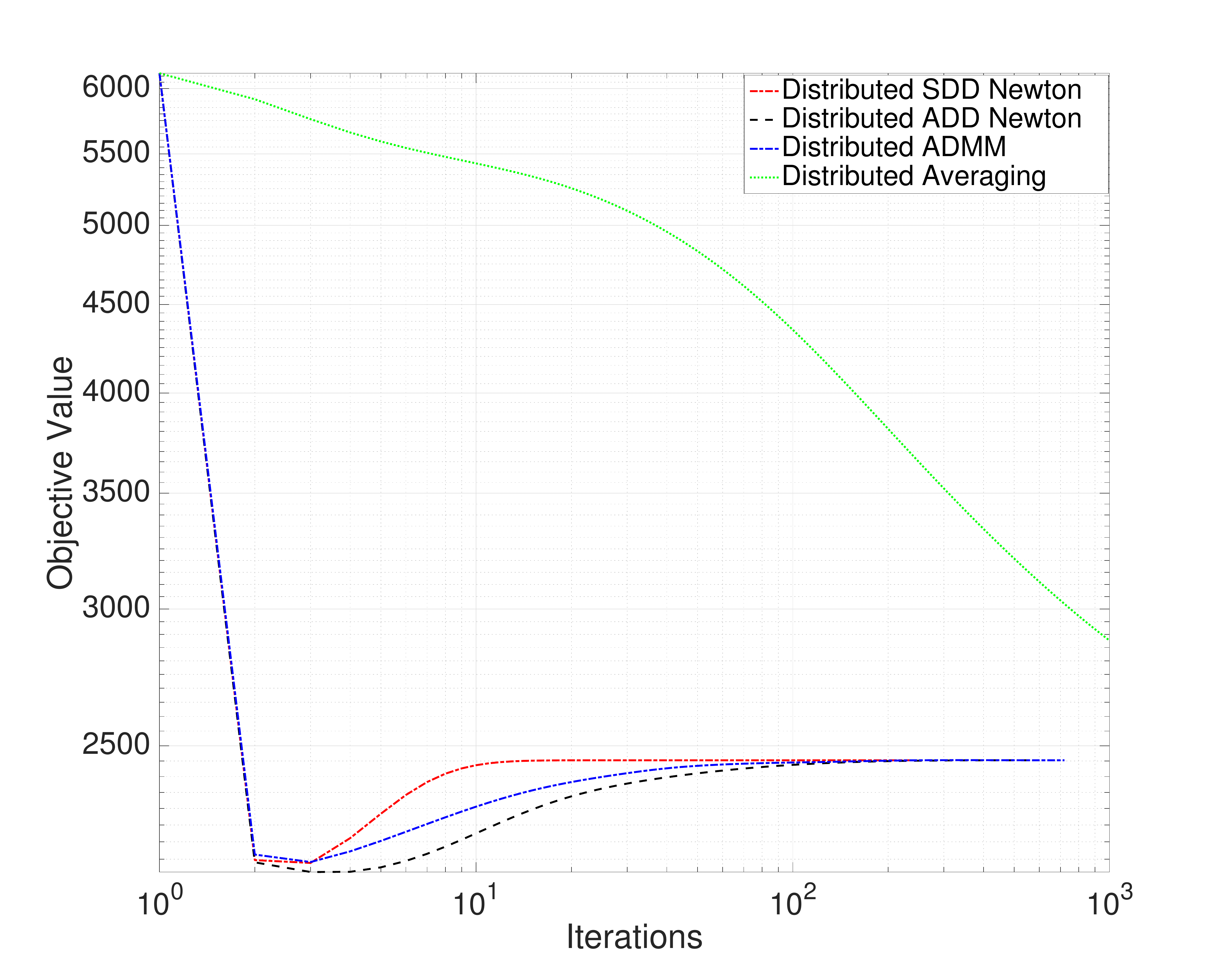}
}
\hfill\hspace{-1.1em}\hfill
\subfigure[Con. MNIST L$_{2}$]{
	\label{fig:ConClassNormal}
\includegraphics[trim = 18mm 20mm 25mm 25mm, clip,height=0.20\textwidth,width=0.32\textwidth]{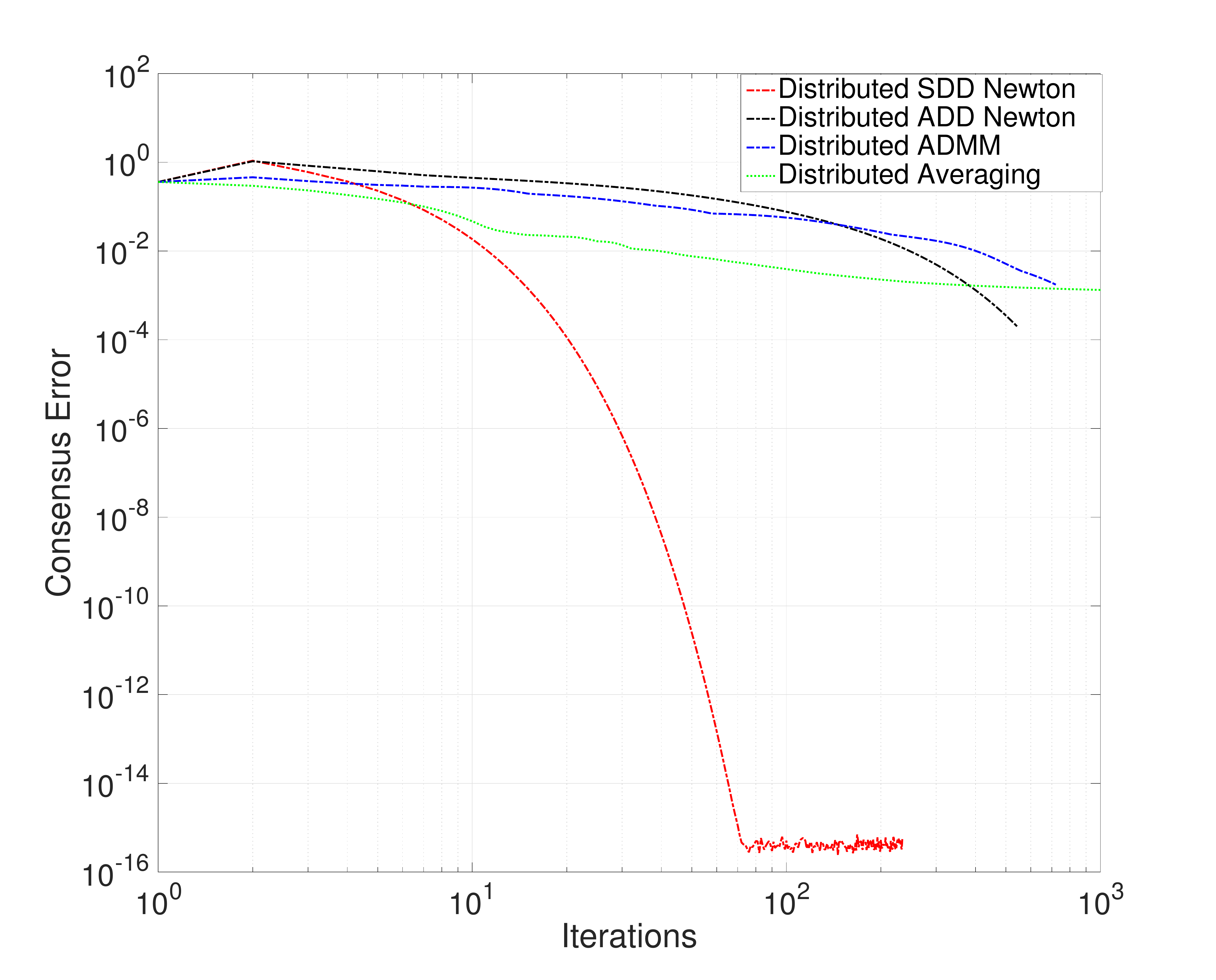}
}
\hfill\hspace{-1.1em}
\subfigure[Obj. MNIST L$_{1}$]{
	\label{fig:ObjClassSparse}
\includegraphics[trim = 18mm 20mm 25mm 25mm, clip,height=0.20\textwidth,width=0.32\textwidth]{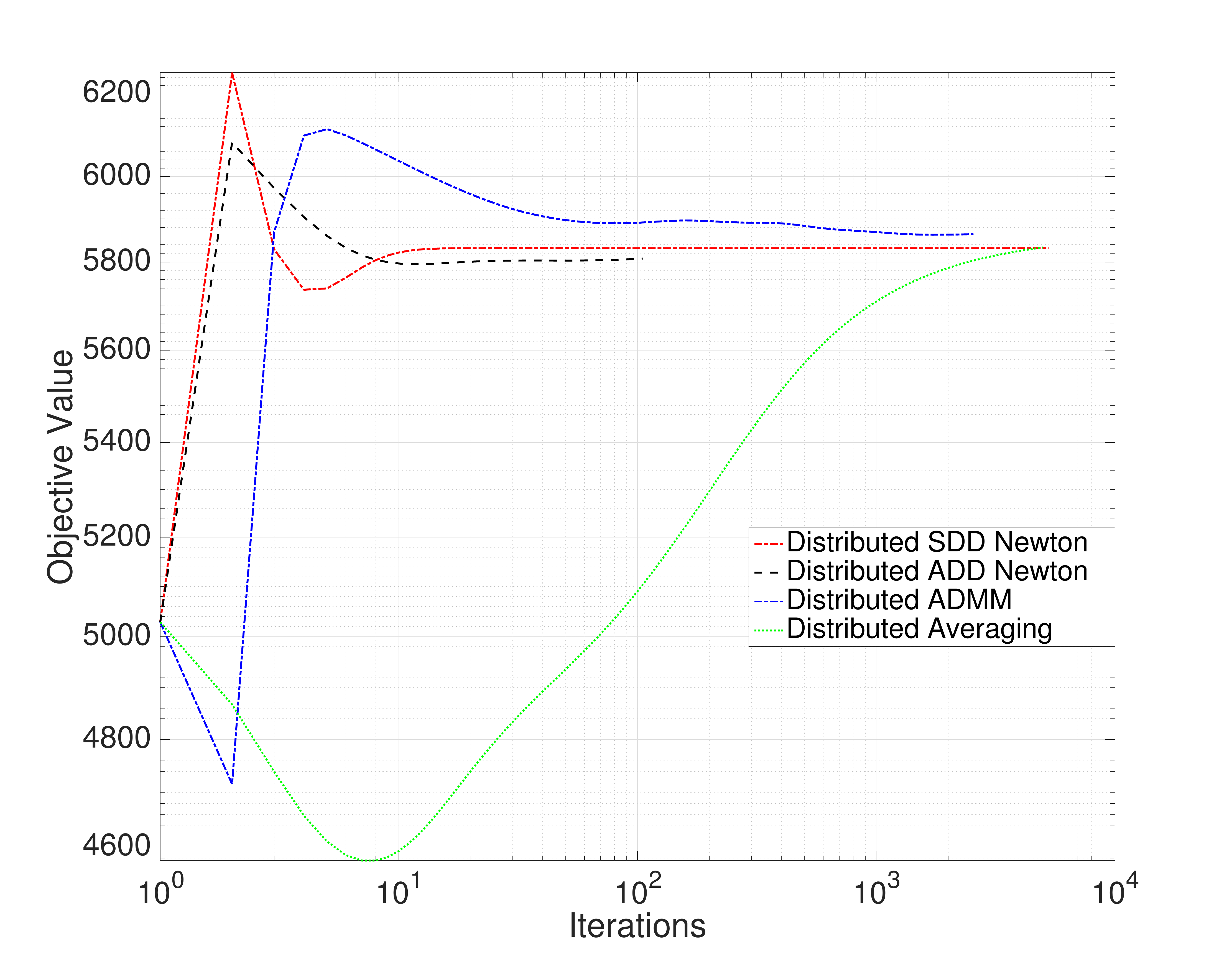}
}
\hfill
\hfill\hspace{-1.4em}\hfill
\subfigure[Con. MNIST L$_{1}$]{
	\label{fig:ConClassSparse}
\includegraphics[trim = 18mm 20mm 25mm 23mm, clip,height=0.20\textwidth,width=0.32\textwidth]{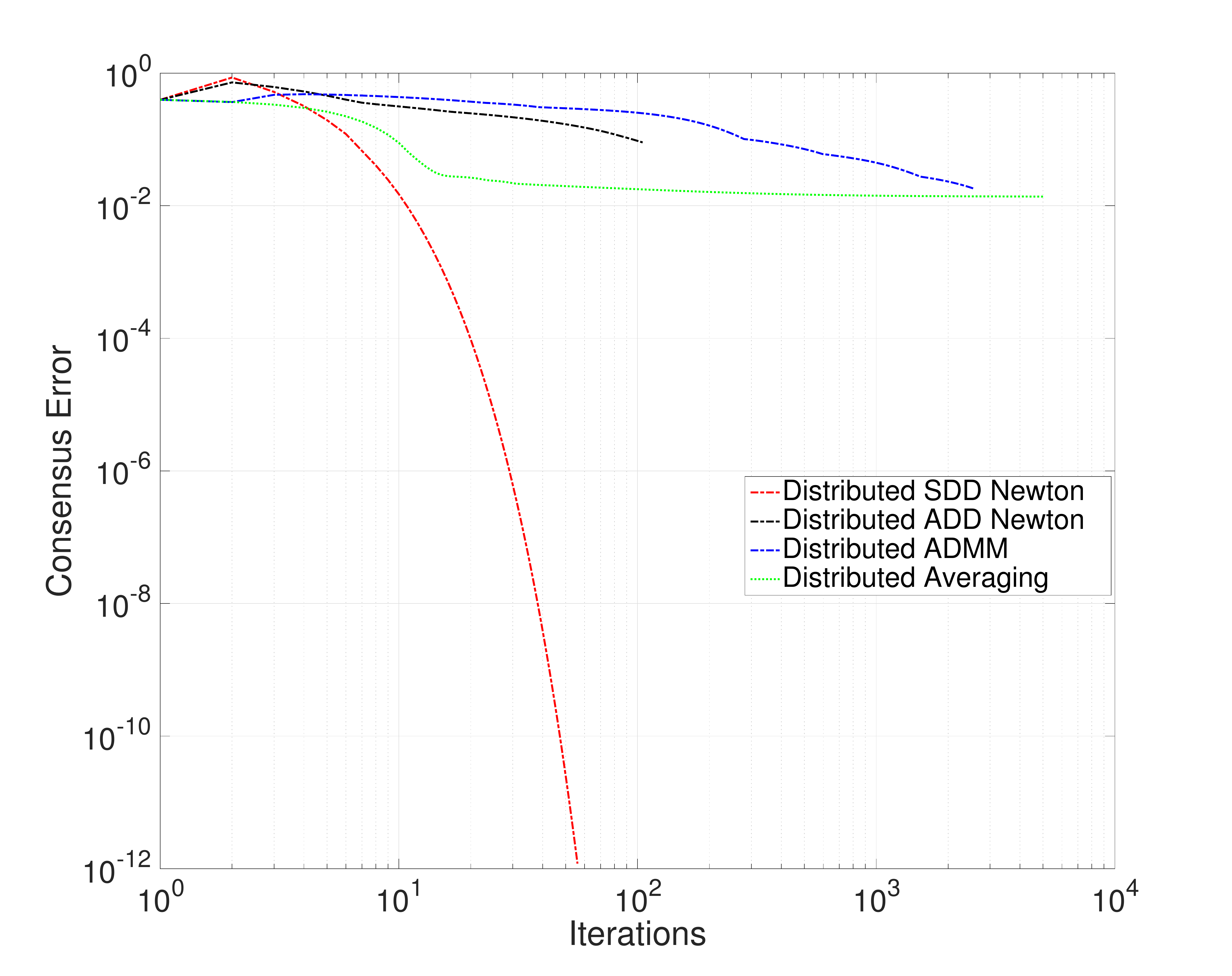}
}
\vspace{-1.1em}
\caption{Figures (a) and (b) report the objective value and consensus error versus iterations on the synthetic regression dataset. Figures (c)-(f) demonstrate the same criteria on the MNIST dataset with both L$_1$ and L$_2$ regularizers. In all these cases, our method outperforms others in literature. }
\end{figure*}

\vspace{-1.5em}
\begin{figure*}
\hfill
\hfill\hspace{-1.4em}\hfill
\subfigure[Obj. fMRI]{
	\label{fig:ObjRL}
\includegraphics[trim = 18mm 15mm 25mm 23mm, clip,height=0.20\textwidth,width=0.25\textwidth]{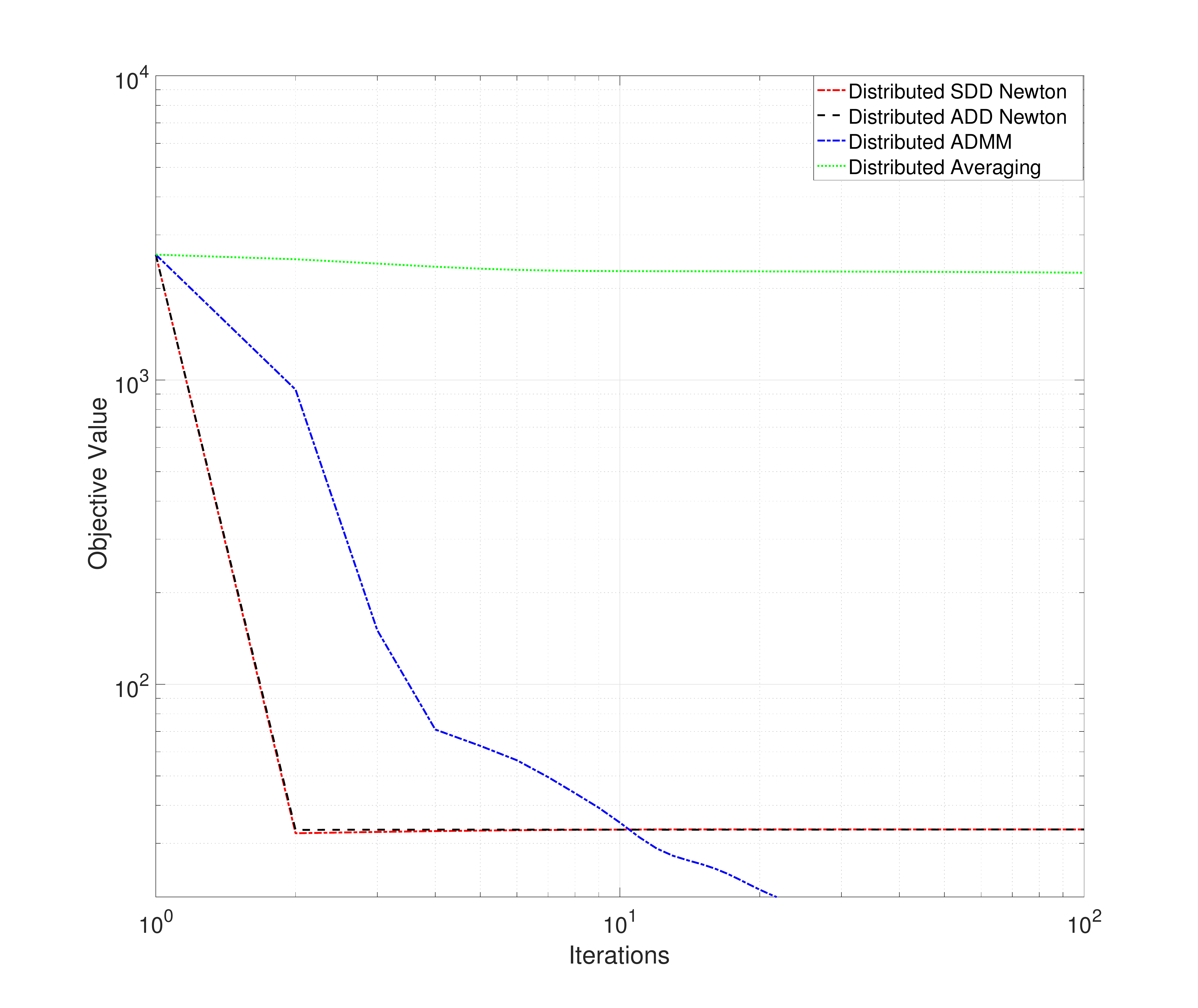}
}
\hfill
\hfill\hspace{-1.4em}\hfill
\subfigure[Con. fMRI]{
	\label{fig:ConRL}
\includegraphics[trim = 18mm 15mm 25mm 23mm, clip, height=0.20\textwidth,width=0.25\textwidth]{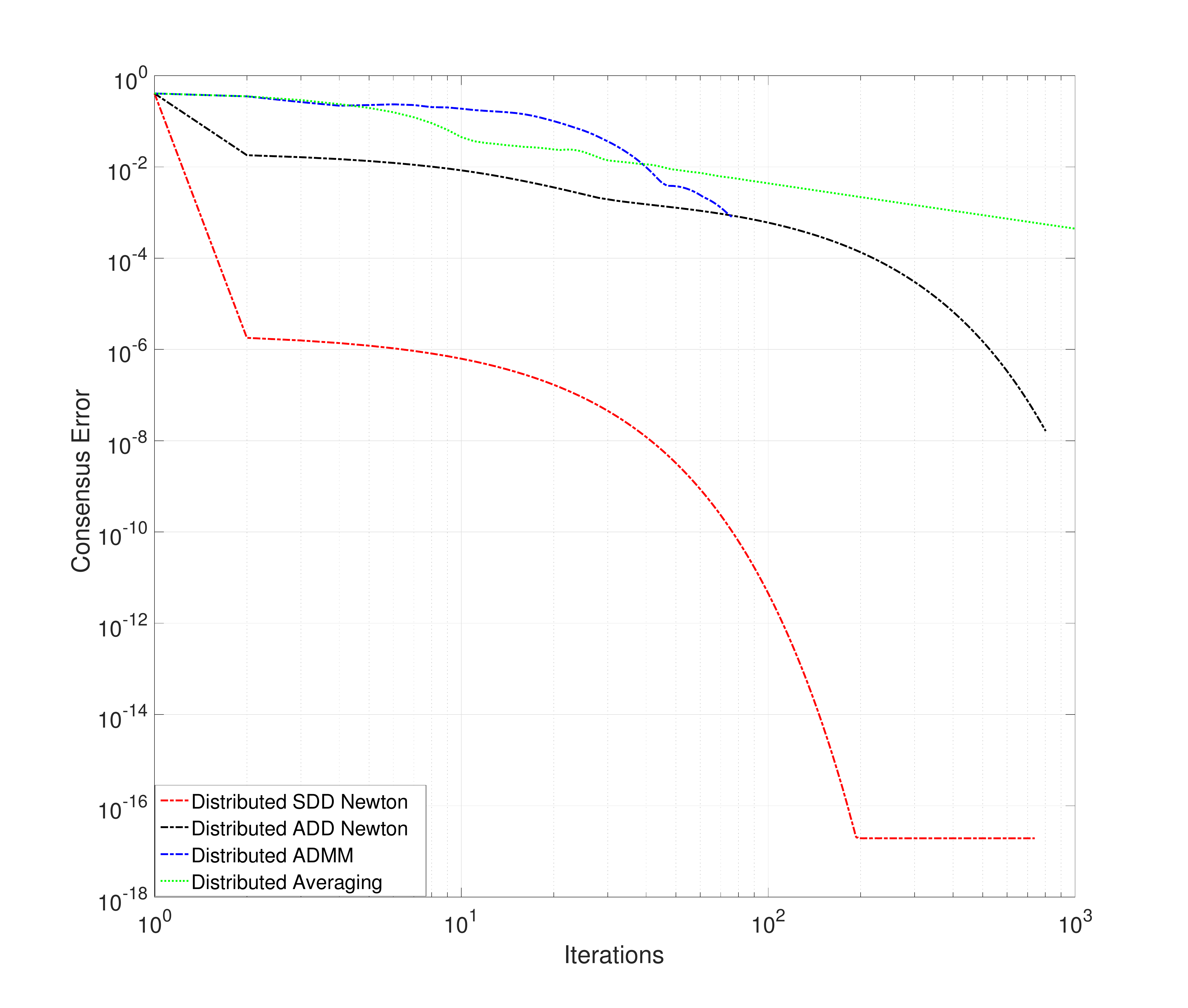}}
\hfill
\hfill\hspace{-1.4em}\hfill
\subfigure[Communication]{
	\label{fig:Communication}
\includegraphics[trim = 18mm 15mm 25mm 15mm, clip,height=0.20\textwidth,width=0.25\textwidth]{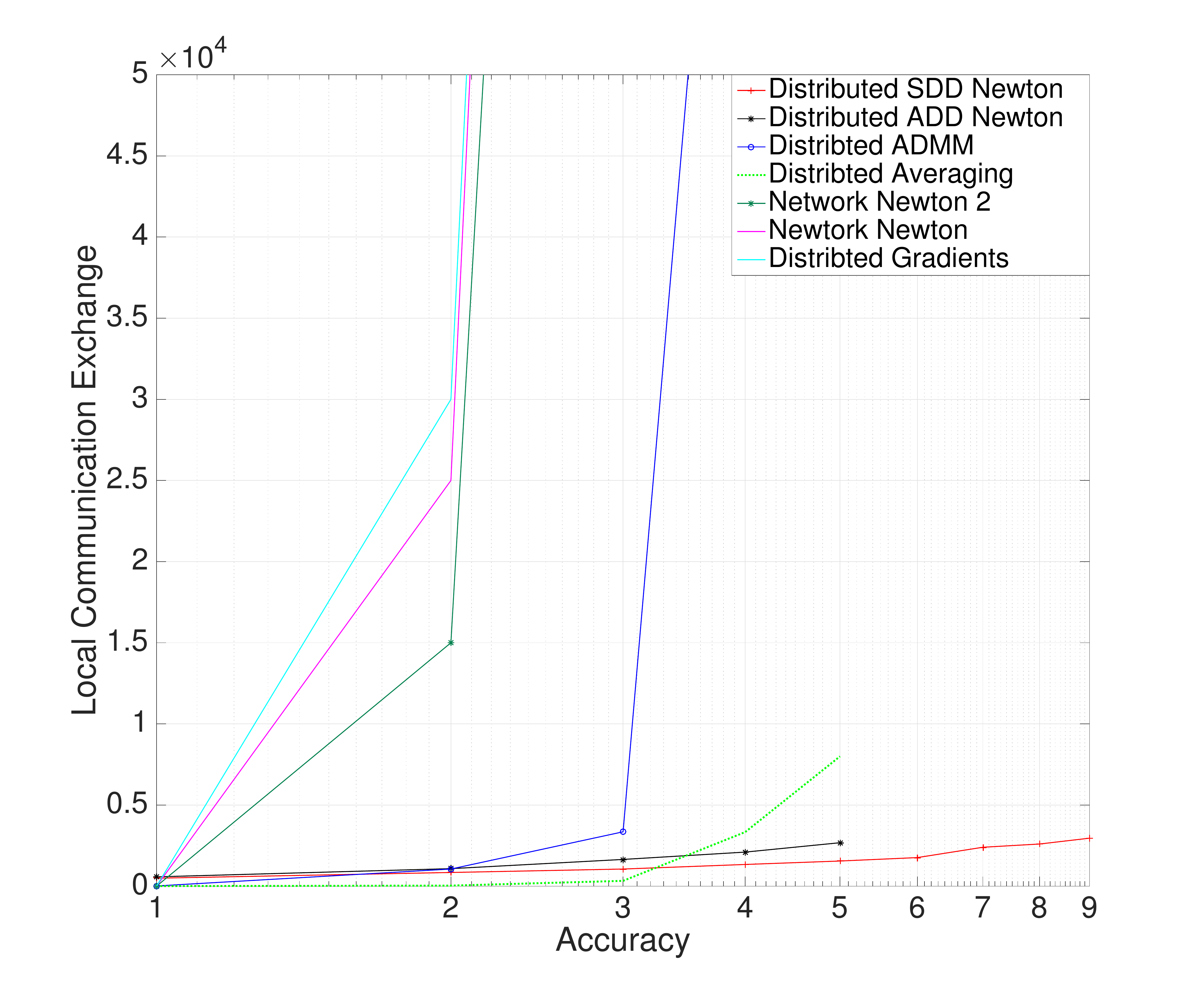}
}
\hfill
\hfill\hspace{-1.4em}\hfill
\subfigure[Times]{
	\label{fig:Communication}
\includegraphics[trim = 15mm 35mm 37mm 52mm, clip,height=0.20\textwidth,width=0.25\textwidth]{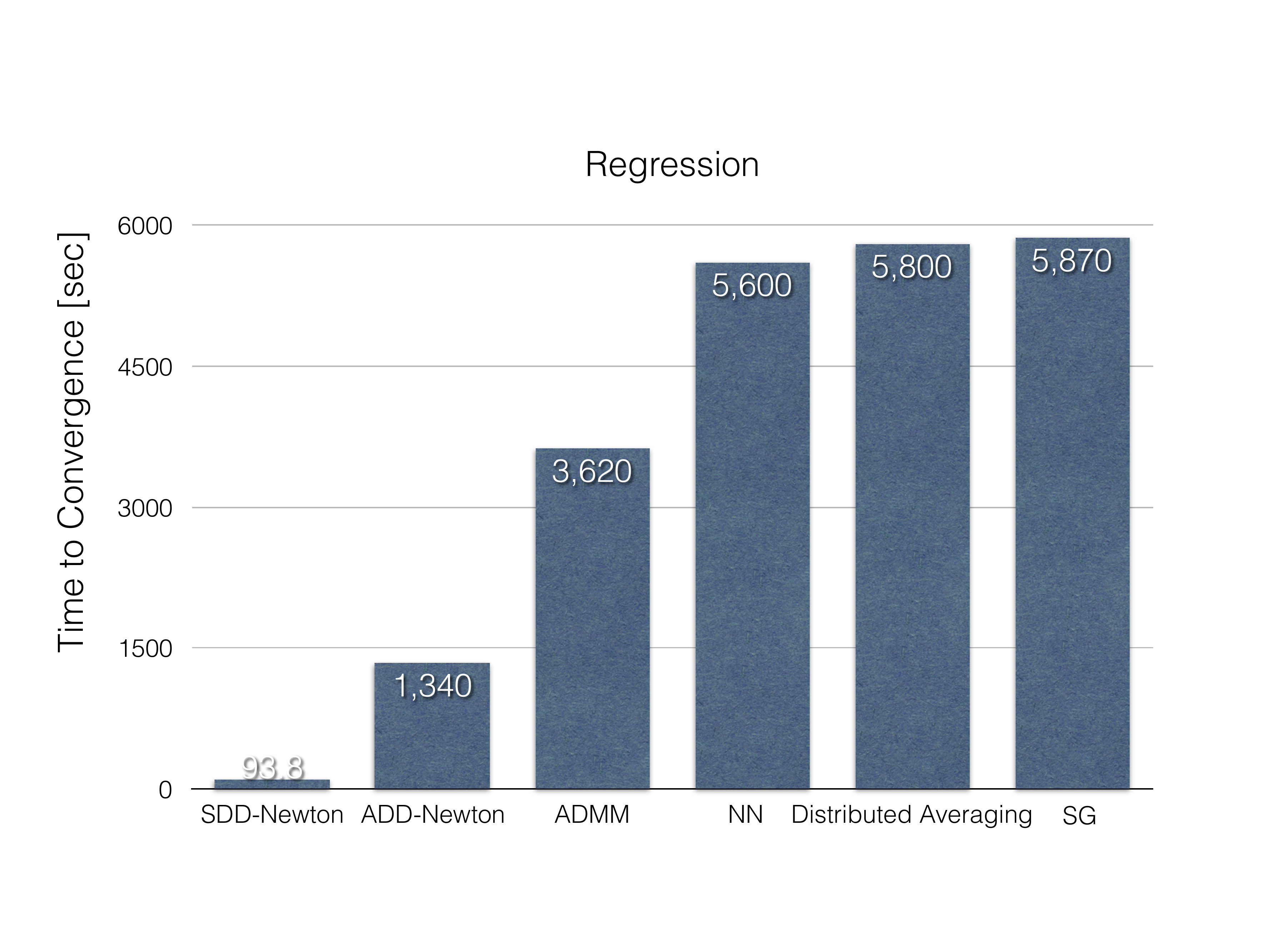}
}
\vspace{-1.1em}
\caption{Figures (a) and (d) consensus errors and objective values on the fMRI dataset showing the our technique outperforms state-of-the-art methods including ADMM.  Figure (c) demonstrates low increase in communication overhead of SDD-Newton compared to other techniques. Figure (d) depicts the overall running times showing that our method converges the fastest.}
\label{fig:NACResultsPostLifelongLearning}
\vspace{-.6em}
\end{figure*}

\subsection{Linear Regression Results} 
\textbf{Synthetic Data:} We randomly distributed the regression objective over a network of 100 nodes and 250 edges. The edges were chosen uniformly at random. An $\epsilon$ of $1/10$ was provided to the SDD solver for determining the approximate Newton direction. Step-sizes were determined separately for each algorithm using a grid-search-like-technique over \{0.01, 0.1, 0.2, 0.3, 0.5, 0.6, 0.9, 1\} to ensure best operating conditions. We used the local objective and the consensus error as performance metrics. Results shown in Figures~\ref{fig:ObjSynth} and~\ref{fig:ConsSynth} demonstrate that our method (titled Distributed SDD Newton) significantly outperforms all other techniques in both objective value and consensus error. Namely, distributed SDD Newton converges to the optimal value in about 40 iterations compared to about 200 for the second-best performing algorithm. It is also interesting to recognize that the worst performing algorithms were distributed gradients and network newton 1 and 2 from~\cite{Aryan}. 
\subsection{Logistic Regression Results} 
We chose the most successful algorithms from previous experiments to perform image classification. We considered both smooth (L$_2$ norms) and non-smooth (L$_{1}$ norms) regularization forms on latent parameters. The processor graph was set to 10 nodes and 20 edges generated uniformly at random. Results depicted in Figures~1(e)-1(f) demonstrate that our algorithm is again capable of outperforming state-of-the-art methods.


\vspace{-1em}
\subsection{fMRI Experiment}
Having shown that our approach outperforms others on relatively dense benchmark datasets, we are now interested in the performance on sparse datasets where the number of features is much larger than the number of inputs. To do so, we used the functional Magnetic Resonance Imaging (fMRI) dataset from~\cite{wang2002detecting}. The goal in these experiments is to classify the cognitive state (i.e., wether looking at a picture or a sentence) of a subject based on fMRI data. Six subjects were considered in total. Each had 40 trials that lasted for 27 seconds attaining in total 54 images per-subject. After preprocessing as described in~\cite{wang2002detecting}, we acquired a sparse data-set with 240 input data points, each having 43,720 features. We then performed logistic regression with an L$_{1}$ regularization and reported objective values and consensus errors. Figures 2(a) and 2(b) demonstrate the objective value and consensus errors on the fMRI dataset. First, it is clear that our approach outperforms others on both criteria. It is worth noting that the second-best performing algorithm to ours is Distributed ADD-Newton; an approach we proposed in this paper for computing the Newton direction. Distributed ADMM and Distributed Averaging perform the worst on such a sparse problem. Second, Figure 2(b) clearly manifests the drawback of ADMM which requires substantial amounts of iterations for converging to the optimal feasible point. Due to the size of the feature set (i.e., 43,720) even small deviations from the optimal model can lead to significant errors in the value of the objective function. This motivates the need for the accurate solutions as acquired by our method. 
\vspace{-1em}
\subsection{Communication Overhead \& Running Times}
It can be argued that our results arrive at a high communication cost between processors. This can be true as our method relies on an SDD-solver while others allow only for few messages per iteration. We conducted a final experiment measuring local communication exchange with respect to accuracy requirements. For that, we chose the London Schools data set as all algorithms performed relatively well. Results reported in Figure 2(c) demonstrate that this increase is negligible compared to other methods. Clearly, as accuracy improves so does the communication overhead of all other algorithms. Distributed SDD-Newton has a growth rate proportional to the condition number of the graph being much slower compared to the exponential growth observed by other techniques. Finally, Figure 2(d) reports running times till convergence on the same dataset. Clearly, our method is the fastest when compared with others. The worst performing algorithms were Network Newton, distributed averaging and sub-gradients.
\vspace{-1em}
\section{Conclusions \& Future Work}
In this paper, we proposed a distributed Newton method for solving general consensus optimization. Our method exploits the SDD property of the dual Hessian leading to an accurate computation of the Newton direction up-to-any arbitrary $\epsilon >0$. We showed that our method exhibits three phases of convergence  with a quadratic phase in the neighborhood of optimal solution. In a set of experiments on standard machine learning benchmarks (including non-smooth cost functions) we demonstrated that our algorithm is capable of outperforming state-of-the-art methods, including ADMM. Finally, we empirically demonstrated that such an improvement arrives at a negligible increase in communication overhead between processors. 

Our next step is to develop incremental versions of this algorithm, and use generalized Hessians to allow for non differentiable cost functions. We also plan on taking such a framework to the lifelong machine learning setting. 

\small
\bibliographystyle{plain}

\newpage{}

\appendix
\section{Synopsis}
We organized appendix as follows. The proofs of Lemmas ~1, 2, 3, 4 are presented in sections B,C,D,E. Theorem 1 is proved is Section F. The experimental result for reinforcement learning and London Schools datasets are presented in section G. Finally, the reductions of standard machine learning problems (regression, classification, reinforcement learning) to global consensus are given in section H.

\section{Proof Primal-Dual Properties}
\begin{lemma}
Let $z_{1} = (\mathcal{L}\bm{\lambda}_{1})_{i}$, $z_{2} = (\mathcal{L}\bm{\lambda}_{2})_{i}$, \dots, $z_{p} = (\mathcal{L}\bm{\lambda}_{p})_{i}$. Under Assumption~\ref{Ass:Two}, the functions $\phi_{1}^{(i)}$, \dots, $\phi^{(i)}_{p}$ exhibit bounded partial derivatives with respect to $z_{1}$, \dots, $z_{p}$. In other words, for any $r=1,\dots, p$: $\left|\frac{\partial \phi_{r}^{(i)}}{\partial z_{1}}\right| \leq \frac{\sqrt{p}}{\gamma} \dots \left|\frac{\partial \phi_{r}^{(i)}}{\partial z_{p}}\right| \leq \frac{\sqrt{p}}{\gamma}$, 
for any $(z_{1},\dots, z_{p}) \in \mathbb{R}^{p}$. 
\end{lemma}
\begin{proof}
Using the definition of $z_1, \ldots z_p$, the primal-dual variable system can be written as:
\begin{equation}\label{inter_prim_system}
   \left\{
     \begin{array}{l}
      \frac{\partial f_i}{\partial \phi^{(i)}_1}  = -z_1 \\
        \frac{\partial f_i}{\partial \phi^{(i)}_2}  = -z_2 \\ 
        \vdots \\
        \frac{\partial f_i}{\partial \phi^{(i)}_p}  = -z_p
     \end{array}
   \right. \hspace{2cm}
\end{equation} 

Taking the derivative of the above system with respect to $z_1$ gives: 

\begin{align*}
   \left\{
     \begin{array}{l}
      \frac{\partial^2 f_i}{\partial (\phi^{(i)}_1)^2}\frac{\partial \phi^{(i)}_1}{\partial z_1} + \frac{\partial^2 f_i}{\partial \phi^{(i)}_1\partial \phi^{(i)}_2}\frac{\partial \phi^{(i)}_2}{\partial z_1} + \ldots + \frac{\partial^2 f_i}{\partial \phi^{(i)}_1\partial \phi^{(i)}_p}\frac{\partial \phi^{(i)}_p}{\partial z_1} = -1 \\
       \frac{\partial^2 f_i}{\partial \phi^{(i)}_2\partial \phi^{(i)}_1}\frac{\partial \phi^{(i)}_1}{\partial z_1} + \frac{\partial^2 f_i}{\partial (\phi^{(i)}_2)^2}\frac{\partial \phi^{(i)}_2}{\partial z_1} + \ldots + \frac{\partial^2 f_i}{\partial \phi^{(i)}_2\partial \phi^{(i)}_p}\frac{\partial \phi^{(i)}_p}{\partial z_1} = 0 \\ 
        \vdots \\
        \frac{\partial^2 f_i}{\partial \phi^{(i)}_p\partial \phi^{(i)}_1}\frac{\partial \phi^{(i)}_1}{\partial z_1} + \frac{\partial^2 f_i}{\partial \phi^{(i)}_p\partial \phi^{(i)}_2}\frac{\partial \phi^{(i)}_2}{\partial z_1} + \ldots + \frac{\partial^2 f_i}{\partial (\phi^{(i)}_p)^2}\frac{\partial \phi^{(i)}_p}{\partial z_1} = 0
     \end{array}
   \right. \hspace{2cm}
\end{align*}
Denoting $\boldsymbol{u}_1 = [\frac{\partial \phi^{(i)}_1}{\partial z_1}, \frac{\partial \phi^{(i)}_2}{\partial z_1}, \ldots, \frac{\partial \phi^{(i)}_p}{\partial z_1}]^{\mathsf{T}}$  we have:
\begin{equation}
[\nabla^2f_i]\boldsymbol{u}_1 = -\boldsymbol{e}_1,
\end{equation}
where $\boldsymbol{e}_1$ is the first vector belonging to the standard basis of $\mathbb{R}^p$. 

Similarly we can show that:
\begin{equation*}
[\nabla^2f_i]\boldsymbol{u}_2 = -\boldsymbol{e}_2 \hspace{0.2cm} [\nabla^2f_i]\boldsymbol{u}_3 = -\boldsymbol{e}_3, \hspace{0.1cm} \ldots \hspace{0.1cm} [\nabla^2f_i]\boldsymbol{u}_p = -\boldsymbol{e}_p,
\end{equation*}
with $\boldsymbol{u}_r = [\frac{\partial \phi^{(i)}_1}{\partial z_r}, \frac{\partial \phi^{(i)}_2}{\partial z_r}, \ldots, \frac{\partial \phi^{(i)}_p}{\partial z_r}]^{\mathsf{T}}$. For convenience, we rewrite the above systems as: 
\begin{equation}\label{matrix_form_equation_derivative}
[\nabla^2f_i]\boldsymbol{U} = -\boldsymbol{I}_{p\times p},
\end{equation}
where 
\begin{equation}
\boldsymbol{U} = \left[\begin{array}{cccc}
\frac{\partial \phi^{(i)}_1}{\partial z_1}&\frac{\partial \phi^{(i)}_1}{\partial z_2}&\cdots &\frac{\partial \phi^{(i)}_1}{\partial z_p}\\
\frac{\partial \phi^{(i)}_2}{\partial z_1}&\frac{\partial \phi^{(i)}_2}{\partial z_2}&\cdots &\frac{\partial \phi^{(i)}_2}{\partial z_p}\\
\vdots & &\ddots &\vdots \\
\frac{\partial \phi^{(i)}_p}{\partial z_1}&\frac{\partial \phi^{(i)}_p}{\partial z_2}&\cdots &\frac{\partial \phi^{(i)}_p}{\partial z_p}\\
\end{array}\right].
\end{equation}
It can be clearly seen that Equation~\ref{matrix_form_equation_derivative} implies: 
\begin{equation*}
\boldsymbol{U} = -[\nabla^2f_i]^{-1}.
\end{equation*}

Hence, using $||\boldsymbol{U}||_2 \le \frac{1}{\gamma}$ we have for each entry of $\boldsymbol{U}$ we have:
\begin{align*}
&|U_{ij}| \le ||\boldsymbol{U}||_{F} \le \sqrt{p}||\boldsymbol{U}||_2 \le \frac{\sqrt{p}}{\gamma}.
\end{align*}
The above finalizes the proof of the Lemma.

\end{proof}

\section{Proof Dual Function Properties}
\begin{lemma}\label{Lemma:Props}
The dual function $q(\bm{\lambda})= q(\bm{\lambda}_{1}, \dots, \bm{\lambda}_{p})$ shares the following characteristics: 
\begin{itemize}
\item The dual Hessian $\bm{H}(\bm{\lambda})$ and gradient $\nabla q(\bm{\lambda})$ are given by:
\begin{equation*}
\bm{H}(\bm{\lambda})= - \bm{M} \left(\nabla^{2} f(\bm{y}(\bm{\lambda})), \right)^{-1}\bm{M}, \ \ \ \ \ \  \nabla q(\bm{\lambda}) = \bm{M} \bm{y}(\bm{\lambda}). 
\end{equation*}
\item The dual Hessian is Lipschitz continuous with respect to $\bm{M}$-weighted norm, where for any $\tilde{\bm{\lambda}}$ and $\bm{\lambda}$: $\left|\left|\bm{H}(\tilde{\bm{\lambda}})- \bm{H}\left({\bm{\lambda}}\right)\right|\right|_{\bm{M}} \leq B \left|\left|\tilde{\bm{\lambda}} - \bm{\lambda}\right|\right|_{\bm{M}}$, with $B=\frac{\delta {p}}{\gamma} \mu_{n}^{2}(\mathcal{L})\sqrt{\mu_{n}(\mathcal{L})}$, where $\mu_{n}(\mathcal{L})$ is the largest eigenvalue of $\mathcal{L}$ and the constants $\gamma$ and $\delta$ are these given in~\ref{Ass:Two}. 
\end{itemize}
\end{lemma}

\begin{proof}
Consider each part separately:
\begin{enumerate}
\item Recall that $\bm{y(\lambda)}$ minimizes the Lagrangian, given by  
$$\boldsymbol{y(\lambda)} = \bm{y}^{+} = \arg\min_{\boldsymbol{y}} f(\boldsymbol{y}) + \boldsymbol{\lambda}^{\mathsf{T}}\boldsymbol{My}.$$ Denote by
\begin{align*}\label{matrix_A}
\centering
&\boldsymbol{M} = \left[\begin{array}{cccc}
m_{11}&m_{12}&\cdots &m_{1np}\\
m_{21}&m_{22}&\cdots &m_{2np}\\
\vdots & &\ddots &\vdots \\
m_{np1}&m_{np2}&\cdots &m_{npnp}\\
\end{array}\right],\hspace{2mm}
\boldsymbol{y}^{+} = \left[\begin{array}{c}
y^{+}_1(\boldsymbol{\lambda})\\
y^{+}_2(\boldsymbol{\lambda})\\
\vdots \\
y^{+}_{np}(\boldsymbol{\lambda})\\
\end{array}\right],\hspace{2mm}
\nabla f(\boldsymbol{y}^+) = \left[\begin{array}{c}
z_1(\boldsymbol{y}^{+})\\
z_2(\boldsymbol{y}^{+})\\
\vdots \\
z_{np}(\boldsymbol{y}^{+})\\
\end{array}\right]. 
\end{align*}

Using conjugate $f^*()$, the dual function can be written as:
\begin{equation*}
q(\boldsymbol{\lambda}) =  - f^{*}(-\boldsymbol{M}\boldsymbol{\lambda}).
\end{equation*}
hence, the dual gradient is given by:
\begin{align}\label{eq_1}
\nabla q(\boldsymbol{\lambda}) = - \nabla f^{*}(-\boldsymbol{M}\boldsymbol{\lambda}).
\end{align} 

Notice that:
\begin{equation}\label{eq_2}
\nabla f(\boldsymbol{y(\lambda)}) + \boldsymbol{M}\boldsymbol{\lambda} = \boldsymbol{0}.
\end{equation}

Denote $\boldsymbol{u} = -\boldsymbol{M}\boldsymbol{\lambda}$, then the $k^{th}$ component of vector $\nabla f^{*}(-\boldsymbol{M}\boldsymbol{\lambda})$ is given by:
\begin{equation*}
[\nabla f^{*}(-\boldsymbol{M}\boldsymbol{\lambda})]_{k} =  \sum_{j=1}^{np} \frac{\partial f^*}{\partial u_j}\frac{\partial u_j}{\partial \lambda_k} = -\left[\begin{array}{cccc}
m_{k1}&m_{k2}&\cdots &m_{knp}\\
\end{array}\right]\left[\begin{array}{c}
\frac{\partial f^*}{\partial u_1}\\
\frac{\partial f^*}{\partial u_2}\\
\vdots \\
\frac{\partial f^*}{\partial u_{np}}\\
\end{array}\right]_{-\boldsymbol{M}\boldsymbol{\lambda}}.
\end{equation*}
Hence, using $(\ref{eq_2})$ and the relation between gradients of a function and its conjugate in the expression for vector $\nabla f^{*}(-\boldsymbol{M}\boldsymbol{\lambda})$ gives:
\begin{align*}
\nabla f^{*}(-\boldsymbol{M}\boldsymbol{\lambda}) &= - \boldsymbol{M}\nabla_{\boldsymbol{u}}f^*(\boldsymbol{u})|_{-\boldsymbol{M}\boldsymbol{\lambda}} = - \boldsymbol{M}\nabla_{\boldsymbol{u}}f^*(-\boldsymbol{M}\boldsymbol{\lambda}) \\\nonumber
&=-\boldsymbol{M}\nabla_{\boldsymbol{u}}f^*(\nabla f(\boldsymbol{y(\lambda)})) = -\boldsymbol{M}\boldsymbol{y}(\boldsymbol{\lambda}).
\end{align*}
Applying this result in (\ref{eq_1}) gives:
\begin{align}\label{eq_3}
&\nabla q(\boldsymbol{\lambda}) =  \boldsymbol{M}\boldsymbol{y}(\boldsymbol{\lambda}).
\end{align}

From (\ref{eq_3}), the dual Hessian is given by:
\begin{align}\label{eq_4}
&\nabla^2q(\boldsymbol{\lambda}) = \boldsymbol{M}\underbrace{\left[\begin{array}{cccc}
\frac{\partial y^{+}_1(\boldsymbol{\lambda})}{\partial \lambda_1}&\frac{\partial y^{+}_1(\boldsymbol{\lambda})}{\partial \lambda_2}&\cdots &\frac{\partial y^{+}_1(\boldsymbol{\lambda})}{\partial \lambda_{np}}\\
\frac{\partial y^{+}_2(\boldsymbol{\lambda})}{\partial \lambda_1}&\frac{\partial y^{+}_2(\boldsymbol{\lambda})}{\partial \lambda_2}&\cdots &\frac{\partial y^{+}_2(\boldsymbol{\lambda})}{\partial \lambda_{np}}\\
\vdots & &\ddots &\vdots \\
\frac{\partial y^{+}_{np}(\boldsymbol{\lambda})}{\partial \lambda_1}&\frac{\partial y^{+}_{np}(\boldsymbol{\lambda})}{\partial \lambda_2}&\cdots &\frac{\partial y^{+}_{np}(\boldsymbol{\lambda})}{\partial \lambda_{np}}\\
\end{array}\right]}_{\boldsymbol{F}(\boldsymbol{y}^+)}. 
\end{align}
In the next step we target matrix $\boldsymbol{F}(\boldsymbol{y}^+)$. Using (\ref{eq_2}):
\begin{equation*}
\nabla f(\boldsymbol{y}^+) = - \boldsymbol{M}\boldsymbol{\lambda}.
\end{equation*} 
Taking the partial derivative $\frac{\partial }{\partial \lambda_1}$ from the both sides of the above equation gives the following for the left and right hand sides.\\
\textbf{Left hand side:}
\begin{align*}
\frac{\partial}{\partial \lambda_1}\nabla f(\boldsymbol{y}^+) &= \left[\begin{array}{c}
\frac{\partial }{\partial \lambda_1}z_1(\boldsymbol{y}^{+}) \\
\frac{\partial}{\partial \lambda_1} z_2(\boldsymbol{y}^{+})\\
\vdots \\
\frac{\partial}{\partial \lambda_1} z_{np}(\boldsymbol{y}^{+})\\ 
\end{array}\right] \\\nonumber
&=\left[\begin{array}{c}
\frac{\partial z_1(\boldsymbol{y}^+)}{\partial y^+_1(\boldsymbol{\lambda})}\frac{\partial y^+_1(\boldsymbol{\lambda})}{\partial \lambda_1} + \frac{\partial z_1(\boldsymbol{y}^+)}{\partial y^+_2(\boldsymbol{\lambda})}\frac{\partial y^+_2(\boldsymbol{\lambda})}{\partial \lambda_1} + \cdots + \frac{\partial z_1(\boldsymbol{y}^+)}{\partial y^+_{np}(\boldsymbol{\lambda})}\frac{\partial y^+_{np}(\boldsymbol{\lambda})}{\partial \lambda_1}\\
\frac{\partial z_2(\boldsymbol{y}^+)}{\partial y^+_1(\boldsymbol{\lambda})}\frac{\partial y^+_1(\boldsymbol{\lambda})}{\partial \lambda_1} + \frac{\partial z_2(\boldsymbol{y}^+)}{\partial y^+_2(\boldsymbol{\lambda})}\frac{\partial y^+_2(\boldsymbol{\lambda})}{\partial \lambda_1} + \cdots + \frac{\partial z_2(\boldsymbol{y}^+)}{\partial y^+_{np}(\boldsymbol{\lambda})}\frac{\partial y^+_{np}(\boldsymbol{\lambda})}{\partial \lambda_1}\\
\vdots\\
\frac{\partial z_{np}(\boldsymbol{y}^+)}{\partial y^+_1(\boldsymbol{\lambda})}\frac{\partial y^+_1(\boldsymbol{\lambda})}{\partial \lambda_1} + \frac{\partial z_{np}(\boldsymbol{y}^+)}{\partial y^+_2(\boldsymbol{\lambda})}\frac{\partial y^+_2(\boldsymbol{\lambda})}{\partial \lambda_1} + \cdots + \frac{\partial z_{np}(\boldsymbol{y}^+)}{\partial y^+_{np}(\boldsymbol{\lambda})}\frac{\partial y^+_{np}(\boldsymbol{\lambda})}{\partial \lambda_1}\\
\end{array}\right]  \\\nonumber
&=\underbrace{\left[\begin{array}{cccc}
\frac{\partial z_1(\boldsymbol{y}^+)}{\partial y^+_1(\boldsymbol{\lambda})}&\frac{\partial z_1(\boldsymbol{y}^+)}{\partial y^+_2(\boldsymbol{\lambda})}&\cdots &\frac{\partial z_1(\boldsymbol{y}^+)}{\partial y^+_{np}(\boldsymbol{\lambda})}\\
\frac{\partial z_2(\boldsymbol{y}^+)}{\partial y^+_1(\boldsymbol{\lambda})}&\frac{\partial z_2(\boldsymbol{y}^+)}{\partial y^+_2(\boldsymbol{\lambda})}&\cdots &\frac{\partial z_2(\boldsymbol{y}^+)}{\partial y^+_{np}(\boldsymbol{\lambda})}\\
\vdots & &\ddots &\vdots \\
\frac{\partial z_{np}(\boldsymbol{y}^+)}{\partial y^+_1(\boldsymbol{\lambda})}&\frac{\partial z_{np}(\boldsymbol{y}^+)}{\partial y^+_2(\boldsymbol{\lambda})}&\cdots &\frac{\partial z_{np}(\boldsymbol{y}^+)}{\partial y^+_{np}(\boldsymbol{\lambda})}\\
\end{array}\right]}_{\nabla^2 f(\boldsymbol{y}^+)}\left[\begin{array}{c}
\frac{\partial y^+_1(\boldsymbol{\lambda})}{\partial \lambda_1} \\
\frac{\partial y^+_2(\boldsymbol{\lambda})}{\partial \lambda_1}\\
\vdots \\
\frac{\partial y^+_{np}(\boldsymbol{\lambda})}{\partial \lambda_1}\\ 
\end{array}\right] = \nabla^2 f(\boldsymbol{y}^+)\left[\begin{array}{c}
\frac{\partial y^+_1(\boldsymbol{\lambda})}{\partial \lambda_1} \\
\frac{\partial y^+_2(\boldsymbol{\lambda})}{\partial \lambda_1}\\
\vdots \\
\frac{\partial y^+_{np}(\boldsymbol{\lambda})}{\partial \lambda_1}\\ 
\end{array}\right].
\end{align*} 
\textbf{Right hand side:}
\begin{align*}
 &\frac{\partial}{\partial \lambda_1}(-\boldsymbol{M}\boldsymbol{\lambda}) = - \left[\begin{array}{c}
m_{11} \\
m_{21}\\
\vdots \\
m_{np1}\\ 
\end{array}\right] \implies \nabla^2 f(\boldsymbol{y}^+)\left[\begin{array}{c}
\frac{\partial y^+_1(\boldsymbol{\lambda})}{\partial \lambda_1} \\
\frac{\partial y^+_2(\boldsymbol{\lambda})}{\partial \lambda_1}\\
\vdots \\
\frac{\partial y^+_{np}(\boldsymbol{\lambda})}{\partial \lambda_1}\\ 
\end{array}\right] = - \left[\begin{array}{c}
m_{11} \\
m_{21}\\
\vdots \\
m_{np1}\\ 
\end{array}\right].
\end{align*}

Repeating the same step for partial derivatives $\frac{\partial }{\partial \lambda_2}, \ldots, \frac{\partial }{\partial \lambda_{np}}$ gives:
\begin{align*}
&\nabla^2f(\boldsymbol{y}^+) \underbrace{\left[\begin{array}{cccc}
\frac{\partial y^+_1(\boldsymbol{\lambda})}{\partial \lambda_1}&\frac{\partial y^+_1(\boldsymbol{\lambda})}{\partial \lambda_2}&\cdots &\frac{\partial y^+_1(\boldsymbol{\lambda})}{\partial \lambda_{np}}\\
\frac{\partial y^+_2(\boldsymbol{\lambda})}{\partial \lambda_1}&\frac{\partial y^+_2(\boldsymbol{\lambda})}{\partial \lambda_2}&\cdots &\frac{\partial y^+_2(\boldsymbol{\lambda})}{\partial \lambda_{np}}\\
\vdots & &\ddots &\vdots \\
\frac{\partial y^+_{np}(\boldsymbol{\lambda})}{\partial \lambda_1}&\frac{\partial y^+_{np}(\boldsymbol{\lambda})}{\partial \lambda_2}&\cdots &\frac{\partial y^+_{np}(\boldsymbol{\lambda})}{\partial \lambda_{np}}\\
\end{array}\right]}_{\boldsymbol{F}(\boldsymbol{y}^+)} = -  \underbrace{\left[\begin{array}{cccc}
m_{11}&m_{12}&\cdots &m_{1np}\\
m_{21}&m_{22}&\cdots &m_{2np}\\
\vdots & &\ddots &\vdots \\
m_{np1}&m_{np2}&\cdots &m_{npnp}\\
\end{array}\right]}_{\boldsymbol{M}}.
\end{align*}
Hence:
\begin{equation*}
\boldsymbol{F}(\boldsymbol{y}^+) = -[\nabla^2 f(\boldsymbol{y}^+)]^{-1}\boldsymbol{M}.
\end{equation*}
Finally, combining this result with (\ref{eq_4}) gives:
\begin{equation*}
\bm{H(\lambda)} = \nabla^2q(\boldsymbol{\lambda}) =  -\boldsymbol{M}[\nabla^2 f(\boldsymbol{y(\lambda)})]^{-1}\boldsymbol{M}.
\end{equation*}

\item To commence, we consider each of the above statements separately. Noting that the proof of the first statement can be found in~\cite{zargham2014accelerated}, we begin with proving the second statement. 
\begin{claim}
The dual Hessian is Lipschitz continuous with respect to $\bm{M}$-weighted norm, where for any $\tilde{\bm{\lambda}}$ and $\bm{\lambda}$: 
\begin{equation*}
\left|\left|\bm{H}(\tilde{\bm{\lambda}})- \bm{H}\left({\bm{\lambda}}\right)\right|\right|_{\bm{M}} \leq B \left|\left|\tilde{\bm{\lambda}} - \bm{\lambda}\right|\right|_{\bm{M}},
\end{equation*}
with $B=\frac{\delta p}{\gamma} \mu_{n}^{2}(\mathcal{L})\sqrt{\mu_{n}(\mathcal{L})}$, where $\mu_{n}(\mathcal{L})$ is the largest eigenvalue of $\mathcal{L}$ and the constants $\gamma$ and $\delta$ are these given in~\ref{Ass:Two}. 
\end{claim}
\begin{proof}
We first remind the reader that weighted norm of a vector $\bm{v}$ and matrix $\bm{A}$ are given by: 
\begin{equation*}
||\bm{v}||_{\bm{M}} = \sqrt{\bm{v}^{\mathsf{T}}\bm{M} \bm{v}}, \ \ \ \ \ \ \ \ \text{and} \ \ \ \ \ \ \ \ \ ||\bm{A}||_{\bm{M}} = \sup_{\bm{v} : \bm{v}^{\mathsf{T}}\bm{M}\bm{v} \neq 0}\frac{||\bm{A}\bm{v}||_{\bm{M}}}{||\bm{v}||_{\bm{M}}}. 
\end{equation*}
Fixing a vector $\bm{v}$ such that $||\bm{v}||_{\bm{M}} \neq 0$, then we have:
\begin{align*}
\left|\left|[\boldsymbol{H}(\bar{\boldsymbol{\lambda}}) - \boldsymbol{H}(\boldsymbol{\lambda})]\boldsymbol{v}\right|\right|^2_{\boldsymbol{M}} &= \left|\left|[\boldsymbol{M}([\nabla^2f(\boldsymbol{y}(\bar{\boldsymbol{\lambda}}))]^{-1} - [\nabla^2f(\boldsymbol{y}(\boldsymbol{\lambda}))]^{-1} )\boldsymbol{M}\boldsymbol{v}]\right|\right|^2_{\boldsymbol{L}} \\
&=\boldsymbol{v}^{\mathsf{T}}\boldsymbol{M}([\nabla^2f(\boldsymbol{y}(\bar{\boldsymbol{\lambda}}))]^{-1} - [\nabla^2f(\boldsymbol{y}(\boldsymbol{\lambda}))]^{-1})\boldsymbol{M^3}([\nabla^2f(\boldsymbol{y}(\bar{\boldsymbol{\lambda}}))]^{-1} - [\nabla^2f(\boldsymbol{y}(\boldsymbol{\lambda}))]^{-1})\boldsymbol{Mv} \\\nonumber
&\le^{(1)} \mu^3_n(\boldsymbol{\mathcal{L}})\boldsymbol{v}^{\mathsf{T}}\boldsymbol{M}([\nabla^2f(\boldsymbol{y}(\bar{\boldsymbol{\lambda}}))]^{-1} - [\nabla^2f(\boldsymbol{y}(\boldsymbol{\lambda}))]^{-1})^2\boldsymbol{Mv}  \\\nonumber
&\le \mu^3_n(\boldsymbol{\mathcal{L}})\mu^2_{\max}(|[\nabla^2f(\boldsymbol{y}(\bar{\boldsymbol{\lambda}}))]^{-1} - [\nabla^2f(\boldsymbol{y}(\boldsymbol{\lambda}))]^{-1}|)\boldsymbol{v}^{\mathsf{T}}\boldsymbol{M}^2\boldsymbol{v}  \\\nonumber
&\le \mu^4_n(\boldsymbol{\mathcal{L}})\mu^2_{\max}(|[\nabla^2f(\boldsymbol{y}(\bar{\boldsymbol{\lambda}}))]^{-1} - [\nabla^2f(\boldsymbol{y}(\boldsymbol{\lambda}))]^{-1}|)||\boldsymbol{v}||^2_{\boldsymbol{M}}.
\end{align*}
Hence, we can immediately write:
\begin{equation}\label{norm_difference}
||\boldsymbol{H}(\bar{\boldsymbol{\lambda}}) - \boldsymbol{H}(\boldsymbol{\lambda})||_{\boldsymbol{M}} \le \mu^2_n(\boldsymbol{\mathcal{L}})\mu_{\max}(|[\nabla^2f(\boldsymbol{y}(\bar{\boldsymbol{\lambda}}))]^{-1} - [\nabla^2f(\boldsymbol{y}(\boldsymbol{\lambda}))]^{-1}|).
\end{equation}
The next step is to upper bound $\mu_{\max}(|[\nabla^2f(\boldsymbol{y}(\bar{\boldsymbol{\lambda}}))]^{-1} - [\nabla^2f(\boldsymbol{y}(\boldsymbol{\lambda}))]^{-1}|)$. For this purpose, we next study the properties of primal Hessian in details and recognize: 
 
 \begin{claim}
The primal Hessian $\nabla^2f(\boldsymbol{y}(\boldsymbol{\lambda}))$ shares the following properties: 
 \begin{equation}\label{primal_hessian_1}
\gamma \preceq \nabla^2f(\boldsymbol{y}(\boldsymbol{\lambda})) \preceq \Gamma,
\end{equation}
and 
\begin{align}\label{primal_hessian_2}
\mu_{\max}(|[\nabla^2f(\boldsymbol{y}(\bar{\boldsymbol{\lambda}}))]^{-1} - [\nabla^2f(\boldsymbol{y}(\boldsymbol{\lambda}))]^{-1}|) \le \delta\max_{i\in\mathbb{V}}\sqrt{\sum_{k=1}^{p}\left(y_k(i)(\bar{\boldsymbol{\lambda}}) - y_k(i)(\boldsymbol{\lambda})  \right)^2},
\end{align}
for any $\bar{\boldsymbol{\lambda}},\boldsymbol{\lambda}\in\mathbb{R}^p$.
\end{claim}
\begin{proof}
Firstly, notice that for any $j\ne i$ and any $r = 1\ldots,p$, we have:
\begin{equation*}
\frac{\partial^2f}{\partial y_1(i)\partial y_r(j)} = \frac{\partial^2f}{\partial y_2(i)\partial y_r(j)} = \ldots = \frac{\partial^2f}{\partial  y_p(i)\partial y_r(j)} = 0
\end{equation*}
Hence, the sparsity pattern of the primal hessian allows the symmetric reordering of  rows and columns that transform $\nabla^2f(\boldsymbol{\boldsymbol{\lambda}})$ into the block diagonal matrix as:
\begin{equation*}
\boldsymbol{W}(\boldsymbol{\lambda}) = \left[\begin{array}{cccc}
\nabla^2f_1(\boldsymbol{\lambda})&\boldsymbol{0}&\cdots &\boldsymbol{0}\\
\boldsymbol{0}&\nabla^2f_2(\boldsymbol{\lambda})&\cdots &\boldsymbol{0}\\
\vdots & &\ddots &\vdots \\
\boldsymbol{0}&\boldsymbol{0}&\cdots &\nabla^2f_p(\boldsymbol{\lambda})\\
\end{array}\right]
\end{equation*}
Note that $\boldsymbol{W}(\boldsymbol{\lambda})$ preserves the important properties of $\nabla^2f(\boldsymbol{\boldsymbol{\lambda}})$. Particularly, the spectrum of these two matrices are the same. That can be easily seen by considering a matrix $\boldsymbol{A}$ and letting $\boldsymbol{T}_{ij}$ to be the operator that swaps $i^{th}$ and $j^{th}$ rows of $\boldsymbol{A}$. Now, consider the matrix $\bar{\boldsymbol{A}}$ resultant of the swapping of the $i^{th}$ and $j^{th}$ row and column of $\boldsymbol{A}$. Then, $\bar{\boldsymbol{A}} = \boldsymbol{T}_{ij}\boldsymbol{A}\boldsymbol{T}_{ij}$, and
\begin{align*}
&\text{det}(\bar{\boldsymbol{A}} - \mu\boldsymbol{I}) = \text{det}(\boldsymbol{T}_{ij}\boldsymbol{A}\boldsymbol{T}_{ij} - \mu\boldsymbol{I}) = \text{det}(\boldsymbol{T}_{ij}(\boldsymbol{A} - \mu\boldsymbol{I})\boldsymbol{T}_{ij} ) =\text{det}(\boldsymbol{A} - \mu\boldsymbol{I})det(\boldsymbol{T}^2_{ij}) = \text{det}(\boldsymbol{A} - \mu\boldsymbol{I}),
\end{align*}
where in the last step, we used the fact that $\boldsymbol{T}^2_{ij} = \boldsymbol{I}$. Since $\boldsymbol{W}(\boldsymbol{\lambda})$ is constructed from $\nabla^2f(\boldsymbol{y}(\boldsymbol{\lambda}))$ by a symmetric reordering of rows and columns, we deduce that $\text{Spectrum}(\boldsymbol{W}(\boldsymbol{\lambda})) = \text{Spectrum}(\nabla^2f(\boldsymbol{y}(\boldsymbol{\lambda})))$. Therefore, we can write: 
\begin{equation*}
\gamma \preceq \boldsymbol{W}(\boldsymbol{\lambda}) \preceq \Gamma,
\end{equation*}
which implies the property in Equation~\ref{primal_hessian_1}. To prove the property in Equation~\ref{primal_hessian_2}, we notice that if $\bar{\boldsymbol{A}} = \boldsymbol{T}_{ij}\boldsymbol{A}\boldsymbol{T}_{ij}$ and $\boldsymbol{A}$ is invertible, then so is $\bar{\boldsymbol{A}}$ and:
\begin{align*}
\text{det}(\bar{\boldsymbol{A}}^{-1} - \mu\boldsymbol{I}) = det(\boldsymbol{T}^{-1}_{ij}\boldsymbol{A}^{-1}\boldsymbol{T}^{-1}_{ij} - \mu\boldsymbol{I}) = \text{det}(\boldsymbol{T}_{ij}(\boldsymbol{A}^{-1} - \mu\boldsymbol{I})\boldsymbol{T}_{ij} ) = 
&\text{det}(\boldsymbol{A}^{-1} - \mu\boldsymbol{I}),
\end{align*}
where we used the fact that $\boldsymbol{T}^{-1}_{ij} = \boldsymbol{T}_{ij}$. Let us denote $\{\boldsymbol{T}_1, \ldots, \boldsymbol{T}_l\}$ to be a collection of operators that swap the rows of $\nabla^2f(\boldsymbol{y}(\boldsymbol{\lambda}))$ to transforming it to $\boldsymbol{W}(\boldsymbol{\lambda})$, i.e.:
\begin{equation*}
\boldsymbol{W}(\boldsymbol{\lambda}) = \boldsymbol{T}_1\cdots\boldsymbol{T}_l\nabla^2f(\boldsymbol{y}(\boldsymbol{\lambda}))\boldsymbol{T}_l\cdots\boldsymbol{T}_1.
\end{equation*}
Then, $[\nabla^2f(\boldsymbol{y}(\boldsymbol{\lambda}))]^{-1} = \boldsymbol{T}_l\cdots\boldsymbol{T}_1\boldsymbol{W}^{-1}(\boldsymbol{\lambda})\boldsymbol{T}_1\cdots\boldsymbol{T}_l$, and: 
\begin{align*}
\mu_{\max}(||[\nabla^2f(\boldsymbol{y}(\bar{\boldsymbol{\lambda}}))]^{-1} - [\nabla^2f(\boldsymbol{y}(\boldsymbol{\lambda}))]^{-1}||) &= \mu_{\max}(\boldsymbol{T}_l\cdots\boldsymbol{T}_1|\boldsymbol{W}^{-1}(\bar{\boldsymbol{\lambda}}) - \boldsymbol{W}^{-1}(\boldsymbol{\lambda})|\boldsymbol{T}_1\cdots\boldsymbol{T}_l)  \\
&\le \mu_{\max}(|\boldsymbol{W}^{-1}(\bar{\boldsymbol{\lambda}}) - \boldsymbol{W}^{-1}(\boldsymbol{\lambda})|)  \\\nonumber
&\le \max_{i\in\mathbb{V}}\mu_{\max}(|[\nabla^2f_i(y_1(i)(\bar{\boldsymbol{\lambda}}), \ldots y_p(i)(\bar{\boldsymbol{\lambda}}))]^{-1} \\
& \hspace{10em}- [\nabla^2f_i(y_1(i)(\boldsymbol{\lambda}), \ldots y_p(i)(\boldsymbol{\lambda}))]^{-1}|) \\\nonumber
&\hspace{-5em}=\max_{i\in\mathbb{V}}||[\nabla^2f_i(y_1(i)(\bar{\boldsymbol{\lambda}}), \ldots y_p(i)(\bar{\boldsymbol{\lambda}}))]^{-1} - [\nabla^2f_i(y_1(i)(\boldsymbol{\lambda}), \ldots y_p(i)(\boldsymbol{\lambda}))]^{-1}||_{2}\ \\\nonumber
&\le\delta\max_{i\in\mathbb{V}}||(y_1(i)(\bar{\boldsymbol{\lambda}}), \ldots y_p(i)(\bar{\boldsymbol{\lambda}})) - (y_1(i)(\boldsymbol{\lambda}), \ldots y_p(i)(\boldsymbol{\lambda}))||_{2}  \\\nonumber
&=\delta\max_{i\in\mathbb{V}}\sqrt{\sum_{k=1}^{p}\left(y_k(i)(\bar{\boldsymbol{\lambda}}) - y_k(i)(\boldsymbol{\lambda})  \right)^2}.
\end{align*}
The above proves the property in Equation~\ref{primal_hessian_2}.
\end{proof}
 Now, consider the term $\left(y_k(i)(\bar{\boldsymbol{\lambda}}) - y_k(i)(\boldsymbol{\lambda})  \right)$. Using the result of Lipschitzness on the solution of the partial differential equations, we can write:
\begin{align*}
\left|y_k(i)(\bar{\boldsymbol{\lambda}}) - y_k(i)(\boldsymbol{\lambda})\right| &= |\phi^{(i)}_k((\boldsymbol{\mathcal{L}}\bar{\boldsymbol{\lambda}}_1)_i, \ldots, (\boldsymbol{\mathcal{L}}\bar{\boldsymbol{\lambda}}_p)_i) - \phi^{i}_k((\boldsymbol{\mathcal{L}\lambda}_1)_i, \ldots, (\boldsymbol{\mathcal{L}\lambda}_p)_i)| \\\nonumber
&\le\frac{\sqrt{p}}{\gamma}\sqrt{\sum_{r=1}^{p}\left((\boldsymbol{\mathcal{L}}\bar{\boldsymbol{\lambda}}_r)_i - (\boldsymbol{\mathcal{L}\lambda}_r)_i  \right)^2}  = \frac{\sqrt{p}}{\gamma}\sqrt{\sum_{r=1}^{p}\left(\boldsymbol{\mathcal{L}}(\bar{\boldsymbol{\lambda}}_r - \boldsymbol{\lambda})\right)^2_i } \le \frac{\sqrt{p}}{\gamma}\sqrt{\sum_{r=1}^{p}||\boldsymbol{\mathcal{L}}(\bar{\boldsymbol{\lambda}}_r - \boldsymbol{\lambda})||^2_2 }  \\\nonumber
&=\frac{\sqrt{p}}{\gamma}\sqrt{\sum_{r=1}^{p}(\bar{\boldsymbol{\lambda}}_r - \boldsymbol{\lambda}_r)^\mathsf{T}\boldsymbol{\mathcal{L}}^2(\bar{\boldsymbol{\lambda}}_r - \boldsymbol{\lambda}_r)} \le \frac{\sqrt{p}}{\gamma}\sqrt{\mu_n(\boldsymbol{\mathcal{L}})\sum_{r=1}^{p}(\bar{\boldsymbol{\lambda}}_r - \boldsymbol{\lambda}_r)^\mathsf{T}\boldsymbol{\mathcal{L}}(\bar{\boldsymbol{\lambda}}_r - \boldsymbol{\lambda}_r)} \\\nonumber
&=\sqrt{\mu_n(\boldsymbol{\mathcal{L}})}\frac{\sqrt{p}}{\gamma}||\bar{\boldsymbol{\lambda}} - \boldsymbol{\lambda}||_{\boldsymbol{M}}.
\end{align*}
In the last step, we used the fact that: 
 \begin{equation*}
\sum_{r=1}^{p}(\bar{\boldsymbol{\lambda}}_r - \boldsymbol{\lambda}_r)^{\mathsf{T}}\boldsymbol{\mathcal{L}}(\bar{\boldsymbol{\lambda}}_r - \boldsymbol{\lambda}_r) = (\bar{\boldsymbol{\lambda}} - \boldsymbol{\lambda})^{\mathsf{T}}\boldsymbol{M}(\bar{\boldsymbol{\lambda}} - \boldsymbol{\lambda}) = ||\bar{\boldsymbol{\lambda}} - \boldsymbol{\lambda}||^2_{\boldsymbol{M}}. 
\end{equation*}
Hence: 
\begin{equation*}
\left(y_k(i)(\bar{\boldsymbol{\lambda}}) - y_k(i)(\boldsymbol{\lambda})\right)^2 \le \mu_n(\boldsymbol{\mathcal{L}})\frac{p}{\gamma^2}||\bar{\boldsymbol{\lambda}} - \boldsymbol{\lambda}||^2_{\boldsymbol{M}}.
\end{equation*}
Combining this result with that from Equation~\ref{primal_hessian_2} gives:
\begin{equation*}
\mu_{max}(||[\nabla^2f(\boldsymbol{y}(\bar{\boldsymbol{\lambda}}))]^{-1} - [\nabla^2f(\boldsymbol{y}(\boldsymbol{\lambda}))]^{-1}||) \le\delta\sqrt{\mu_n(\boldsymbol{\mathcal{L}})}\frac{p}{\gamma}||\bar{\boldsymbol{\lambda}} - \boldsymbol{\lambda}||_{\boldsymbol{M}}.
\end{equation*}
Applying the previous equation to that in Equation~\ref{norm_difference} gives: 
\begin{equation*}
||\boldsymbol{H}(\bar{\boldsymbol{\lambda}}) - \boldsymbol{H}(\boldsymbol{\lambda})||_{\boldsymbol{M}} \le \frac{\delta p}{\gamma}\mu^2_n(\boldsymbol{\mathcal{L}})\sqrt{\mu_n(\boldsymbol{\mathcal{L}})}||\bar{\boldsymbol{\lambda}} - \boldsymbol{\lambda}||_{\boldsymbol{M}} = B||\bar{\boldsymbol{\lambda}} - \boldsymbol{\lambda}||_{\boldsymbol{M}}.
\end{equation*}
\end{proof}
The above finalizes the statement of the claim and consequently that of the lemma. 
\end{enumerate}

\end{proof}


\section{Proof Approximation Accuracy}

\begin{lemma}
Let $\tilde{\bm{d}}_{1}^{[k]}, \dots, \tilde{\bm{d}}_{p}^{[k]}$ be the $\epsilon_{0}$-approximate solution to Equation~\ref{Eq:Seventeen}, then $\tilde{\bm{d}}^{[k]}$ is an $\epsilon$-approximate solution to~\ref{Eq:NewtonSystem} with $\epsilon = \epsilon_{0} \sqrt{\frac{\Gamma}{\gamma} \frac{\mu_{n}(\mathcal{L})}{\mu_{2}(\mathcal{L})}}\left[1 + \epsilon_{0} \frac{\mu_{n}(\mathcal{L})}{\mu_{2}(\mathcal{L})}\sqrt{\frac{\Gamma}{\gamma}} + \sqrt{\frac{\mu_{n}(\mathcal{L})}{\mu_{2}(\mathcal{L})}}\right]$.
\end{lemma} 
\begin{proof}
Let $\boldsymbol{\tilde{z}}^{[k]}_1, \ldots,\boldsymbol{\tilde{z}}^{[k]}_p $  be $\epsilon_0$ approximate solutions of the first collection of systems in Equation (9), and let $\boldsymbol{\tilde{b}}^{[k]} = [\nabla^2f(\boldsymbol{y}^{[k]})]^{-1}\boldsymbol{\tilde{z}}^{[k]}$. Denote vectors $\boldsymbol{\hat{d}}^{[k]}_1, \ldots, \boldsymbol{\hat{d}}^{[k]}_p$ be the exact solution of:
\begin{equation*}
\begin{cases} \boldsymbol{\mathcal{L}d}^{[k]}_1 = \boldsymbol{\tilde{b}}^{[k]}_1 \\ 
\boldsymbol{\mathcal{L}d}^{[k]}_2 = \boldsymbol{\tilde{b}}^{[k]}_2 \\
\vdots \\
\boldsymbol{\mathcal{L}d}^{[k]}_p = \boldsymbol{\tilde{b}}^{[k]}_p.
\end{cases} 
\end{equation*}
We next denote the exact solution of the system in Equation~8 as:
\begin{equation}\label{d_exact}
\boldsymbol{d}^{*[k]} = \left[
\begin{array}{c}
\boldsymbol{d}^{*[k]}_1\\
\boldsymbol{d}^{*[k]}_2\\
\vdots\\
\boldsymbol{d}^{*[k]}_p
\end{array}
\right].
\end{equation}
Second, we let $\boldsymbol{z}^{*[k]} = \left[\nabla^2f(\boldsymbol{y}^{[k]})\right]^{-1}\boldsymbol{M}\boldsymbol{d}^{*[k]} = [(\boldsymbol{z}^{*[k]}_1)^{\mathsf{T}}, \ldots, (\boldsymbol{z}^{*[k]}_p)^{\mathsf{T}}]^{\mathsf{T}}$ be the corresponding vector $\boldsymbol{z}^{[k]}$. Hence: 
\begin{equation}\label{exact_z_decomposition}
\begin{cases} \boldsymbol{\mathcal{L}z}^{*[k]}_1 =  \boldsymbol{\mathcal{L}}\boldsymbol{y}^{[k]}_1 \\ 
\boldsymbol{\mathcal{L}z}^{*[k]}_2 =  \boldsymbol{\mathcal{L}}\boldsymbol{y}^{[k]}_2 \\
\vdots \\
\boldsymbol{\mathcal{L}z}^{*[k]}_p =  \boldsymbol{\mathcal{L}}\boldsymbol{y}^{[k]}_p. 
\end{cases}
\end{equation} 

Now, let $\boldsymbol{\tilde{z}}^{[k]}_1, \ldots, \boldsymbol{\tilde{z}}^{[k]}_p$ be the $\epsilon_0$-approximate solutions of (\ref{exact_z_decomposition}), and $\boldsymbol{\hat{d}}^{[k]}_1, \ldots, \boldsymbol{\hat{d}}^{[k]}_p$ and $\tilde{\boldsymbol{d}}^{[k]}_1, \ldots, \tilde{\boldsymbol{d}}^{[k]}_p$ be the exact and $\epsilon_0-$approximate solutions of the following systems:

\begin{equation}\label{approx_decomposition}
\begin{cases} 
\boldsymbol{\mathcal{L}d}^{[k]}_1 = \boldsymbol{\tilde{b}}^{[k]}_1 \\ 
\boldsymbol{\mathcal{L}d}^{[k]}_2 = \boldsymbol{\tilde{b}}^{[k]}_2 \\
\vdots \\
\boldsymbol{\mathcal{L}d}^{[k]}_p = \boldsymbol{\tilde{b}}^{[k]}_p, 
\end{cases}
\end{equation}
with $\boldsymbol{\tilde{b}}^{[k]} = [\nabla^2f(\boldsymbol{y}^{[k]})]^{-1}\boldsymbol{\tilde{z}}^{[k]} = [(\boldsymbol{\tilde{b}}^{[k]}_1)^{\mathsf{T}}, \ldots,  (\boldsymbol{\tilde{b}}^{[k]}_p)^{\mathsf{T}}]^{\mathsf{T}}$.

Now, we prove the following: 
\begin{enumerate}
\item Notice that 
\begin{align*}
||\boldsymbol{\tilde{z}}^{[k]}_1 - \boldsymbol{z}^{*[k]}_1||_{\boldsymbol{\mathcal{L}}} &\le \epsilon_0||\boldsymbol{z}^{*[k]}_1||_{\boldsymbol{\mathcal{L}}}\\\nonumber
||\boldsymbol{\tilde{z}}^{[k]}_2 - \boldsymbol{z}^{*[k]}_2||_{\boldsymbol{\mathcal{L}}} &\le \epsilon_0||\boldsymbol{z}^{*[k]}_2||_{\boldsymbol{\mathcal{L}}}\\\nonumber
&\vdots\\\nonumber
||\boldsymbol{\tilde{z}}^{[k]}_p - \boldsymbol{z}^{*[k]}_p||_{\boldsymbol{\mathcal{L}}} &\le \epsilon_0||\boldsymbol{z}^{*[k]}_p||_{\boldsymbol{\mathcal{L}}}.
\end{align*}
Hence: 
\begin{equation}\label{addit_equa_z_vectors}
||\boldsymbol{\tilde{z}}^{[k]} - \boldsymbol{z}^{*[k]}||^2_{\boldsymbol{M}} = \sum_{i=1}^{p}||\boldsymbol{\tilde{z}}^{[k]}_i - \boldsymbol{z}^{*[k]}_i||^2_{\boldsymbol{\mathcal{L}}} \le \epsilon^2_0\sum_{i=1}^p||\boldsymbol{z}^{*[k]}_i||_{\boldsymbol{\mathcal{L}}} = \epsilon^2_0||\boldsymbol{z}^{*[k]}||^2_{\boldsymbol{M}}.
\end{equation}
The next step is to rewrite the right and left hand sides of Equation~\ref{addit_equa_z_vectors} in terms of $\boldsymbol{\hat{d}}^{[k]}$ and $\boldsymbol{d}^{*[k]}$: 
\begin{align*}
||\boldsymbol{z}^{*[k]}||^2_{\boldsymbol{M}} &= ||[\nabla^2f(\boldsymbol{y}^{[k]})]^{-1}\boldsymbol{M}\boldsymbol{d}^{*[k]}||^2_{\boldsymbol{M}} = (\boldsymbol{d}^{*[k]})^{\mathsf{T}}\boldsymbol{M}[\nabla^2f(\boldsymbol{y}^{[k]})]^{-1}\boldsymbol{M}[\nabla^2f(\boldsymbol{y}^{[k]})]^{-1}\boldsymbol{M}\boldsymbol{d}^{*[k]}  \\\nonumber
&\le \mu_n(\boldsymbol{\mathcal{L}})(\boldsymbol{d}^{*[k]})^{\mathsf{T}}\boldsymbol{M}[\nabla^2f(\boldsymbol{y}^{[k]})]^{-2}\boldsymbol{M}\boldsymbol{d}^{*[k]} \le \frac{\mu_n(\boldsymbol{\mathcal{L}})}{\gamma}(\boldsymbol{d}^{*[k]})^{\mathsf{T}}\boldsymbol{M}[\nabla^2f(\boldsymbol{y}^{[k]})]^{-1}\boldsymbol{M}\boldsymbol{d}^{*[k]}  \\\nonumber
&=\frac{\mu_n(\boldsymbol{\mathcal{L}})}{\gamma}||\boldsymbol{d}^{*[k]}||^2_{\boldsymbol{H}^{[k]}}.
\end{align*}
Similarly:
\begin{align*}
||\boldsymbol{\tilde{z}}^{[k]} - \boldsymbol{z}^{*[k]}||^2_{\boldsymbol{M}} &= ||[\nabla^2f(\boldsymbol{y}^{[k]})]^{-1}\boldsymbol{M}\boldsymbol{\hat{d}}^{[k]} - [\nabla^2f(\boldsymbol{y}^{[k]})]^{-1}\boldsymbol{M}\boldsymbol{d}^{*[k]}||^2_{\boldsymbol{M}}   \\\nonumber
&=(\boldsymbol{\hat{d}}^{[k]} -\boldsymbol{d}^{*[k]})^{\mathsf{T}}\boldsymbol{M}[\nabla^2f(\boldsymbol{y}^{[k]})]^{-1} \boldsymbol{M} [\nabla^2f(\boldsymbol{y}^{[k]})]^{-1} \boldsymbol{M}(\boldsymbol{\hat{d}}^{[k]} -\boldsymbol{d}^{*[k]}) \\\nonumber
& \ge\frac{1}{\mu_n(\boldsymbol{\mathcal{L}})}(\boldsymbol{\hat{d}}^{[k]} -\boldsymbol{d}^{*[k]})^{\mathsf{T}}[\boldsymbol{H^{[k]}}]^{2}(\boldsymbol{\hat{d}}^{[k]} -\boldsymbol{d}^{*[k]}) \ge \frac{1}{\Gamma}\frac{\mu^2_2(\boldsymbol{\mathcal{L}})}{\mu_n(\boldsymbol{\mathcal{L}})}||\boldsymbol{\hat{d}}^{[k]} -\boldsymbol{d}^{*[k]}||^2_{\boldsymbol{H}^{[k]}},
\end{align*}
where we use the fact that $\boldsymbol{M}[\nabla^2f(\boldsymbol{y}^{[k]})]^{-1}\boldsymbol{M}[\nabla^2f(\boldsymbol{y}^{[k]})]^{-1}\boldsymbol{M} \succeq \frac{1}{\mu_n(\boldsymbol{\mathcal{L}})}[\boldsymbol{H}^{[k]}]^2$. The last transition follows from the fact that $\ker(\boldsymbol{H}^{[k]}) = \ker(\boldsymbol{M})$ and, therefore: 
\begin{equation*}
[\boldsymbol{H}^{[k]}]^2 = [\boldsymbol{H}^{[k]}]^{\frac{1}{2}}\boldsymbol{H}^{[k]}[\boldsymbol{H}^{[k]}]^{\frac{1}{2}} \succeq \frac{1}{\Gamma}\boldsymbol{H}^{[k]}]^{\frac{1}{2}}\boldsymbol{M}^{2}[\boldsymbol{H}^{[k]}]^{\frac{1}{2}} \succeq \frac{\mu^2_2(\boldsymbol{\mathcal{L}})}{\Gamma}\boldsymbol{H}^{[k]}.
\end{equation*}
Combining the above results for (\ref{addit_equa_z_vectors}) immediately gives:
\begin{align}\label{intermediate_result_21_0}
&||\boldsymbol{\hat{d}}^{[k]} -\boldsymbol{d}^{*[k]}||^2_{\boldsymbol{H}^{[k]}} \le \epsilon^2_0\frac{\mu^2_n(\boldsymbol{\mathcal{L}})}{\mu^2_2(\boldsymbol{\mathcal{L}})}\frac{\Gamma}{\gamma}||\boldsymbol{d}^{*[k]}||^2_{\boldsymbol{H}^{[k]}} = \epsilon^2_1||\boldsymbol{d}^{*[k]}||^2_{\boldsymbol{H}^{[k]}}
\end{align}
with $\epsilon_1 = \epsilon_0\frac{\mu_n(\boldsymbol{\mathcal{L}})}{\mu_2(\boldsymbol{\mathcal{L}})}\sqrt{\frac{\Gamma}{\gamma}}$. 
\item Now, we use the triangular inequality:
\begin{equation}\label{addit_ineq_triangul}
||\boldsymbol{\tilde{d}}^{[k]} - \boldsymbol{d}^{*[k]}||_{\boldsymbol{H}^{[k]}} \le ||\boldsymbol{\tilde{d}}^{[k]} - \boldsymbol{\hat{d}}^{[k]}||_{\boldsymbol{H}^{[k]}} + ||\boldsymbol{\hat{d}}^{[k]} - \boldsymbol{d}^{*[k]}||_{\boldsymbol{H}^{[k]}}.
\end{equation} 
Due to the definition of $\boldsymbol{\tilde{d}}^{[k]}_1, \ldots, \boldsymbol{\tilde{d}}^{[k]}_p$, we can write:
\begin{align*}
||\boldsymbol{\tilde{d}}^{[k]} - \boldsymbol{d}^{*[k]}||^2_{\boldsymbol{H}^{[k]}} &= (\boldsymbol{\tilde{d}}^{[k]} - \boldsymbol{d}^{*[k]})^{\mathsf{T}}\boldsymbol{M}[\nabla^2f(\boldsymbol{y}^{[k]})]^{-1}\boldsymbol{M}   (\boldsymbol{\tilde{d}}^{[k]} - \boldsymbol{d}^{*[k]})  \\\nonumber
&\le\frac{1}{\gamma}(\boldsymbol{\tilde{d}}^{[k]} - \boldsymbol{d}^{*[k]})^{\mathsf{T}}\boldsymbol{M}^2(\boldsymbol{\tilde{d}}^{[k]} - \boldsymbol{d}^{*[k]}) \le \frac{\mu_n(\boldsymbol{\mathcal{L}})}{\gamma}||\boldsymbol{\tilde{d}}^{[k]} - \boldsymbol{d}^{*[k]}||^2_{\boldsymbol{M}} \le \epsilon^2_0\frac{\mu_n(\boldsymbol{\mathcal{L}})}{\gamma}||\boldsymbol{\hat{d}}^{[k]}||^2_{\boldsymbol{M}} \\\nonumber
& \le \epsilon^2_0\frac{\mu_n(\boldsymbol{\mathcal{L}})}{\gamma}\frac{\Gamma}{\mu_2(\boldsymbol{\mathcal{L}})}||\boldsymbol{\hat{d}}^{[k]}||^2_{\boldsymbol{H}^{[k]}},
\end{align*}
where we used $\boldsymbol{H}^{[k]} \succeq \frac{\Gamma}{\mu_2(\boldsymbol{\mathcal{L}})}\boldsymbol{M}$. Hence: 
\begin{equation}\label{intermediate_result_21_1}
||\boldsymbol{\tilde{d}}^{[k]} - \boldsymbol{d}^{*[k]}||_{\boldsymbol{H}^{[k]}} \le \epsilon_0\sqrt{\frac{\mu_n(\boldsymbol{\mathcal{L}})}{\mu_2(\boldsymbol{\mathcal{L}})}\frac{\Gamma}{\gamma}}||\boldsymbol{\hat{d}}^{[k]}||_{\boldsymbol{H}^{[k]}}.
\end{equation}
Notice that from Equation~\ref{intermediate_result_21_0}, it follows that:
\begin{equation}\label{intermediate_result_21_2}
||\boldsymbol{\hat{d}}^{[k]}||_{\boldsymbol{H}^{[k]}} \le (1 + \epsilon_1)||\boldsymbol{d}^{*[k]}||_{\boldsymbol{H}^{[k]}}.
\end{equation}
Therefore, combining Equations~\ref{intermediate_result_21_0}, \ref{intermediate_result_21_1}, and \ref{intermediate_result_21_2}, in Equation~\ref{addit_ineq_triangul}, yields:
\begin{align*}
||\boldsymbol{\tilde{d}}^{[k]} - \boldsymbol{d}^{*[k]}||_{\boldsymbol{H}^{[k]}} &\le  \epsilon_0\sqrt{\frac{\mu_n(\boldsymbol{\mathcal{L}})}{\mu_2(\boldsymbol{\mathcal{L}})}\frac{\Gamma}{\gamma}}(1 + \epsilon_1)||\boldsymbol{d}^{*[k]}||_{\boldsymbol{H}^{[k]}} +  \epsilon_1||\boldsymbol{d}^{*[k]}||_{\boldsymbol{H}^{[k]}}  \\\nonumber
&= \epsilon_0\sqrt{\frac{\mu_n(\boldsymbol{\mathcal{L}})}{\mu_2(\boldsymbol{\mathcal{L}})}\frac{\Gamma}{\gamma}}\left[ 1 + \epsilon_1 + \sqrt{\frac{\mu_n(\boldsymbol{\mathcal{L}})}{\mu_2(\boldsymbol{\mathcal{L}})}}\right]||\boldsymbol{d}^{*[k]}||_{\boldsymbol{H}^{[k]}} = \epsilon||\boldsymbol{d}^{*[k]}||_{\boldsymbol{H}^{[k]}}.
\end{align*}
\end{enumerate} 
The above finalizes the statement of the lemma. 
\end{proof}


\section{Proof Dual Gradient Change}

\begin{lemma}
Let $\bm{g}^{[k]} = \nabla q\left(\bm{\lambda}^{[k]}\right)$ be the dual gradient at the $k^{th}$ iteration. Then: 
\begin{align*}
\left|\left|\bm{g}^{[k+1]}\right|\right|_{\bm{M}} &\leq \left[1 -\alpha_{k} + \epsilon \alpha_{k} \sqrt{\frac{\Gamma}{\gamma} \frac{\mu_{n}^{3}(\mathcal{L})}{\mu_{2}^{3}(\mathcal{L})}}\right]\left|\left|\bm{g}^{[k]}\right|\right|_{\bm{M}} +\frac{B(\alpha_{k} (1+\epsilon))^{2}}{2\mu_{2}^{4}(\mathcal{L})} \left|\left|\bm{g}^{[k]}\right|\right|_{\bm{M}}.
\end{align*}
\end{lemma} 

\begin{proof}
We start with the following claim, which plays a crucial role in our analysis:
\begin{claim}\label{claim_1}
Let $\nabla q(\boldsymbol{\lambda})$ be the dual gradient and $\boldsymbol{H}(\boldsymbol{\lambda})$ its Hessian, then for any $\bar{\boldsymbol{\lambda}}, \boldsymbol{\lambda}$:
\begin{equation}\label{obj_fun_taylor_expan_claim}
||\nabla q(\bar{\boldsymbol{\lambda}}) - \nabla q(\boldsymbol{\lambda}) - \boldsymbol{H}(\boldsymbol{\lambda})(\bar{\boldsymbol{\lambda}} - \boldsymbol{\lambda})||_{\boldsymbol{M}} \le \frac{B}{2}||\bar{\boldsymbol{\lambda}} - \boldsymbol{\lambda}||^2_{\boldsymbol{M}}.
\end{equation}
\begin{proof}
We apply the Fundamental Theorem of Calculus for the gradient $\nabla q$ that implies for any two vectors $\bar{\boldsymbol{\lambda}}$ and $\boldsymbol{\lambda}$ in $\mathbb{R}^{np}$ we have:
\begin{equation}\label{fundamental_result2}
\nabla q(\bar{\boldsymbol{\lambda}})= \nabla q(\boldsymbol{\lambda})+\int_{0}^1 \boldsymbol{H}(\boldsymbol{\lambda}+t(\bar{\boldsymbol{\lambda}}-\boldsymbol{\lambda}))(\bar{\boldsymbol{\lambda}}-\boldsymbol{\lambda})\ dt.
\end{equation}
We proceed by adding and subtracting $\boldsymbol{H}(\boldsymbol{\lambda})(\bar{\boldsymbol{\lambda}}-\boldsymbol{\lambda})$ to the integral in the right hand side of Equation~\ref{fundamental_result2}:
\begin{align}\label{taylor_first_two terms_44}
\nabla q(\bar{\boldsymbol{\lambda}})= \nabla q(\boldsymbol{\lambda})+\int_{0}^1 \left[\boldsymbol{H}(\boldsymbol{\lambda}+t(\bar{\boldsymbol{\lambda}}-\boldsymbol{\lambda}))-\boldsymbol{H}(\boldsymbol{\lambda})\right] (\bar{\boldsymbol{\lambda}}-\boldsymbol{\lambda})+\boldsymbol{H}(\boldsymbol{\lambda})(\bar{\boldsymbol{\lambda}}-\boldsymbol{\lambda})\ dt.
\end{align}
Consequently, we can separate the integral in Equation~\ref{taylor_first_two terms_44} as: 
\begin{align}\label{taylor_first_two terms_45}
\nabla q(\bar{\boldsymbol{\lambda}})= \nabla q(\boldsymbol{\lambda})+\int_{0}^1 \left[\boldsymbol{H}(\boldsymbol{\lambda}+t(\bar{\boldsymbol{\lambda}}-\boldsymbol{\lambda}))-\boldsymbol{H}(\boldsymbol{\lambda})\right] (\bar{\boldsymbol{\lambda}}-\boldsymbol{\lambda})\ dt+\int_{0}^1 \boldsymbol{H}(\boldsymbol{\lambda})(\bar{\boldsymbol{\lambda}}-\boldsymbol{\lambda})\ dt.
\end{align}
The second integral on the right hand side of Equation~\ref{taylor_first_two terms_45} is independent of $t$. Therefore, we can simplify the integral as $ \boldsymbol{H}(\boldsymbol{\lambda})(\bar{\boldsymbol{\lambda}}-\boldsymbol{\lambda})$, which implies:
\begin{align}\label{taylor_first_two terms_46}
\nabla q(\bar{\boldsymbol{\lambda}})= \nabla q(\boldsymbol{\lambda})+\boldsymbol{H}(\boldsymbol{\lambda})(\bar{\boldsymbol{\lambda}}-\boldsymbol{\lambda})+\int_{0}^1 \left[\boldsymbol{H}(\boldsymbol{\lambda}+t(\bar{\boldsymbol{\lambda}}-\boldsymbol{\lambda}))-\boldsymbol{H}(\boldsymbol{\lambda})\right] (\bar{\boldsymbol{\lambda}}-\boldsymbol{\lambda})\ dt.
\end{align}
By rearranging the terms in Equation~\ref{taylor_first_two terms_46} and taking the norm of both sides we obtain: 
\begin{align}\label{taylor_first_two terms_47}
&||\nabla q(\bar{\boldsymbol{\lambda}})- \nabla q(\boldsymbol{\lambda})-\boldsymbol{H}(\boldsymbol{\lambda})(\bar{\boldsymbol{\lambda}}-\boldsymbol{\lambda})||_{\boldsymbol{M}}=\\\nonumber
&\left|\left|\int_{0}^1 \left[\boldsymbol{H}(\boldsymbol{\lambda}+t(\bar{\boldsymbol{\lambda}}-\boldsymbol{\lambda}))-\boldsymbol{H}(\boldsymbol{\lambda})\right] (\bar{\boldsymbol{\lambda}}-\boldsymbol{\lambda})\ dt\right|\right|_{\boldsymbol{M}}
\end{align}
Considering the inequality in Equation~\ref{taylor_first_two terms_47} and the fact that norm of integral is less than the integral of the norms, we can write: 
\begin{align}\label{taylor_first_two terms_48}
&||\nabla q(\bar{\boldsymbol{\lambda}})- \nabla q(\boldsymbol{\lambda})-\boldsymbol{H}(\boldsymbol{\lambda})(\bar{\boldsymbol{\lambda}}-\boldsymbol{\lambda})||_{\boldsymbol{M}} \le \int_{0}^1 \left|\left|\left[\boldsymbol{H}(\boldsymbol{\lambda}+t(\bar{\boldsymbol{\lambda}}-\boldsymbol{\lambda}))-\boldsymbol{H}(\boldsymbol{\lambda})\right] (\bar{\boldsymbol{\lambda}}-\boldsymbol{\lambda}) \right|\right|_{\boldsymbol{M}}\ dt.
\end{align}
Now, we can prove the following claim: 
\begin{claim}
Let $\boldsymbol{H}(\boldsymbol{\lambda})$ be the Hessian of the dual function $q(\boldsymbol{\lambda}) = q(\boldsymbol{\lambda_1}, \ldots, \boldsymbol{\lambda}_p)$. Then for any $\bm{v}\in\mathbb{R}^{np}$:
\begin{align}\label{claim_result_2}
&||[\boldsymbol{H}(\bar{\boldsymbol{\lambda}}) - \boldsymbol{H}(\boldsymbol{\lambda})]\boldsymbol{v}||_{\boldsymbol{M}} \le ||\boldsymbol{H}(\bar{\boldsymbol{\lambda}}) - \boldsymbol{H}(\boldsymbol{\lambda})||_{\boldsymbol{M}}||\bm{v}||_{\boldsymbol{M}}
\end{align}
\end{claim}
\begin{proof}
Lemma~7 gives:
\begin{equation*}
\boldsymbol{H}(\boldsymbol{\lambda}) = \boldsymbol{M}[\nabla^2f(\boldsymbol{y}(\boldsymbol{\lambda}))]^{-1}\boldsymbol{M}.
\end{equation*}
Now, let us consider the following three cases: 
\begin{enumerate}
\item $\boldsymbol{v}\in \ker\{\boldsymbol{M}\}^{\perp}$: In this case Equation~\ref{claim_result_2} follows immediately from the definition: $||\boldsymbol{A}||_{\boldsymbol{M}} = \sup_{\boldsymbol{v}: \boldsymbol{v} \notin \ker\{\boldsymbol{M}\}}\frac{||\boldsymbol{Av}||_{\boldsymbol{M}}}{||\boldsymbol{v}||_{\boldsymbol{M}}}$.
\item $\boldsymbol{v}\in \ker\{\boldsymbol{M}\}$: In this case $||\boldsymbol{v}||_{\boldsymbol{M}} = 0$, and $[\boldsymbol{H}(\bar{\boldsymbol{\lambda}}) - \boldsymbol{H}(\boldsymbol{\lambda})]\boldsymbol{v} = \boldsymbol{0} - \boldsymbol{0} = \boldsymbol{0}$.
\item $\boldsymbol{v} = \boldsymbol{u}_1 + \boldsymbol{u}_2$, where $\boldsymbol{u}_1\in\ker\{\boldsymbol{M}\}^{\perp}, \boldsymbol{u}_2\in \ker{\boldsymbol{M}}$. In this case $||\boldsymbol{v}||_{\boldsymbol{M}} = ||\boldsymbol{u}_1||_{\boldsymbol{M}}$, and using the first case result for $\boldsymbol{u}_1\in\ker\{\boldsymbol{M}\}^{\perp}$, we have:
\begin{align*}
&||[\boldsymbol{H}(\bar{\boldsymbol{\lambda}}) - \boldsymbol{H}(\boldsymbol{\lambda})]\boldsymbol{v}||_{\boldsymbol{M}} = ||[\boldsymbol{H}(\bar{\boldsymbol{\lambda}}) - \boldsymbol{H}(\boldsymbol{\lambda})]\boldsymbol{u}_1||_{\boldsymbol{M}} \le \\\nonumber
&||\boldsymbol{H}(\bar{\boldsymbol{\lambda}}) - \boldsymbol{H}(\boldsymbol{\lambda})||_{\boldsymbol{M}}||\boldsymbol{u}_1||_{\boldsymbol{M}} = ||\boldsymbol{H}(\bar{\boldsymbol{\lambda}}) - \boldsymbol{H}(\boldsymbol{\lambda})||_{\boldsymbol{M}}||\boldsymbol{v}||_{\boldsymbol{M}}
\end{align*} 
\end{enumerate}
This finishes the proof of the claim. 
\end{proof}
Applying the above result to Equation~\ref{taylor_first_two terms_48} gives:
\begin{align*}
||\nabla q(\bar{\boldsymbol{\lambda}})- \nabla q(\boldsymbol{\lambda})-\boldsymbol{H}(\boldsymbol{\lambda})(\bar{\boldsymbol{\lambda}}-\boldsymbol{\lambda})||_{\boldsymbol{M}} &\le \int_{0}^1 \left|\left|\left[\boldsymbol{H}(\boldsymbol{\lambda}+t(\bar{\boldsymbol{\lambda}}-\boldsymbol{\lambda}))-\boldsymbol{H}(\boldsymbol{\lambda})\right]\right|\right|_{\boldsymbol{M}}|| (\bar{\boldsymbol{\lambda}}-\boldsymbol{\lambda})||_{\boldsymbol{M}}\ dt\\\nonumber
&\le B||\bar{\boldsymbol{\lambda}} - \boldsymbol{\lambda}||^2_{\boldsymbol{M}}\int_{0}^1tdt = \frac{B}{2}||\boldsymbol{\bar{\boldsymbol{\lambda}}} - \boldsymbol{\lambda}||^2_{\boldsymbol{M}}.
\end{align*}
This finishes the proof of claim \ref{claim_1}.
\end{proof}
\end{claim}
Applying the result of claim~\ref{claim_1} to $\boldsymbol{\lambda}^{[k+1]}$ and $\boldsymbol{\lambda}^{[k]}$ gives:
\begin{equation*}
||\boldsymbol{g}^{[k+1]} - \boldsymbol{g}^{[k]} - \alpha_k\boldsymbol{H}^{[k]}\tilde{\boldsymbol{d}}^{[k]}||_{\boldsymbol{M}} \le \frac{B\alpha^2_k}{2}||\tilde{\boldsymbol{d}}^{[k]}||^2_{\boldsymbol{M}}.
\end{equation*}
Applying the triangular inequality, we have:
\begin{equation}\label{gradient_norm_inequality_interm}
||\boldsymbol{g}^{[k+1]}||_{\boldsymbol{M}} \le  ||\boldsymbol{g}^{[k]} + \alpha_k\boldsymbol{H}^{[k]}\tilde{\boldsymbol{d}}^{[k]}||_{\boldsymbol{M}} + \frac{B\alpha^2_k}{2}||\tilde{\boldsymbol{d}}^{[k]}||^2_{\boldsymbol{M}}.
\end{equation}
The next step is to evaluate $||\tilde{\boldsymbol{d}}^{[k]}||^2_{\boldsymbol{M}}$ and $||\boldsymbol{g}^{[k]} + \alpha_k\boldsymbol{H}^{[k]}\tilde{\boldsymbol{d}}^{[k]}||_{\boldsymbol{M}}$:
\begin{enumerate}
\item To upper bound $||\tilde{\boldsymbol{d}}^{[k]}||^2_{\boldsymbol{M}}$ notice that: 
\begin{equation}\label{hessian_primal_lower_upper_bounds}
\frac{\mu_2(\boldsymbol{\mathcal{L}})}{\Gamma}\boldsymbol{M} \preceq \boldsymbol{H}^{[k]}\preceq \frac{\mu_n(\boldsymbol{\mathcal{L}})}{\gamma}\boldsymbol{M}.
\end{equation}
Hence:
\begin{align*}
||\tilde{\boldsymbol{d}}^{[k]}||^2_{\boldsymbol{M}} &= (\tilde{\boldsymbol{d}}^{[k]})^{\mathsf{T}}\boldsymbol{M}\tilde{\boldsymbol{d}}^{[k]} \le \frac{\Gamma}{\mu_2(\boldsymbol{\mathcal{L}})}(\tilde{\boldsymbol{d}}^{[k]})^{\mathsf{T}}\boldsymbol{H}^{[k]}\tilde{\boldsymbol{d}}^{[k]}  \\\nonumber
&\le(1 + \epsilon)^2\frac{\Gamma}{\mu_2(\boldsymbol{\mathcal{L}})}(\boldsymbol{d}^{*[k]})^{\mathsf{T}}\boldsymbol{H}^{[k]}\boldsymbol{d}^{*[k]} = (1 + \epsilon)^2\frac{\Gamma}{\mu_2(\boldsymbol{\mathcal{L}})}(\boldsymbol{g}^{[k]})^{\mathsf{T}}(\boldsymbol{H}^{[k]})^{\dagger}\boldsymbol{H}^{[k]}(\boldsymbol{H}^{[k]})^{\dagger}\boldsymbol{g}^{[k]} \\\nonumber
&=(1 + \epsilon)^2\frac{\Gamma}{\mu_2(\boldsymbol{\mathcal{L}})}(\boldsymbol{g}^{[k]})^{\mathsf{T}}(\boldsymbol{H}^{[k]})^{\dagger}\boldsymbol{g}^{[k]}.
\end{align*}
Because $\boldsymbol{g}^{[k]} \in \ker\{\boldsymbol{H}^{[k]}\} = \ker\{\boldsymbol{M}\}$ and using the result in Equation~\ref{hessian_primal_lower_upper_bounds}, it follows that:
\begin{align}\label{approx_newton_direction_bound}
||\tilde{\boldsymbol{d}}^{[k]}||^2_{\boldsymbol{M}} \le (1 + \epsilon)^2\frac{\Gamma}{\mu_2(\boldsymbol{\mathcal{L}})}\frac{\Gamma}{\mu_2(\boldsymbol{\mathcal{L}})\mu^2_{\min}(\boldsymbol{M})}||\boldsymbol{g}^{[k]}||^2_{\boldsymbol{M}} = (1 + \epsilon)^2\frac{\Gamma^2}{\mu^4_2(\boldsymbol{\mathcal{L}})}||\boldsymbol{g}^{[k]}||^2_{\boldsymbol{M}}
\end{align} 
where $\mu_{\min}(\boldsymbol{M})$ is the second smallest eigenvalue of $\boldsymbol{M}$ which is equal to $\mu_2(\boldsymbol{\mathcal{L}})$.

\item Let us denote $\boldsymbol{c}^{[k]} = \tilde{\boldsymbol{d}}^{[k]} - \boldsymbol{d}^{*[k]}$, then:
\begin{equation*}
||\boldsymbol{c}^{[k]}||_{\boldsymbol{H}^{[k]}} \le \epsilon||\boldsymbol{d}^{*[k]}||_{\boldsymbol{H}^{[k]}}
\end{equation*}
Therefore, for the term $||\boldsymbol{g}^{[k]} + \alpha_k\boldsymbol{H}^{[k]}\tilde{\boldsymbol{d}}^{[k]}||_{\boldsymbol{M}}$ we can write:
\begin{align}\label{intermediate_equation_for_case_2}
||\boldsymbol{g}^{[k]} + \alpha_k\boldsymbol{H}^{[k]}\tilde{\boldsymbol{d}}^{[k]}||_{\boldsymbol{M}} &= ||\boldsymbol{g}^{[k]} + \alpha_k\boldsymbol{H}^{[k]}(\boldsymbol{d}^{*[k]} + \boldsymbol{c}^{[k]})||_{\boldsymbol{M}} \\\nonumber
&=  ||\boldsymbol{g}^{[k]} - \alpha_k\boldsymbol{g}^{[k]} + \alpha_k\boldsymbol{H}^{[k]}\boldsymbol{c}^{[k]}||_{\boldsymbol{M}} \le (1 - \alpha_k)||\boldsymbol{g}^{[k]}||_{\boldsymbol{M}} + \alpha_k||\boldsymbol{H}^{[k]}\boldsymbol{c}^{[k]}||_{\boldsymbol{M}}.
\end{align}
Therefore, our goal now is to upper bound the term $||\boldsymbol{H}^{[k]}\boldsymbol{c}^{[k]}||_{\boldsymbol{M}}$. Using Equation~\ref{hessian_primal_lower_upper_bounds} and the fact that $\boldsymbol{M}\preceq \mu_n(\boldsymbol{\mathcal{L}})\boldsymbol{I}_{np\times np}$, we have: 
\begin{align*}
||\boldsymbol{H}^{[k]}\boldsymbol{c}^{[k]}||^2_{\boldsymbol{M}} &= (\boldsymbol{c}^{[k]})^{\mathsf{T}}\boldsymbol{H}^{[k]}\boldsymbol{M}\boldsymbol{H}^{[k]}\boldsymbol{c}^{[k]} \le \mu_n(\boldsymbol{\mathcal{L}}) (\boldsymbol{c}^{[k]})^{\mathsf{T}}(\boldsymbol{H}^{[k]})^2\boldsymbol{c}^{[k]}  \\\nonumber
&\le\mu_n(\boldsymbol{\mathcal{L}})\frac{\mu^2_n(\boldsymbol{\mathcal{L}})}{\gamma}(\boldsymbol{c}^{[k]})^{\mathsf{T}}\boldsymbol{H}^{[k]}\boldsymbol{c}^{[k]} \le \epsilon^2\frac{\mu^3_n(\boldsymbol{\mathcal{L}})}{\gamma}||\boldsymbol{d}^{*[k]}||^2_{\boldsymbol{H}^{[k]}} \\\nonumber
&=\epsilon^2\frac{\mu^3_n(\boldsymbol{\mathcal{L}})}{\gamma}(\boldsymbol{g}^{[k]})^{\mathsf{T}}(\boldsymbol{H}^{[k]})^{\dagger}\boldsymbol{g}^{[k]} \le \epsilon^2\frac{\mu^3_n(\boldsymbol{\mathcal{L}})}{\gamma} \frac{\Gamma}{•\mu^3_2(\boldsymbol{\mathcal{L}})}||\boldsymbol{g}^{[k]}||^2_{\boldsymbol{M}}  
\end{align*}
Therefore:
\begin{equation*}
||\boldsymbol{H}^{[k]}\boldsymbol{c}^{[k]}||_{\boldsymbol{M}} \le \epsilon\sqrt{\frac{\Gamma}{\gamma}\frac{\mu^3_n(\boldsymbol{\mathcal{L}})}{\mu^3_2(\boldsymbol{\mathcal{L}})}}||\boldsymbol{g}^{[k]}||_{\boldsymbol{M}}.
\end{equation*}
Applying this result in Equation~\ref{intermediate_equation_for_case_2} gives:
\begin{equation}\label{gradient_plus_approx_newton_bound}
||\boldsymbol{g}^{[k]} + \alpha_k\boldsymbol{H}^{[k]}\tilde{\boldsymbol{d}}^{[k]}||_{\boldsymbol{M}} \le \left[1 - \alpha_k + \epsilon\alpha_k\sqrt{\frac{\Gamma}{\gamma}\frac{\mu^3_n(\boldsymbol{\mathcal{L}})}{\mu^3_2(\boldsymbol{\mathcal{L}})}}\right]||\boldsymbol{g}^{[k]}||_{\boldsymbol{M}}. 
\end{equation}
\end{enumerate}
Applying the results of Equations~\ref{approx_newton_direction_bound} and~\ref{gradient_plus_approx_newton_bound}, in Equation~\ref{gradient_norm_inequality_interm}:
\begin{align*}
||\boldsymbol{g}^{[k+1]}||_{\boldsymbol{M}} \le \left[1 - \alpha_k + \epsilon\alpha_k\sqrt{\frac{\Gamma}{\gamma}\frac{\mu^3_n(\boldsymbol{\mathcal{L}})}{\mu^3_2(\boldsymbol{\mathcal{L}})}}\right]||\boldsymbol{g}^{[k]}||_{\boldsymbol{M}} + \frac{B(\alpha_k(1 + \epsilon)\Gamma)^2}{2\mu^4_2(\boldsymbol{\mathcal{L}})}||\boldsymbol{g}^{[k]}||^2_{\boldsymbol{M}}.
\end{align*}

\end{proof}


\section{Proof Convergence Phases}
\begin{theorem}
Let $\Gamma$, $\gamma$, be the constants defined in Assumption~\ref{Ass:Two}, $\mu_{n}(\mathcal{L})$, $\mu_{2}(\mathcal{L})$ be the largest and the second smallest eigenvalues of $\mathcal{L}$, respectively, and $\epsilon \in \left(0, \frac{\mu_{2}(\mathcal{L})}{\mu_{n}(\mathcal{L})}\sqrt{\frac{\Gamma}{\gamma}\frac{\mu_{2}(\mathcal{L})}{\mu_{n}(\mathcal{L})}}\right]$. Consider the following iteration scheme: $\bm{\lambda}^{[k+1]}=\bm{\lambda}^{[k]} + \alpha^{\star} \tilde{\bm{d}}^{[k]}$, where $\alpha^{\star}= \left(\frac{\gamma}{\Gamma}\right)^{2}\left(\frac{\mu_{2}(\mathcal{L})}{\mu_{n}(\mathcal{L})}\right)^{4}\frac{1-\epsilon}{(1+\epsilon)^{2}}$. Then the distributed Newton algorithm exhibits the following convergence phases: 
\begin{itemize}
\item \textbf{Strict Decrease Phase:} while $\left|\left|\bm{g}^{[k]}\right|\right|_{\bm{M}} \geq \eta_{1}$: $q\left(\bm{\lambda}^{[k+1]}\right)-q\left(\bm{\lambda}^{[k]}\right) \leq -\frac{\gamma^{3}}{\Gamma^{2}}\left(\frac{1-\epsilon}{1+\epsilon}\right)^{2}\frac{\mu_{2}^{4}(\mathcal{L})}{\mu_{n}^{7}(\mathcal{L})} \eta_{1}^{2}$,
\item \textbf{Quadratic Decrease Phase:} while $\eta_{0} \leq \left|\left|\bm{g}^{[k]}\right|\right|_{\bm{M}} \leq \eta_{1}$: $
\left|\left|\bm{g}^{[k+1]}\right|\right|_{\bm{M}} \leq \frac{1}{\eta_{1}}\left|\left|\bm{g}^{[k]}\right|\right|_{\bm{M}}^{2}$,
\item \textbf{Terminal Phase:} while $\left|\left|\bm{g}^{[k]}\right|\right|_{\bm{M}} \leq \eta_{0}$: $
\left|\left|\bm{g}^{[k+1]}\right|\right|_{\bm{M}} \leq \zeta \left|\left|\bm{g}^{[k]}\right|\right|_{\bm{M}}$,
where $\eta_{0} = \frac{\zeta (1 - \zeta)}{\xi}$, $\eta_{1} = \frac{1-\zeta}{\xi}$, and 
$\zeta = \sqrt{\left[1-\alpha_{k} + \epsilon \alpha_{k} \sqrt{\frac{\Gamma}{\gamma}\frac{\mu_{n}^{3}(\mathcal{L})}{\mu_{2}^{3}(\mathcal{L})}}\right]}, \ \ \ \xi= \frac{B(\alpha_{k}\Gamma(1+\epsilon))^{2}}{2\mu_{2}^{4}(\mathcal{L})}$. 
\end{itemize}
\end{theorem}
\begin{proof}
We will proof each phase separately. We start with phase one when $\left|\left|\bm{g}^{[k]}\right|\right|_{\bm{M}} \geq \eta$. Taking the Taylor expansion of the dual function gives: 
\begin{align*}
&q\left(\bm{\lambda}^{[k+1]}\right)= q\left(\bm{\lambda}^{[k]}\right)+ \left(\bm{g}^{[k]}\right)^{\mathsf{T}}\left(\bm{\lambda}^{[k+1]} - \bm{\lambda}^{[k]}\right) + \frac{1}{2} \left(\bm{\lambda}^{[k+1]} - \bm{\lambda}^{[k]}\right)^{\mathsf{T}}\bm{H}(\bm{z})\left(\bm{\lambda}^{[k+1]} - \bm{\lambda}^{[k]}\right) \\
&\leq q\left(\bm{\lambda}^{[k]}\right) + \alpha_{k} \bm{g}^{[k],\mathsf{T}}\tilde{\bm{d}}^{[k]}+ \frac{\alpha_{k}^{2}}{2} \frac{\mu_{n}(\mathcal{L})}{\gamma}\tilde{\bm{d}}^{[k],\mathsf{T}}\bm{M} \tilde{\bm{d}}^{[k]} = q\left(\bm{\lambda}^{[k]}\right) + \alpha_{k}\bm{g}^{[k],\mathsf{T}}\tilde{\bm{d}}^{[k]}  + \frac{\alpha_{k}^{2}}{2}\frac{\mu_{n}(\mathcal{L})}{\gamma} \left|\left|\tilde{\bm{d}}^{[k]}\right|\right|_{\bm{M}}^{2}. 
\end{align*}
Therefore, we can prove that: 
\begin{equation*}
\left|\left|\tilde{\bm{d}}^{[k]}\right|\right|_{\bm{M}}^{2} \leq (1+\epsilon)^{2} \frac{\Gamma^{2}}{\mu_{2}^{4}(\mathcal{L})}\left|\left|\bm{g}^{[k]}\right|\right|_{\bm{M}}^{2}. 
\end{equation*}
To proceed, the goal now is to bound $\left(\bm{g}^{[k]}\right)^{\mathsf{T}}\tilde{\bm{d}}^{[k]}$. Noticing that $\left(\bm{g}^{[k]}\right)^{\mathsf{T}}\tilde{\bm{d}}^{[k]}= - \left(\tilde{\bm{d}}^{[k]}\right)^{\mathsf{T}}\bm{H}^{[k]}\bm{d}^{\star, [k]}$ and using that $\tilde{\bm{d}}^{[k]}$ is the $\epsilon$-approximate solution, we have: 
\begin{align*}
-2 \tilde{\bm{d}}^{[k],\mathsf{T}}\bm{H}^{[k]}\bm{d}^{\star,[k]} &\leq - (1-\epsilon)^{2}||\bm{d}^{\star,[k]}||_{\bm{H}^{[k]}}^{2} - ||\tilde{\bm{d}}^{[k]}||_{\bm{H}^{[k]}}^{2}.
\end{align*}
Now, notice that: 
\begin{align*}
\left|\left|\bm{d}^{\star,[k]}\right|\right|_{\bm{H}^{[k]}}^{2} = \left(\bm{g}^{[k]}\right)^{\mathsf{T}}\left(\bm{H}^{[k]}\right)^{\dagger}\bm{g}^{[k]}\geq \frac{\gamma}{\mu_{n}^{3}(\mathcal{L})}\left|\left|\bm{g}^{[k]}\right|\right|_{\bm{M}}^{2},
\end{align*}
and 
\begin{align*}
\left|\left|\tilde{\bm{d}}^{[k]}\right|\right|_{\bm{H}^{[k]}}^{2} \leq (1-\epsilon)^{2} \frac{\gamma}{\mu_{n}^{3} (\mathcal{L})} \left|\left|\bm{g}^{[k]}\right|\right|_{\bm{M}}^{2}.
\end{align*}
Consequently, we can write: 
\begin{align*}
\left(\bm{g}^{[k]}\right)^{\mathsf{T}} \tilde{\bm{d}}^{[k]} \leq - (1-\epsilon) \frac{\gamma}{\mu_{n}^{3}(\mathcal{L})} \left|\left|\bm{g}^{[k]}\right|\right|_{\bm{M}}^{2}. 
\end{align*}
Therefore
\begin{align*}
q\left(\bm{\lambda}^{[k+1]}\right)- q\left(\bm{\lambda}^{[k]}\right) &\leq - \Bigg[ (1-\epsilon) \frac{\gamma}{\mu_{n}^{3}(\mathcal{L})}\alpha_{k}   - \frac{\mu_{n}(\mathcal{L})\Gamma^{2} (1+\epsilon)^{2}}{2\gamma \mu_{2}^{4}(\mathcal{L})}\alpha_{k}^{2}
\Bigg]\left|\left|\bm{g}^{[k]}\right|\right|_{\bm{M}}^{2}.
\end{align*}
Hence, choosing ${\alpha}_{k} = \alpha^{\star}= \left(\frac{\gamma}{\Gamma}\right)^{2}\left(\frac{\mu_{2}(\mathcal{L})}{\mu_{n}(\mathcal{L})}\right)^{4}\frac{1-\epsilon}{(1+\epsilon)^{2}}$ and using $\left|\left|\bm{g}^{[k]}\right|\right|_{\bm{M}}^{2} \geq \eta_{1}$, we arrive at the strict decrease phase. To prove the quadratic decrease phase, we let $\eta_{0} \leq \left|\left|\bm{g}^{[k]}\right|\right|_{\bm{M}}^{2} < \eta$. It immediately follows that: 
\begin{align*}
\left|\left|\bm{g}^{[k+1]}\right|\right|_{\bm{M}} \leq \xi \left(\frac{\zeta}{1-\zeta} +1\right) \left|\left|\bm{g}^{[k]}\right|\right|_{\bm{M}}^{2}  = \frac{1}{\eta_1} \left|\left|\bm{g}^{[k]}\right|\right|_{\bm{M}}^{2}.
\end{align*}

Finally for $\left|\left|\bm{g}^{[k]}\right|\right|_{\bm{M}}^{2} < \eta_{0}$, we have 
\begin{equation*}
\left|\left|\bm{g}^{[k+1]}\right|\right|_{\bm{M}} \leq \zeta \left|\left|\bm{g}^{[k]}\right|\right|_{\bm{M}}.
\end{equation*}
The above finalizes the statement of the theorem. 
\end{proof}
\section{London Schools \& Reinforcement Learning Results}
\subsection{London Schools Data} The London Schools data set consists of examination scores from 15,362 students in 139 schools. This is a benchmark regression task with a goal of predicting examination scores of each student. We use the same feature encoding used in~\cite{daume12gomtl}, where four school-specific categorical variables along with three student-specific categorical variables are encoded as a collection of binary features. In addition, we use the examination year and a bias term as additional features, giving each data instance 27 features.
\subsection{Reinforcement Learning} We considered the policy search framework to control a double cart-pole system (DCP). As detailed in~\cite{Bou-AmmarELR15}, the DCP adds a second inverted pendulum to the standard cart-pole system, with six parameters and six state features. The goal is to balance both poles upright. We generated 20,000 rollouts each with a length of 150 time steps.

\begin{figure*}[tb!]
\centering
\hfill\hspace{-1.1em}
\subfigure[Obj. London Sch.]{
	\label{fig:ObjLondon}
\includegraphics[height=0.35\textwidth,width=0.45\textwidth]{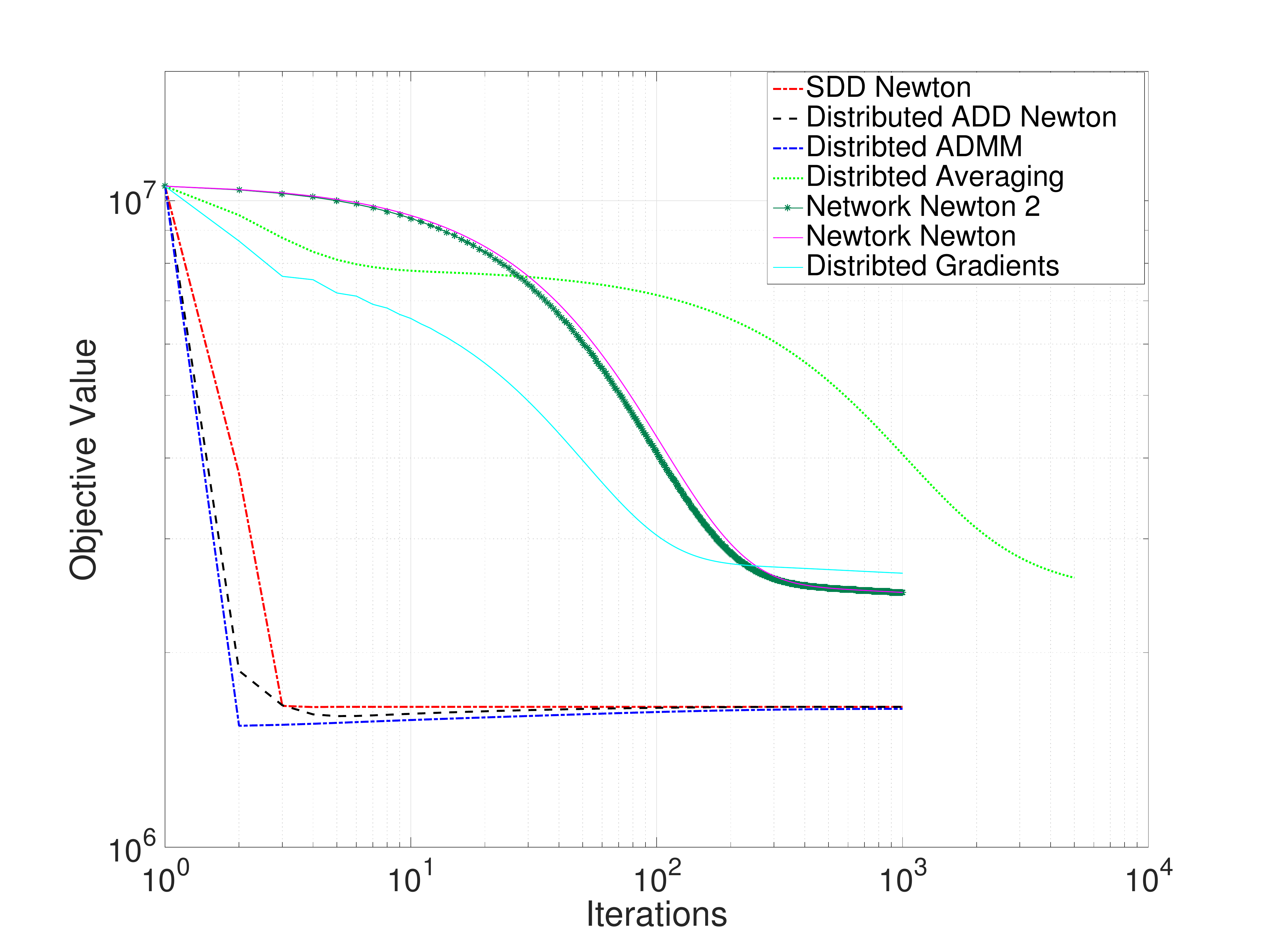}
}
\hfill
\hfill\hspace{-1.4em}\hfill
\subfigure[Con. London Sch.]{
	\label{fig:ConLondon}
\includegraphics[height=0.35\textwidth,width=0.45\textwidth]{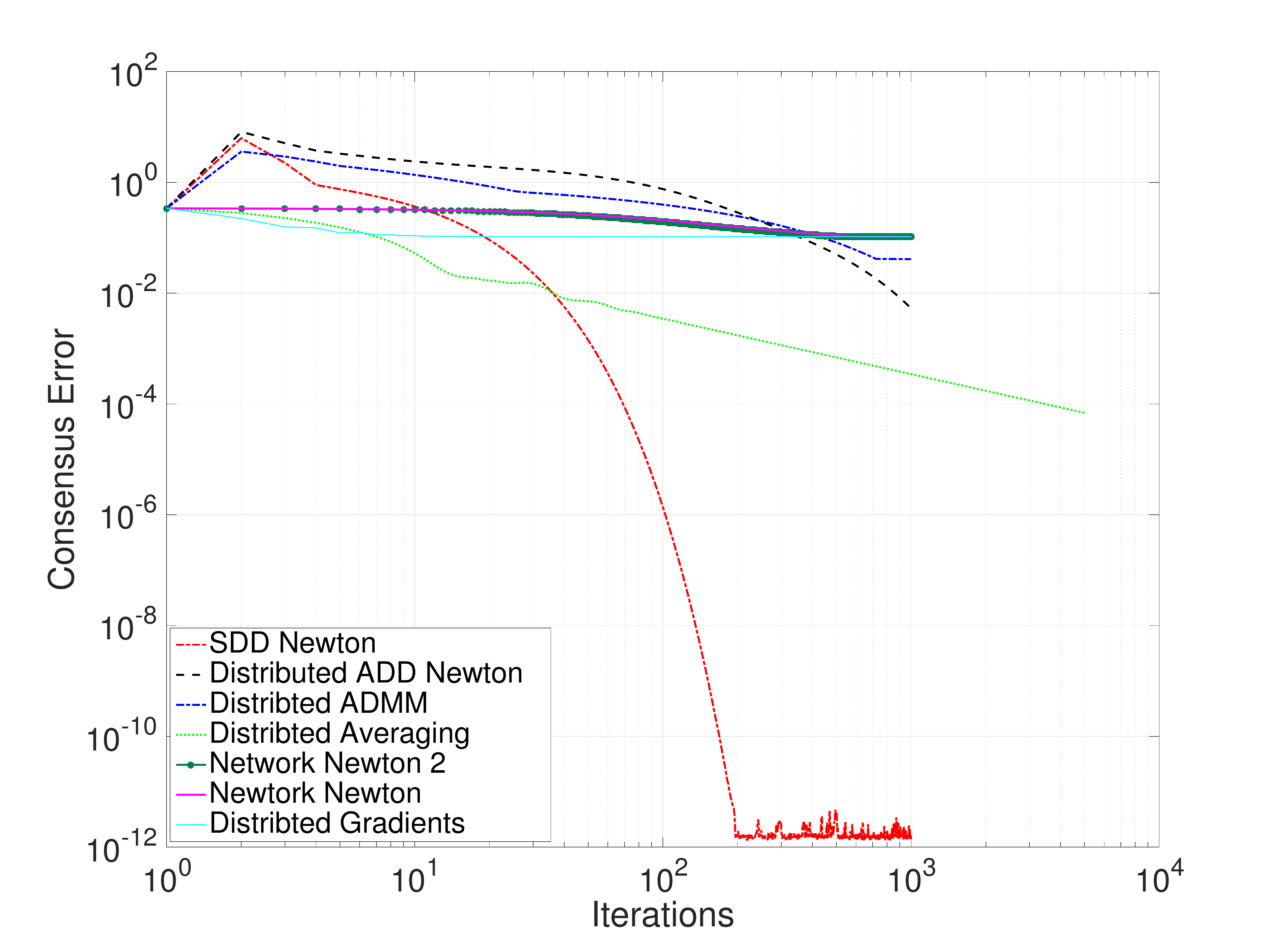}
}
\hfill
\subfigure[Obj. RL]{
	\label{fig:ObjClassNormal}
\includegraphics[trim = 18mm 20mm 25mm 25mm, clip,height=0.30\textwidth,width=0.45\textwidth]{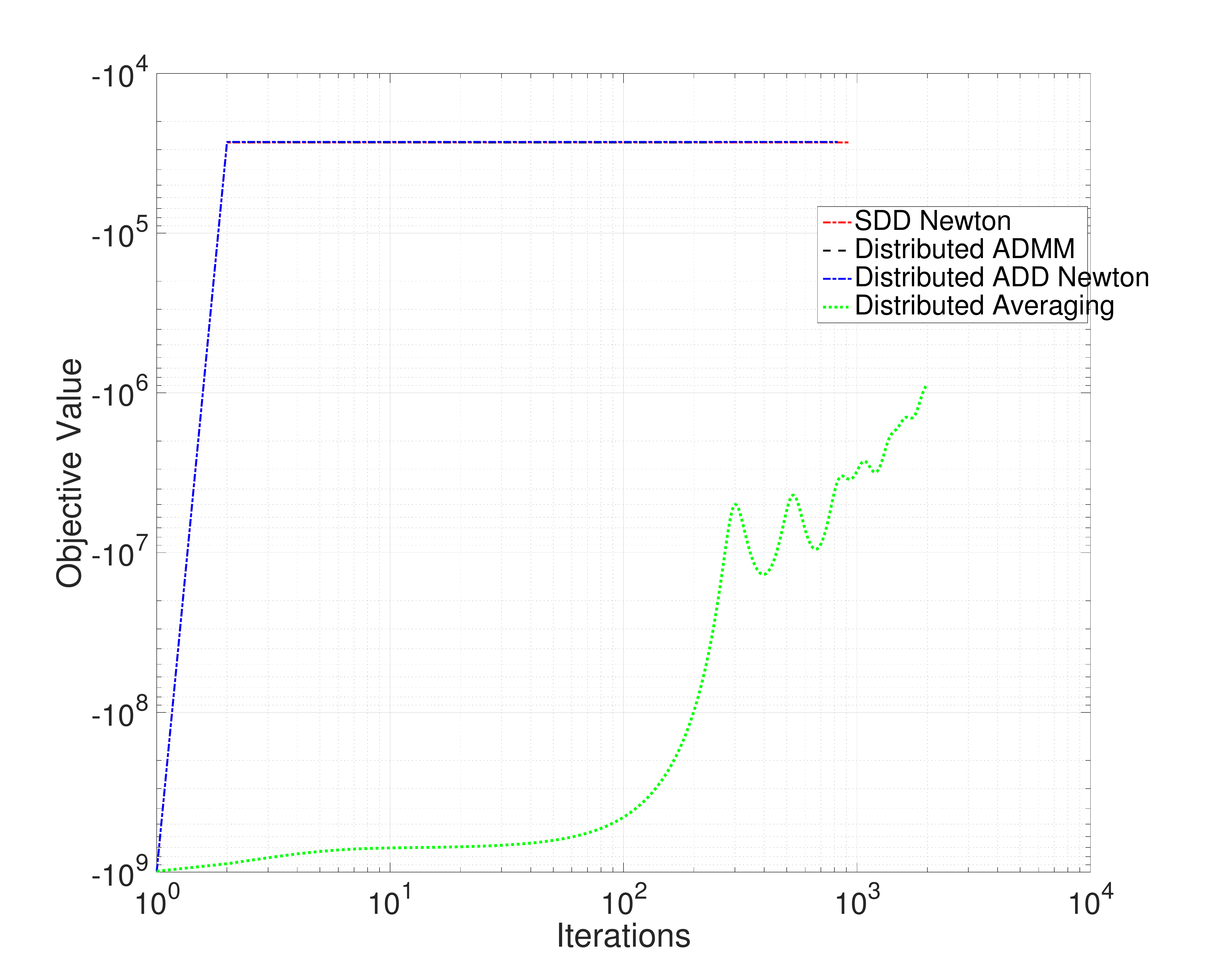}
}
\hfill\hspace{-1.1em}\hfill
\subfigure[Con. RL]{
	\label{fig:ConClassNormal}
\includegraphics[trim = 18mm 20mm 25mm 25mm, clip,height=0.30\textwidth,width=0.45\textwidth]{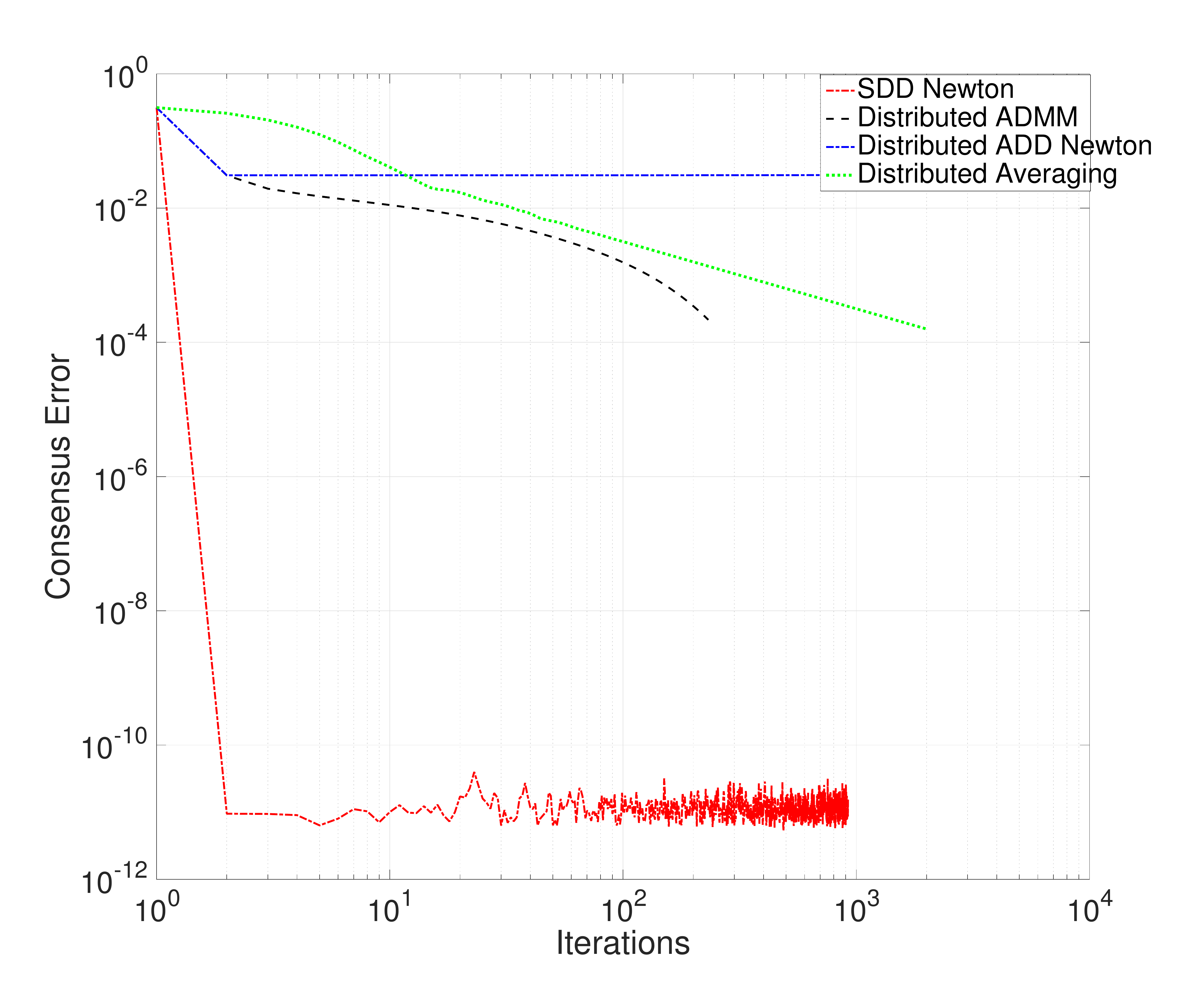}
}
\vspace{-1.1em}
\caption{Figures (a) and (b) report the objective value and consensus error on the London Schools dataset. Figures (c) and (d) demonstrate the same criteria on the reinforcement learning benchmarks. These results agains how that our method outperforms others in literature. }
\end{figure*}
\section{Reductions to Global Consensus}
\subsection{Linear Regression}
In linear regression, we assume a data set $\mathcal{D}= \{a_{i}, \bm{x}_{i}\}_{i=1}^{m}$, with $a_{i} \in \mathbb{R} $ being the dependent variable and $\bm{x}_{i} \in \mathbb{R}^{d}$ denoting the independent vector. We further assume that the input data points have been transformed using a relevant feature extractor, leading to $\bm{b}_{i} = \bm{\Phi}(\bm{x}_{i}) \in \mathbb{R}^{p}, \ \ \forall i =1, \dots, m$. Distributing the standard sum-of-squares-error objective, the goal in distributed linear regression is to determine a set of latent parameters $\bm{\theta}_{1}, \dots, \bm{\theta}_{n}$ which minimize: 
\begin{align}
&\min_{\bm{\theta}_{1}:\bm{\theta}_{n}} \sum_{i=1}^{n} f_{i}(\bm{\theta}_{i}) \\ \nonumber 
&\text{s.t.} \ \ \bm{\theta}_{1} = \dots = \bm{\theta}_{n}, 
\end{align}
with $\bm{\theta}_{i} \in \mathbb{R}^{p}$ being a latent parameter determined by each processor, and 
\begin{equation}
\label{Eq:ChunkLinear}
f_{i}(\bm{\theta}_{i}) = \sum_{j=1}^{m_{i}} \left(a_{j} - \bm{\theta}_{i}^{\mathsf{T}}\bm{b}_{j}\right)^{2} + \mu_{i} m_{i} ||\bm{\theta}_{i}||_{2}^{2}, 
\end{equation}
with $\mu_{i}$ being a regularization parameter\footnote{In our experiments $\mu_{i}$'s were fixed for all nodes. These were set for values between \{0.01, 0.02, 0.05, 0.06, 0.1\} depending on the size of input dataset.}, and $\sum_{i=1}^{n} m_{i} = m$. To simplify the computations, next we rewrite Equation~\ref{Eq:ChunkLinear} in an equivalent matrix-vector form: 

\begin{align*}
&f_i(\boldsymbol{\theta}_i) = \boldsymbol{\theta}^{\mathsf{T}}_i\underbrace{\left[\mu_im_i\boldsymbol{I}_{p\times p} + \sum_{j=1}^{m_i}\boldsymbol{b}_j\boldsymbol{b}^{\mathsf{T}}_j\right]}_{\boldsymbol{P}_i}\boldsymbol{\theta}_i - 2\left(\underbrace{\sum_{j=1}^{m_i}a_j\boldsymbol{b}_j}_{\boldsymbol{c}_i}\right)^{\mathsf{T}}\boldsymbol{\theta}_i + \underbrace{\sum_{j=1}^{m_i}a^2_j}_{u_i} = \boldsymbol{\theta}^{\mathsf{T}}_i\boldsymbol{P}_i\boldsymbol{\theta}_i - 2\boldsymbol{c}^{\mathsf{T}}_i\boldsymbol{\theta}_i + u_i.
\end{align*}
Consequently: 

\begin{align}\label{lin_reg_params_expressions}
&\boldsymbol{P}_i = \boldsymbol{B}_i\boldsymbol{B}^{\mathsf{T}}_i + \mu_im_i\boldsymbol{I}_{p\times p},\\\nonumber
&\boldsymbol{c}_i = \boldsymbol{B}_i\boldsymbol{a}_i,\\\nonumber
&u_i = \boldsymbol{a}^{\mathsf{T}}_i\boldsymbol{a}_i,
\end{align}
where 
\begin{equation*}
\boldsymbol{B}_i=
\begin{blockarray}{cccc}
\begin{block}{[cccc]}
| & | &  & | \\
\boldsymbol{b}_1 & \boldsymbol{b}_2 & \ldots & \boldsymbol{b}_{m_i} \\
| & | &  & | \\
\end{block}
\end{blockarray}\quad \in \mathbb{R}^{p\times m_i} \ \ \ \ \ \ \ \ \text{and} \ \ \ \ \ \ \ \ \boldsymbol{a}_i =  \left[
\begin{array}{c}
a_1\\
a_2\\
\vdots\\
a_{m_i}
\end{array}
\right]\in\mathbb{R}^{m_i}.
\end{equation*}

Introducing $\boldsymbol{y}_1, \ldots, \boldsymbol{y}_p$, we can write the linear regression problem as: 

\begin{align*}
& \min_{\boldsymbol{y}_1,\ldots \boldsymbol{y}_p}\sum_{i=1}^nf_i(y_1(i), y_2(i), \ldots, y_p(i))\\\nonumber
&\text{s.t.} \hspace{0.2cm} \mathcal{L} \bm{y}_{1} = \mathcal{L} \bm{y}_{1} = \ldots = \mathcal{L} \bm{y}_{1} = \boldsymbol{0}, 
\end{align*}
where $\boldsymbol{\theta}_i = [y_1(i), y_2(i), \ldots, y_p(i)]^{\mathsf{T}}$, and:
\begin{align*}
&f_i(y_1(i), y_2(i), \ldots, y_p(i)) = \sum_{k=1,l=1}^{p,p}[\boldsymbol{P}_i]_{kl}y_k(i)y_l(i) - 2\sum_{k=1}^p[\boldsymbol{c}_i]_ky_k(i) + u^2_i.
\end{align*}
Hence, the Lagrangian of the problem can be written as:
\begin{align*}
&\mathcal{L}(\boldsymbol{y}_1, \ldots, \boldsymbol{y}_p, \boldsymbol{\lambda}_1, \ldots, \boldsymbol{\lambda}_p) = \sum_{i=1}^{n}\left[f_i(y_1(i), y_2(i), \ldots, y_p(i)) + y_1(i)(\boldsymbol{L\lambda}_1)_i + y_2(i)(\boldsymbol{L\lambda}_2)_i + \cdots + y_p(i)(\boldsymbol{L\lambda}_p)_i \right].
\end{align*}
Therefore, the primal variables can be recovered using: 
\begin{equation*}
\left[
\begin{array}{c}
y_1(i)\\
y_2(i)\\
\vdots\\
y_p(i)
\end{array}
\right]_{[\boldsymbol{\lambda}_1, \ldots, \boldsymbol{\lambda}_p]} = \boldsymbol{P}^{-1}_i\left[\boldsymbol{c}_i - \frac{1}{2}(\boldsymbol{L\Lambda})(i,:)]^{\mathsf{T}}\right],
\end{equation*}
where $(\boldsymbol{L\Lambda})(i,:)$ is $i^{th}$ row of $\mathcal{L}\bm{\Lambda}$. Note, that to apply the symmetric diagonally dominant the second time the Hessian of the local objective function $f_r(y_1(r), y_2(r), \ldots, y_p(r))$ must be computed. It is easy to see, that  $\nabla^2f_r(y_1(r), y_2(r), \ldots, y_p(r)) = 2\boldsymbol{P}_i$. 
\subsubsection{Linear Regression via ADMM}
In this section, we will describe distributed ADMM method for linear regression. Recall, that in distributed ADMM each node implements the following instructions:
\begin{enumerate}
\item \textbf{{Initialization:}} Chose arbitrary $\boldsymbol{\theta}^{[0]}_{i} \in\mathbb{R}^{p}$ and $\boldsymbol{\lambda}^{[0]}_{ji}\in\mathbb{R}^p$ for $j \in P(i)$ for $i=1,\ldots,n$.
\item \textbf{{For}} $k \ge 0$
\begin{enumerate}
\item Each agent $i$ updates its estimate of $\boldsymbol{\theta}^{[k]}_i$ in a sequential order with
\begin{align}\label{Serega_dalbich_111}
&\boldsymbol{\theta}^{[k+1]}_i = \arg\min_{\boldsymbol{\theta}^{}_i}\left(\underbrace{f_i(\boldsymbol{\theta}^{}_i) + \frac{\beta}{2}\sum_{j\in P(i)}\left|\left|\boldsymbol{\theta}^{[k+1]}_j - \boldsymbol{\theta}^{}_i - \frac{1}{\beta}\boldsymbol{\lambda}^{[k]}_{ji}  \right|\right|^2 + \frac{\beta}{2}\sum_{j\in S(i)}\left|\left|\boldsymbol{\theta}_i - \boldsymbol{\theta}^{[k]}_j - \frac{1}{\beta}\boldsymbol{\lambda}^{[k]}_{ij}\right|\right|^2}_{\xi_i}\right).
\end{align}
\item Each agent updates $\boldsymbol{\lambda}_{ji}$ for $j\in P(i)$ as follows:
\begin{equation*}
\boldsymbol{\lambda}^{[k+1]}_{ji} = \boldsymbol{\lambda}^{[k]}_{ji} - \beta(\boldsymbol{\theta}^{[k+1]}_j - \boldsymbol{\theta}^{[k+1]}_i).
\end{equation*}
\end{enumerate}
\end{enumerate}
We can get the closed form solution for (\ref{Serega_dalbich_111}):
\begin{align*}
&\nabla\zeta_i(\boldsymbol{\theta}_i) = \nabla f_i(\boldsymbol{\theta}_i) + \beta d(i)\boldsymbol{\theta}_i - \beta\left(\sum_{j\in S(i)}\left[\boldsymbol{\theta}^{[k]}_j  + \frac{1}{\beta}\boldsymbol{\lambda}^{[k]}_{ij}\right] + \sum_{j\in P(i)}\left[\boldsymbol{\theta}^{[k+1]}_{j} - \frac{1}{\beta}\boldsymbol{\lambda}^{[k]}_{ji}\right] \right) = \\\nonumber
&2\boldsymbol{P}_i\boldsymbol{\theta}_i - 2\boldsymbol{c}_i + \beta d(i)\boldsymbol{\theta}_i - \beta\left(\sum_{j\in S(i)}\left[\boldsymbol{\theta}^{[k]}_j  + \frac{1}{\beta}\boldsymbol{\lambda}^{[k]}_{ij}\right] + \sum_{j\in P(i)}\left[\boldsymbol{\theta}^{[k+1]}_{j} - \frac{1}{\beta}\boldsymbol{\lambda}^{[k]}_{ji}\right] \right).
\end{align*}
Hence, for the iterative rule (\ref{Serega_dalbich_11}) :
\begin{equation*}
\boldsymbol{\theta}^{[k+1]}_i = \left[\boldsymbol{P}_i + \frac{\beta d(i)}{2}\boldsymbol{I}_{p\times p}\right]^{-1}\left(\boldsymbol{c}_i + \frac{\beta}{2}\left(\sum_{j\in S(i)}\left[\boldsymbol{\theta}^{[k]}_j  + \frac{1}{\beta}\boldsymbol{\lambda}^{[k]}_{ij}\right] + \sum_{j\in P(i)}\left[\boldsymbol{\theta}^{[k+1]}_{j} - \frac{1}{\beta}\boldsymbol{\lambda}^{[k]}_{ji}\right] \right) \right).
\end{equation*} 
\subsubsection{Linear Regression via Distributed Averaging}
In distributed averaging, each node keeps three variables:
\begin{enumerate}
\item \textbf{{Initialization:}} For each node $i\in\mathbb{V}$ initialize $\boldsymbol{\theta}_i(1)\in\mathbb{R}^p$ and set
\begin{equation*}
\boldsymbol{z}_i(1) = \boldsymbol{\omega}_i(1) = \boldsymbol{\theta}_i(1).
\end{equation*}
\item Each node $i\in\mathbb{V}$ implements the following instructions:
\begin{align*}
&\boldsymbol{\omega}_i(t+1) = \boldsymbol{\theta}_i(t) + \frac{1}{2}\sum_{j\in\mathbb{N}(i)}\frac{\boldsymbol{\theta}_j(t) - \boldsymbol{\theta}_i(t)}{\max\{d(i), d(j)\}} - \beta \boldsymbol{g}_i(t)\\\nonumber
&\boldsymbol{z}_{i}(t+1) = \boldsymbol{w}_i(t) - \beta \boldsymbol{g}_i(t)\\\nonumber
&\boldsymbol{\theta}_i(t+1) = \boldsymbol{\omega}_i(t+1) + \left(1 - \frac{2}{9n + 1}\right)(\boldsymbol{\omega}_i(t+1) - \boldsymbol{z}_i(t+1)),
\end{align*}
\end{enumerate}
where $\beta$ is a step-size and $\boldsymbol{g}_i(t)$ is the sub-gradient of $f_i$ evaluated at $\boldsymbol{w}_i(t)$, i.e.:
\begin{align*}
&\boldsymbol{g}_i(t) = \nabla f_i(\boldsymbol{w}_i(t)) 
=2\boldsymbol{P}_i\boldsymbol{w}_i(t) - 2\boldsymbol{c}_i.
\end{align*}
After $T$ iterations the node $i\in\mathbb{V}$ computes the average
\begin{equation}
\bar{\boldsymbol{w}}_i = \frac{1}{T}\sum_{t=1}^{T}\boldsymbol{w}_i(t),
\end{equation}
as a solution.

\subsection{Logistic Regression}
Next, we repeat the above for logistic regression considering both smooth and non-smooth regularizers. We consider a data set defined by the following collection $\left(a_{i}, \bm{b}_{i}\right)_{i=1}^{m}$ with $a_{i} \in \{0,1\}$ representing a class and $\bm{b}_{i} = \bm{\Phi}(\bm{x}_{i}) \in \mathbb{R}^{p}$. Similar to the linear regression case, the goal is to determine the solution of the following objective: 
\begin{align}\label{gcp_log_regression}
&\min_{\bm{\theta}_{1}: \bm{\theta}_{n}} \sum_{i=1}^{n}f_i(\boldsymbol{\theta}_i)\\\nonumber
&\text{s.t.} \hspace{0.2cm} \boldsymbol{\theta}_1 = \ldots =\boldsymbol{\theta}_n \in\mathbb{R}^p.
\end{align}
Here, however, each local cost is given as a logistic loss defined as: 

\begin{equation}\label{local_objective_common}
f_i(\boldsymbol{\theta}_i) = -\sum_{j=1}^{m_i}\left[a_j\log\frac{1}{1 + e^{-\boldsymbol{\theta}_i\boldsymbol{b}_j}} + (1 - a_j)\log\left(1 - \frac{1}{1 + e^{-\boldsymbol{\theta}_i\boldsymbol{b}_j}}\right)\right] + \mu_i m_i\Psi(\boldsymbol{\theta}_i),
\end{equation}
with $\mu_{i}$ being a regularization parameter and $\sum_{i=1}^{n} m_{i} = m$. $\Psi(\bm{\theta}_{i})$ is a regularization function. Next, we consider two such cases when $\Psi$ is both smooth and non-smooth. 
\subsubsection{Smooth Regularizers}
When $\Psi(\bm{\theta}_{i}) = ||\bm{\theta}_{i} ||_{2}^{2}$, the local objective in Equation~\ref{local_objective_common} can be written as: 

\begin{equation}\label{local_objective_11}
f_i(\boldsymbol{\theta}_i) = -\sum_{j=1}^{m_i}\left[a_j\log\frac{1}{1 + e^{-\boldsymbol{\theta}_i\boldsymbol{b}_j}} + (1 - a_j)\log\left(1 - \frac{1}{1 + e^{-\boldsymbol{\theta}_i\boldsymbol{b}_j}}\right)\right] + \mu_i m_i||\boldsymbol{\theta}_i||^2_2.
\end{equation}
It is easy to see that Equation~\ref{local_objective_11} can be simplified as:
\begin{align*}
f_i(\boldsymbol{\theta}_i) &= -\sum_{j=1}^{m_i}\left[a_j\log\frac{1}{1 + e^{-\boldsymbol{\theta}_i\boldsymbol{b}_j}} + (1 - a_j)\log\left(1 - \frac{1}{1 + e^{-\boldsymbol{\theta}_i\boldsymbol{b}_j}}\right)\right] + \mu_i m_i||\boldsymbol{\theta}_i||^2_2  \\\nonumber
&=-\sum_{j=1}^{m_i}\left[a_j\boldsymbol{\theta}^{\mathsf{T}}_i\boldsymbol{b}_j - \log(1 + e^{\boldsymbol{\theta}^{\mathsf{T}}_i\boldsymbol{b}_j})\right] + \mu_i m_i||\boldsymbol{\theta}_i||^2_2  \\
&= -\left(\sum_{j=1}^{m_i}a_j\boldsymbol{b}_j\right)^{\mathsf{T}}\boldsymbol{\theta}_i + \sum_{j=1}^{m_i}\log(1 + e^{\boldsymbol{\theta}^{\mathsf{T}}_i\boldsymbol{b}_j}) + \mu_im_i||\boldsymbol{\theta}_i||^2_2.
\end{align*}
Introducing
\begin{equation}\label{y_vectors}
\boldsymbol{y}_1 = \left[
\begin{array}{c}
\theta_1(1)\\
\theta_2(1)\\
\vdots\\
\theta_n(1)
\end{array}
\right]\hspace{0.2cm}
\boldsymbol{y}_2 = \left[
\begin{array}{c}
\theta_1(2)\\
\theta_2(2)\\
\vdots\\
\theta_n(2)
\end{array}
\right]\hspace{0.1cm}\ldots
\boldsymbol{y}_p = \left[
\begin{array}{c}
\theta_1(p)\\
\theta_2(p)\\
\vdots\\
\theta_n(p)
\end{array}
\right]
\end{equation}
and 
\begin{equation}\label{b_vectors}
\boldsymbol{b}_1 = \left[
\begin{array}{c}
b_1(1)\\
b_1(2)\\
\vdots\\
b_1(p))
\end{array}
\right]\hspace{0.2cm}
\boldsymbol{b}_2 = \left[
\begin{array}{c}
b_2(1)\\
b_2(2)\\
\vdots\\
b_2(p)
\end{array}
\right]\hspace{0.1cm}\ldots
\boldsymbol{b}_{m_i} = \left[
\begin{array}{c}
b_{m_i}(1)\\
b_{m_i}(2)\\
\vdots\\
b_{m_i}(p)
\end{array}
\right],
\end{equation}
the above problem can be rewritten as: 
\begin{align*}
& \min_{\boldsymbol{y}_1,\ldots \boldsymbol{y}_p}\sum_{i=1}^nf_i(y_1(i), y_2(i), \ldots, y_p(i))\\\nonumber
&\text{s.t.}\hspace{0.2cm} \mathcal{L}\bm{y}_1 = \mathcal{L}\bm{y}_2 = \ldots = \mathcal{L}\bm{y}_p = \boldsymbol{0}, 
\end{align*}
where $\boldsymbol{\theta}_i = [y_1(i), y_2(i), \ldots, y_p(i)]^{\mathsf{T}}$. Using Equation~\ref{local_objective_11}:
\begin{align*}
&f_i(y_1(i), y_2(i), \ldots, y_p(i)) = -\sum_{r=1}^p\left(\sum_{j =1}^{m_i}a_jb_j(r)\right)y_r(i) + \sum_{j=1}^{m_i}\log\left(1 + e^{\sum_{r=1}^{p}y_r(i)b_j(r)}\right) + \mu_im_i\sum_{r=1}^{p}y^2_r(i).
\end{align*}
Hence, the Lagrangian of the above problem can be written as:
\begin{align*}
\mathcal{L}(\boldsymbol{y}_1, \ldots, \boldsymbol{y}_p, \boldsymbol{\lambda}_1, \ldots, \boldsymbol{\lambda}_p) = 
\sum_{i=1}^{n}\left[f_i(y_1(i), y_2(i), \ldots, y_p(i)) + y_1(i)(\boldsymbol{L\lambda}_1)_i + y_2(i)(\boldsymbol{L\lambda}_2)_i + \cdots + y_p(i)(\boldsymbol{L\lambda}_p)_i \right].
\end{align*}
To define the relation between primal an dual variables, each node $i\in\mathbb{V}$ needs to solve the corresponding optimization problem of the form:
\begin{align}\label{primal_dual_relation_1}
&\min_{y_1(i), \ldots, y_p(i)}\underbrace{f_i(y_1(i), y_2(i), \ldots, y_p(i)) + \sum_{r=1}^{p}y_r(i)(\boldsymbol{L\lambda}_r)_i}_{\zeta_i}   
\end{align}
Applying the standard Newton method for Equation~\ref{primal_dual_relation_1}:
\begin{equation}\label{grad_descent}
\left[
\begin{array}{c}
y_1(i)\\
y_2(i)\\
\vdots\\
y_p(i)
\end{array}
\right]_{(t+1)} = 
\hspace{0.2cm}
\left[
\begin{array}{c}
y_1(i)\\
y_2(i)\\
\vdots\\
y_p(i)
\end{array}
\right]_{(t)} -
\hspace{0.1cm}
\alpha_t\boldsymbol{y}^{[i]}_{Newton}|_{(t)},
\end{equation}
where $t$ is the iteration count, and $\boldsymbol{y}^{[i]}_{Newton}|_{(t)}$ is Newton direction, evaluated at $[y_1(i), \ldots, y_p(i)]_{(t)}$, and given by the solution of the following system:
\begin{equation}\label{Newton_direction}
\boldsymbol{H}_{i}|_{(t)}\boldsymbol{y}^{[i]}_{Newton}|_{(t)} = - \nabla\zeta_i|_{(t)},
\end{equation}
with $\boldsymbol{H}_{i}|_{(t)}$ and $\nabla\zeta_i|_{(t)}$ being the Hessian and noting that the gradient of $\zeta_i$ is evaluated at $[y_1(i), \ldots, y_p(i)]_{(t)}$. 
The components of the gradient $\nabla\zeta_i|_{(t)}$ can be computed as:
\begin{align}\label{gradient_zeta_equation}
&\frac{\partial \zeta_i}{\partial y_1(i)_{(t)}} =  \sum_{j=1}^{m_i}\left[-a_j + \frac{e^{\sum_{r=1}^{p}y_r(i)_{(t)}b_j(r)}}{1 + e^{\sum_{r=1}^{p}y_r(i)_{(t)}b_j(r)}}\right]b_j(1) + 2\mu_im_iy_1(i)_{(t)} + (\boldsymbol{L\lambda}_1)_i \\\nonumber
&\frac{\partial \zeta_i}{\partial y_2(i)_{(t)}} =  \sum_{j=1}^{m_i}\left[-a_j + \frac{e^{\sum_{r=1}^{p}y_r(i)_{(t)}b_j(r)}}{1 + e^{\sum_{r=1}^{p}y_r(i)_{(t)}b_j(r)}}\right]b_j(2) + 2\mu_im_iy_2(i)_{(t)} + (\boldsymbol{L\lambda}_2)_i \\\nonumber
&\vdots\\\nonumber
&\frac{\partial \zeta_i}{\partial y_p(i)_{(t)}} =  \sum_{j=1}^{m_i}\left[-a_j + \frac{e^{\sum_{r=1}^{p}y_r(i)_{(t)}b_j(r)}}{1 + e^{\sum_{r=1}^{p}y_r(i)_{(t)}b_j(r)}}\right]b_j(p) + 2\mu_im_iy_p(i)_{(t)} + (\boldsymbol{L\lambda}_p)_i
\end{align}
This can be written in the following matrix-vector form: 
\begin{equation}\label{gradient_zeta_vector_equation}
\nabla\zeta_i|_{(t)} = \boldsymbol{B}_i\boldsymbol{\delta}_{(t)} + 2\mu_im_i\left[
\begin{array}{c}
y_1(i)\\
y_2(i)\\
\vdots\\
y_p(i)
\end{array}
\right]_{(t)} + [(\boldsymbol{L\Lambda})(i,:)]^{\mathsf{T}}
\end{equation}
where  $(\boldsymbol{L\Lambda})(i,:)$ is $i^{th}$ row of matrix $\boldsymbol{L\Lambda}$ and:
\begin{equation}\label{B_matrix}
\boldsymbol{B}_i=
\begin{blockarray}{cccc}
\begin{block}{[cccc]}
| & | &  & | \\
\boldsymbol{b}_1 & \boldsymbol{b}_2 & \ldots & \boldsymbol{b}_{m_i} \\
| & | &  & | \\
\end{block}
\end{blockarray}\quad \in \mathbb{R}^{p\times m_i}
\end{equation}
\begin{equation}\label{Lambda_matrix}
\boldsymbol{\Lambda}=
\begin{blockarray}{cccc}
\begin{block}{[cccc]}
| & | &  & | \\
\boldsymbol{\lambda}_1 & \boldsymbol{\lambda}_2 & \ldots & \boldsymbol{\lambda}_{p} \\
| & | &  & | \\
\end{block}
\end{blockarray}\quad \in \mathbb{R}^{n\times p}.
\end{equation}
Finally, we note that: 
\begin{equation}\label{delta_vector}
\boldsymbol{\delta}|_{(t)} = \left[
\begin{array}{c}
\delta_1|_{(t)}\\
\delta_2|_{(t)}\\
\vdots\\
\delta_{m_i}|_{(t)}
\end{array}
\right] \hspace{0.3cm} \text{with} \hspace{0.3cm} \delta_j|_{(t)}= -a_j + \frac{e^{\sum_{r=1}^{p}y_r(i)_{(t)}b_j(r)}}{1 + e^{\sum_{r=1}^{p}y_r(i)_{(t)}b_j(r)}}. 
\end{equation}

The Hessian of $\zeta_i$ can be immediately written as:
\begin{equation}\label{Hessian_zeta_i_matrix_equation_1}
\boldsymbol{H}_i|_{(t)} = 2\mu_im_i\boldsymbol{I}_{p\times p} + \boldsymbol{B}_i\boldsymbol{D}_{i}|_{(t)}\boldsymbol{B}^{\mathsf{T}}_i,
\end{equation}
where $\boldsymbol{D}_{i}|_{(t)}$ is diagonal $m_i\times m_i$ matrix, such that
\begin{equation*}
[\boldsymbol{D}_{i}|_{(t)}]_{jj} = \frac{e^{\sum_{l=1}^py_l(i)_{(t)}b_j(l)}}{\left(1 + e^{\sum_{l=1}^py_l(i)_{(t)}b_j(l)}\right)^2}.
\end{equation*}

It is again easy to see that to apply the SDDM-solver the Hessian of the local objective is needed. This can be derived as: 
\begin{equation*}
\nabla^2f_r = 2\mu_rm_r\boldsymbol{I}_{p\times p} + \boldsymbol{B}_r\boldsymbol{D}_{r}\boldsymbol{B}^{\mathsf{T}}_r
\end{equation*}
where $\boldsymbol{D}_{r}$ is diagonal matrix, given by:
\begin{equation*}
[\boldsymbol{D}_{r}]_{jj} = \frac{e^{\sum_{l=1}^py_l(y)b_j(l)}}{\left(1 + e^{\sum_{l=1}^py_l(r)b_j(l)}\right)^2}
\end{equation*}

\paragraph{Smooth Regularizers for ADMM}
The derivations, so-far, are appropriate for the proposed distributed Newton method. To be able to compare against ADMM, corresponding mathematics needs to be developed. This section details such constructs. 
We start by recalling that each node in distributed ADMM implements the following instructions:
\begin{enumerate}
\item \textbf{{Initialization:}} Chose $\boldsymbol{\theta}^{[0]}_{i} \in\mathbb{R}^{p}$ and $\boldsymbol{\lambda}^{[0]}_{ji}\in\mathbb{R}^p$ for $j \in P(i)$ for $i=1,\ldots,n$ with $P(i)$ being the set of predecessors of node $i$.
\item \textbf{{For}} $k \ge 0$
\begin{enumerate}
\item Each agent $i$ updates its estimate of $\boldsymbol{\theta}^{[k]}_i$ in a sequential order with:
\begin{align}\label{Serega_dalbich_1}
\boldsymbol{\theta}^{[k+1]}_i = \arg\min_{\boldsymbol{\theta}^{}_i}\left(\underbrace{f_i(\boldsymbol{\theta}^{}_i) + \frac{\beta}{2}\sum_{j\in P(i)}\left|\left|\boldsymbol{\theta}^{[k+1]}_j - \boldsymbol{\theta}^{}_i - \frac{1}{\beta}\boldsymbol{\lambda}^{[k]}_{ji}  \right|\right|^2 + \frac{\beta}{2}\sum_{j\in S(i)}\left|\left|\boldsymbol{\theta}_i - \boldsymbol{\theta}^{[k]}_j - \frac{1}{\beta}\boldsymbol{\lambda}^{[k]}_{ij}\right|\right|^2}_{\xi_i}\right).
\end{align}
\item Each agent updates $\boldsymbol{\lambda}_{ji}$ for $j\in P(i)$ as follows:
\begin{equation}\label{Serega_dalbich_2}
\boldsymbol{\lambda}^{[k+1]}_{ji} = \boldsymbol{\lambda}^{[k]}_{ji} - \beta(\boldsymbol{\theta}^{[k+1]}_j - \boldsymbol{\theta}^{[k+1]}_i).
\end{equation}
\end{enumerate}
\end{enumerate}
For solving the optimization problem in Equation~\ref{Serega_dalbich_1}, standard Newton is used for function $\xi_i$:
\begin{equation}\label{newton_method_admm}
\boldsymbol{\theta}^{[k+1]}_i(t+1) = \boldsymbol{\theta}^{[k+1]}_i(t) - \alpha_t\boldsymbol{\theta}^{[i]}_{Newton}|_{(t)}, 
\end{equation}
where $\alpha_t$ is a step-size and  $\boldsymbol{\theta}^{[i]}_{Newton}|_{(t)}$ is the Newton Direction, evaluated at $\boldsymbol{\theta}^{[k+1]}_i(t)$, and given by the solution to the following system:
\begin{equation}\label{dist_admm_newtom_direction}
\boldsymbol{H}_{i}|_{(t)}\boldsymbol{\theta}^{[i]}_{Newton}|_{(t)} = - \nabla\xi_i(\boldsymbol{\theta}^{[k+1]}_i(t)),
\end{equation}
with $\boldsymbol{H}_{i}|_{(t)}$ and $\nabla\xi_i(\boldsymbol{\theta}^{[k+1]}_i(t))$ being the Hessian and the gradient of function $\xi_i$ evaluated at vector $\boldsymbol{\theta}^{[k+1]}_i(t)$. Clearly, the gradient can be written in a vector form as: 
\begin{align*}
&\nabla\xi_i(\boldsymbol{\theta}^{[k+1]}_i(t)) = \boldsymbol{B}_i\boldsymbol{\delta}|_{(t)} + (2\mu_im_i + \beta d(i))\boldsymbol{\theta}^{[k+1]}_i(t)  - \\\nonumber
&\beta\left(\sum_{j\in S(i)}\left[\boldsymbol{\theta}^{[k]}_j + \frac{1}{\beta}\boldsymbol{\lambda}^{[k]}_{ij}\right] + \sum_{j\in P(i)}\left[\boldsymbol{\theta}^{[k+1]}_j - \frac{1}{\beta}\boldsymbol{\lambda}^{[k]}_{ji}\right]\right),
\end{align*}
where $d(i)$ is the degree of node $i\in\mathbb{V}$, $\boldsymbol{B}_i$ is given in Equation~\ref{B_matrix}, and
\begin{equation}\label{delta_vector_admm}
\boldsymbol{\delta}|_{(t)} = \left[
\begin{array}{c}
\delta_1|_{(t)}\\
\delta_2|_{(t)}\\
\vdots\\
\delta_{m_i}|_{(t)}
\end{array}
\right] \hspace{0.3cm} \text{with} \hspace{0.3cm} \delta_j|_{(t)}= -a_j + \frac{e^{\boldsymbol{b}^{\mathsf{T}}_j\boldsymbol{\theta}^{[k+1]}_i(t)}}{1 + e^{\boldsymbol{b}^{\mathsf{T}}_j\boldsymbol{\theta}^{[k+1]}_i(t)}}. 
\end{equation}
The Hessian of $\xi_i$ can be immediately written as:
\begin{align*}
&\boldsymbol{H}_{i}|_{(t)} = \nabla^2\xi_i(\boldsymbol{\theta}^{[k+1]}_i(t)) = \nabla^2f_i(\boldsymbol{\theta}^{[k+1]}_i(t)) + \beta d(i)\boldsymbol{I}_{p\times p} =  \boldsymbol{B}_i\boldsymbol{D}_{i}|_{(t)}\boldsymbol{B}^{\mathsf{T}}_i + (2\mu_im_i + \beta d(i))\boldsymbol{I}_{p\times p},
\end{align*}
where $\boldsymbol{D}_{i}|_{(t)}$ is diagonal $m_i \times m_i$ matrix, such that
\begin{equation}\label{D_i_matrix_admm}
[\boldsymbol{D}_{i}|_{(t)}]_{jj} = \frac{e^{\boldsymbol{b}^{\mathsf{T}}_j\boldsymbol{\theta}^{[k+1]}_i(t)}}{\left(1 + e^{\boldsymbol{b}^{\mathsf{T}}_j\boldsymbol{\theta}^{[k+1]}_i(t)}\right)^2} .
\end{equation}
\paragraph{Smooth Regularizers for Distributed Averaging}
In this section, we describe the necessary mathematics needed for distributed averaging. We recall that distributed averaging operates as: 
\begin{enumerate}
\item \textbf{{Initialization:}} For each node $i\in\mathbb{V}$, initialize $\boldsymbol{\theta}_i(1)\in\mathbb{R}^p$ and set
\begin{equation*}
\boldsymbol{z}_i(1) = \boldsymbol{\omega}_i(1) = \boldsymbol{\theta}_i(1)
\end{equation*}
\item Each node $i\in\mathbb{V}$ implements the following instructions:
\begin{align}\label{Olshevsky_iter_scheme}
&\boldsymbol{\omega}_i(t+1) = \boldsymbol{\theta}_i(t) + \frac{1}{2}\sum_{j\in\mathbb{N}(i)}\frac{\boldsymbol{\theta}_j(t) - \boldsymbol{\theta}_i(t)}{\max\{d(i), d(j)\}} - \beta \boldsymbol{g}_i(t),\\\nonumber
&\boldsymbol{z}_{i}(t+1) = \boldsymbol{w}_i(t) - \beta \boldsymbol{g}_i(t),\\\nonumber
&\boldsymbol{\theta}_i(t+1) = \boldsymbol{\omega}_i(t+1) + \left(1 - \frac{2}{9n + 1}\right)(\boldsymbol{\omega}_i(t+1) - \boldsymbol{z}_i(t+1)).
\end{align}
\end{enumerate}
where $\beta$ is a step-size and $\boldsymbol{g}_i(t)$ is the sub-gradient of $f_i$ evaluated at $\boldsymbol{w}_i(t)$, i.e.:
\begin{align*}
&\boldsymbol{g}_i(t) = \nabla f_i(\boldsymbol{w}_i(t)) = \boldsymbol{B}_i\boldsymbol{\delta}_{(t)} + 2\mu_im_i\boldsymbol{w}_i(t),
\end{align*}
where 

\begin{equation}\label{delta_vector_Olshevsky}
\boldsymbol{\delta}|_{(t)} = \left[
\begin{array}{c}
\delta_1|_{(t)}\\
\delta_2|_{(t)}\\
\vdots\\
\delta_{m_i}|_{(t)}
\end{array}
\right] \hspace{0.3cm} \text{with} \hspace{0.3cm} \delta_j|_{(t)}= -a_j + \frac{e^{\boldsymbol{b}^{\mathsf{T}}_j\boldsymbol{w}_i(t)}}{1 + e^{\boldsymbol{b}^{\mathsf{T}}_j\boldsymbol{w}_i(t)}}. 
\end{equation}

After $T$ iterations, node $i\in\mathbb{V}$ computes the average:
\begin{equation}
\bar{\boldsymbol{w}}_i = \frac{1}{T}\sum_{t=1}^{T}\boldsymbol{w}_i(t),
\end{equation}
as the solution.

\subsubsection{Non-Smooth Regularizers}
Now, we consider the case when the local objective is defined by: 
\begin{equation}\label{local_objective_1}
f_i(\boldsymbol{\theta}_i) = -\sum_{j=1}^{m_i}\left[a_j\log\frac{1}{1 + e^{-\boldsymbol{\theta}_i\boldsymbol{b}_j}} + (1 - a_j)\log\left(1 - \frac{1}{1 + e^{-\boldsymbol{\theta}_i\boldsymbol{b}_j}}\right)\right] + \mu_i m_i||\boldsymbol{\theta}_i||_1.
\end{equation}
It is easy to see that (\ref{local_objective_1}) can be simplified as:
\begin{align*}
&f_i(\boldsymbol{\theta}_i) = -\sum_{j=1}^{m_i}\left[a_j\log\frac{1}{1 + e^{-\boldsymbol{\theta}_i\boldsymbol{b}_j}} + (1 - a_j)\log\left(1 - \frac{1}{1 + e^{-\boldsymbol{\theta}_i\boldsymbol{b}_j}}\right)\right] + \mu_i m_i||\boldsymbol{\theta}_i||_1 = \\\nonumber
&-\sum_{j=1}^{m_i}\left[a_j\boldsymbol{\theta}^{\mathsf{T}}_i\boldsymbol{b}_j - \log(1 + e^{\boldsymbol{\theta}^{\mathsf{T}}_i\boldsymbol{b}_j})\right] + \mu_i m_i||\boldsymbol{\theta}_i||_1 = \\\nonumber
&-\left(\sum_{j=1}^{m_i}a_j\boldsymbol{b}_j\right)^{\mathsf{T}}\boldsymbol{\theta}_i + \sum_{j=1}^{m_i}\log(1 + e^{\boldsymbol{\theta}^{\mathsf{T}}_i\boldsymbol{b}_j}) + \mu_im_i||\boldsymbol{\theta}_i||_1.
\end{align*}
Similar to Equations~\ref{y_vectors} and~\ref{b_vectors}, we introduce $\boldsymbol{y}_1, \ldots, \boldsymbol{y}_p$ and $\boldsymbol{b}_1, \ldots, \boldsymbol{b}_{m_i}$. Then, the problem can be written as:
\begin{align}
& \min_{\boldsymbol{y}_1,\ldots \boldsymbol{y}_p}\sum_{i=1}^nf_i(y_1(i), y_2(i), \ldots, y_p(i))\\\nonumber
& \text{s.t.} \hspace{0.2cm} \mathcal{L}\bm{y}_1 = \mathcal{L}\bm{y}_2 = \ldots = \mathcal{L}\bm{y}_p = \boldsymbol{0}, 
\end{align}
where $\boldsymbol{\theta}_i = [y_1(i), y_2(i), \ldots, y_p(i)]^{\mathsf{T}}$, and using Equation~\ref{local_objective_1} we have:
\begin{align*}
&f_i(y_1(i), y_2(i), \ldots, y_p(i)) = -\sum_{r=1}^p\left(\sum_{j =1}^{m_i}a_jb_j(r)\right)y_r(i) + \sum_{j=1}^{m_i}\log\left(1 + e^{\sum_{r=1}^{p}y_r(i)b_j(r)}\right) + \mu_im_i\sum_{r=1}^{p}|y_r(i)|.
\end{align*}

Hence, the Lagrangian can be written as:
\begin{align*}
&\mathcal{L}(\boldsymbol{y}_1, \ldots, \boldsymbol{y}_p, \boldsymbol{\lambda}_1, \ldots, \boldsymbol{\lambda}_p) = \\\nonumber
&\sum_{i=1}^{n}\left[f_i(y_1(i), y_2(i), \ldots, y_p(i)) + y_1(i)(\boldsymbol{L\lambda}_1)_i + y_2(i)(\boldsymbol{L\lambda}_2)_i + \cdots + y_p(i)(\boldsymbol{L\lambda}_p)_i \right].
\end{align*}
In order to define the relation between primal an dual variables, each node $i\in\mathbb{V}$ needs to solve the corresponding optimization problem of the form:
\begin{align}\label{primal_dual_relation}
&\min_{y_1(i), \ldots, y_p(i)}\underbrace{f_i(y_1(i), y_2(i), \ldots, y_p(i)) + \sum_{r=1}^{p}y_r(i)(\boldsymbol{L\lambda}_r)_i}_{\zeta_i} .  
\end{align}
Clearly, $\zeta_i$ is not smooth. To proceed, we use the smooth approximations of the L$_{1}$ norm, given by:
\begin{equation}\label{smooth_approx_1}
|x|_{(\alpha)} = \frac{1}{\alpha}\left[\log(1 + e^{-\alpha x}) + \log(1 + e^{\alpha x})\right],
\end{equation}
where $\alpha$ is parameter controlling the approximation's quality. Therefore, the local objective can be written as:
\begin{align*}
&f_i(y_1(i), y_2(i), \ldots, y_p(i)) = \\\nonumber
&-\sum_{r=1}^p\left(\sum_{j =1}^{m_i}a_jb_j(r)\right)y_r(i) + \sum_{j=1}^{m_i}\log\left(1 + e^{\sum_{r=1}^{p}y_r(i)b_j(r)}\right) + \\\nonumber
&\frac{1}{\alpha}\mu_im_i\sum_{r=1}^{p}\left[\log(1 + e^{-\alpha y_r(i)}) + \log(1 + e^{\alpha y_r(i)})\right] = \\\nonumber
&-\sum_{r=1}^p\left(\sum_{j =1}^{m_i}a_jb_j(r)\right)y_r(i) + \sum_{j=1}^{m_i}\log\left(1 + e^{\sum_{r=1}^{p}y_r(i)b_j(r)}\right) + \\\nonumber
&\frac{1}{\alpha}\mu_im_i\sum_{r=1}^{p}\left[2\log(1 + e^{\alpha y_r(i)}) - \alpha y_r(i)\right].
\end{align*}

Applying standard Newton for Equation~\ref{primal_dual_relation} gives:
\begin{equation}\label{grad_descent}
\left[
\begin{array}{c}
y_1(i)\\
y_2(i)\\
\vdots\\
y_p(i)
\end{array}
\right]_{(t+1)} = 
\hspace{0.2cm}
\left[
\begin{array}{c}
y_1(i)\\
y_2(i)\\
\vdots\\
y_p(i)
\end{array}
\right]_{(t)} -
\hspace{0.1cm}
\gamma_t\boldsymbol{y}^{[i]}_{Newton}|_{(t)},
\end{equation}
where $t$ is the iteration count, and $\boldsymbol{y}^{[i]}_{Newton}|_{(t)}$ is Newton direction, evaluated at vector $[y_1(i), \ldots, y_p(i)]_{(t)}$, and given by the solution to the following system:
\begin{equation}\label{Newton_direction}
\boldsymbol{H}_{i}|_{(t)}\boldsymbol{y}^{[i]}_{Newton}|_{(t)} = - \nabla\zeta_i|_{(t)},
\end{equation}
with $\boldsymbol{H}_{i}|_{(t)}$ and $\nabla\zeta_i|_{(t)}$ being the Hessian and the gradient of $\zeta_i$ evaluated at vector $[y_1(i), \ldots, y_p(i)]_{(t)}$. 
The components of the gradient $\nabla\zeta_i|_{(t)}$ can be computed as:
\begin{align}\label{gradient_zeta_equation}
&\frac{\partial \zeta_i}{\partial y_1(i)_{(t)}} =  \sum_{j=1}^{m_i}\left[-a_j + \frac{e^{\sum_{r=1}^{p}y_r(i)_{(t)}b_j(r)}}{1 + e^{\sum_{r=1}^{p}y_r(i)_{(t)}b_j(r)}}\right]b_j(1) + \mu_im_i\frac{e^{\alpha y_1(i)_{(t)}} - 1}{ 1+ e^{\alpha y_1(i)_{(t)}}}  + (\boldsymbol{L\lambda}_1)_i \\\nonumber
&\frac{\partial \zeta_i}{\partial y_2(i)_{(t)}} =  \sum_{j=1}^{m_i}\left[-a_j + \frac{e^{\sum_{r=1}^{p}y_r(i)_{(t)}b_j(r)}}{1 + e^{\sum_{r=1}^{p}y_r(i)_{(t)}b_j(r)}}\right]b_j(2) + \mu_im_i\frac{e^{\alpha y_2(i)_{(t)}} - 1}{ 1+ e^{\alpha y_2(i)_{(t)}}} + (\boldsymbol{L\lambda}_2)_i \\\nonumber
&\vdots\\\nonumber
&\frac{\partial \zeta_i}{\partial y_p(i)_{(t)}} =  \sum_{j=1}^{m_i}\left[-a_j + \frac{e^{\sum_{r=1}^{p}y_r(i)_{(t)}b_j(r)}}{1 + e^{\sum_{r=1}^{p}y_r(i)_{(t)}b_j(r)}}\right]b_j(p) + \mu_im_i\frac{e^{\alpha y_p(i)_{(t)}} - 1}{ 1+ e^{\alpha y_p(i)_{(t)}}} + (\boldsymbol{L\lambda}_p)_i.
\end{align}
The above can be written in a vector-matrix form as: 
\begin{equation}\label{gradient_zeta_vector_equation}
\nabla\zeta_i|_{(t)} = \boldsymbol{B}_i\boldsymbol{\delta}|_{(t)} + \mu_im_i \boldsymbol{\rho}|_{(t)}+ [(\boldsymbol{L\Lambda})(i,:)]^{\mathsf{T}},
\end{equation}
where $\boldsymbol{B}_i, \boldsymbol{\Lambda}, \boldsymbol{\delta}|_{(t)}$ are defined in (\ref{B_matrix}),(\ref{Lambda_matrix}), (\ref{delta_vector}) respectively, $(\boldsymbol{L\Lambda})(i,:)$ is the $i^{th}$ row of matrix $\boldsymbol{L\Lambda}$,  and 
\begin{equation}\label{rho_vector}
\boldsymbol{\rho}|_{(t)} = \left[
\begin{array}{c}
\rho_1|_{(t)}\\
\rho_2|_{(t)}\\
\vdots\\
\rho_p|_{(t)}
\end{array}
\right] \hspace{0.3cm} \textit{with} \hspace{0.3cm} \rho_j|_{(t)} = \frac{e^{\alpha y_j(i)_{(t)}} - 1}{ 1+ e^{\alpha y_j(i)_{(t)}}}.
\end{equation}

The Hessian of $\zeta_i$ can be immediately written as:
\begin{equation}\label{Hessian_zeta_i_new}
\boldsymbol{H}_i|_{(t)} = \boldsymbol{B}_i\boldsymbol{D}_i|_{(t)}\boldsymbol{B}^{\mathsf{T}}_i + 2\alpha\mu_im_i\boldsymbol{\Delta}_i|_{(t)},
\end{equation} 
where $\boldsymbol{D}_{i}|_{(t)}$ is a diagonal $m_i\times m_i$ matrix, such that
\begin{equation*}
[\boldsymbol{D}_{i}|_{(t)}]_{jj} = \frac{e^{\sum_{l=1}^py_l(i)_{(t)}b_j(l)}}{\left(1 + e^{\sum_{l=1}^py_l(i)_{(t)}b_j(l)}\right)^2},
\end{equation*}
and $\boldsymbol{\Delta}_i|_{(t)}$ is diagonal $p\times p$ matrix, such that 
\begin{equation*}
[\boldsymbol{\Delta}_i|_{(t)}]_{jj} = \frac{e^{\alpha y_j(i)_{(t)}}}{(1 + e^{\alpha y_j(i)_{(t)}})^2}.
\end{equation*}
To apply SDDM-solver, the Hessian of the local objective function $f_r(y_1(r), y_2(r), \ldots, y_p(r))$ must be computed. It is easy to see that:
\begin{equation*}
\nabla^2f_r = \boldsymbol{B}_r\boldsymbol{D}_{r}\boldsymbol{B}^{\mathsf{T}}_r + 2\alpha\mu_rm_r\boldsymbol{\Delta}_r,
\end{equation*}
where $\boldsymbol{D}_r$ and $\boldsymbol{\Delta}_{r}$ are diagonal matrices, given by:
\begin{equation*}
[\boldsymbol{D}_{r}]_{jj} = \frac{e^{\sum_{l=1}^py_l(i)b_j(l)}}{\left(1 + e^{\sum_{l=1}^py_l(i)b_j(l)}\right)^2}. 
\end{equation*}
\paragraph{Non-smooth Regularization for ADMM}
In this section, we describe ADMM for logistic regression problems with $L1$ regularization. The approach is very similar to L$_2$ case, with the following differences:
\begin{enumerate}
\item The gradient of function $\xi_i$ is given as:
\begin{align}\label{admm_gradient_equation}
&\nabla\xi_i(\boldsymbol{\theta}^{[k+1]}_i(t)) = \boldsymbol{B}_i\boldsymbol{\delta}|_{(t)} + \mu_im_i \boldsymbol{\rho}|_{(t)} +   \beta d(i)\boldsymbol{\theta}^{[k+1]}_i(t)  - \beta\left(\sum_{j\in S(i)}\left[\boldsymbol{\theta}^{[k]}_j + \frac{1}{\beta}\boldsymbol{\lambda}^{[k]}_{ij}\right] + \sum_{j\in P(i)}\left[\boldsymbol{\theta}^{[k+1]}_j - \frac{1}{\beta}\boldsymbol{\lambda}^{[k]}_{ji}\right]\right),
\end{align}
where $d(i)$ is the degree of node $i\in\mathbb{V}$, vector $\boldsymbol{\delta}|_{(t)}$ is given in (\ref{delta_vector_admm}), and
\begin{equation*}
\boldsymbol{\rho}|_{(t)} = \left[
\begin{array}{c}
\rho_1|_{(t)}\\
\rho_2|_{(t)}\\
\vdots\\
\rho_{p}|_{(t)}
\end{array}
\right]\hspace{0.3cm} \textit{with} \hspace{0.3cm} \rho_j|_{(t)}= \frac{1 - e^{-\alpha [\boldsymbol{\theta}^{[k+1]}_i(t)]_{j}  }}{ 1+ e^{-\alpha [\boldsymbol{\theta}^{[k+1]}_i(t)]_{j}  }},
\end{equation*}
where $[\boldsymbol{\theta}^{[k+1]}_i(t)]_{j}$ is $j^{th}$ component of vector $\boldsymbol{\theta}^{[k+1]}_{i}(t)$. \\

\item The Hessian of function $\xi_i$ can be written:
\begin{align}\label{hessian_admm_method}
&\boldsymbol{H}_{i}|_{(t)} = \nabla^2\xi_i(\boldsymbol{\theta}^{[k+1]}_i(t)) = \nabla^2f_i(\boldsymbol{\theta}^{[k+1]}_i(t)) + \beta d(i)\boldsymbol{I}_{p\times p} = \boldsymbol{B}_i\boldsymbol{D}_{i}|_{(t)}\boldsymbol{B}^{\mathsf{T}}_i + \beta d(i)\boldsymbol{I}_{p\times p} + 2\alpha\mu_im_i\boldsymbol{\Delta}_i|_{(t)},
\end{align}
where $\boldsymbol{D}_{i}|_{(t)}$ is diagonal $m_i \times m_i$ matrix, given in (\ref{D_i_matrix_admm}), and $\boldsymbol{\Delta}_i|_{(t)}$ is diagonal $p\times p$ matrix such that
\begin{equation*}
[\boldsymbol{\Delta}_i|_{(t)}]_{jj} = \frac{e^{\alpha [\boldsymbol{\theta}^{[k+1]}_i(t)]_j}}{(1 + e^{\alpha [\boldsymbol{\theta}^{[k+1]}_i(t)]_j})^2},
\end{equation*}
where $[\boldsymbol{\theta}^{[k+1]}_i(t)]_{j}$ is $j^{th}$ component of vector $\boldsymbol{\theta}^{[k+1]}_{i}(t)$.
\end{enumerate}

\paragraph{Non-Smooth Regularization for Distributed Averaging}
In this section, we describe the mathematics needed for distributed averaging. Again, these are similar to the L$_{2}$ regularization setting with the following differences: 

The sub-gradient of $f_i$ can be written as :
\begin{align}\label{subgradient_Olshevsky_method_1}
&\boldsymbol{g}_i(t) = \nabla f_i(\boldsymbol{w}_i(t)) = \boldsymbol{B}_i\boldsymbol{\delta}|_{(t)} + \mu_im_i \boldsymbol{\rho}|_{(t)},
\end{align}
where $\boldsymbol{\delta}|_{(t)}$ is given in (\ref{delta_vector_Olshevsky}) and 
\begin{equation*}
\boldsymbol{\rho}|_{(t)} = \left[
\begin{array}{c}
\rho_1|_{(t)}\\
\rho_2|_{(t)}\\
\vdots\\
\rho_{p}|_{(t)}
\end{array}
\right]\hspace{0.3cm} \text{with} \hspace{0.3cm} \rho_j|_{(t)}= \frac{1 - e^{-\alpha [\boldsymbol{w}_i(t)]_{j}  }}{ 1+ e^{-\alpha [\boldsymbol{w}_i(t)]_{j}  }},
\end{equation*}
where $[\boldsymbol{w}_i(t)]_j$ is $j^{th}$ component of vector $\boldsymbol{w}_i(t)$.
\subsection{Reinforcement Learning}
We consider the policy search framework for reinforcement learning with a uni-variate Gaussian policy. Here, the data is represented as a collection of trajectories $\{\boldsymbol{\tau}_i\}^{m}_{i=1}$, where
\begin{align*}
&\boldsymbol{\tau}_i = [\boldsymbol{x}^{(i)}_1, a^{(i)}_1, \ldots, \boldsymbol{x}^{(i)}_T, a^{(i)}_T] \\\nonumber
& \hspace{0.3cm} \boldsymbol{x}^{(i)}_t\in\mathbb{R}^{d}, \hspace{0.3cm} a^{(i)}_t\in\mathbb{R}.
\end{align*}
For each trajectory, we define the reward $\mathcal{R}(\boldsymbol{\tau}_i)$ given by function $\mathcal{R}(): \mathbb{R}^{2T}\to \mathbb{R}^{+}$. We consider the feature representation of vectors $\boldsymbol{x}^{(i)}_t$ given by map $\Phi(): \mathbb{R}^{d} \to \mathbb{R}^p $ such that $\boldsymbol{b}^{(i)}_t = \Phi(\boldsymbol{x}^{(i)}_t)\in\mathbb{R}^p$. Moreover, we assume that this data set is distributed among the nodes of $\mathbb{G}$. The goal, is to fined the solution of the following optimization problem:
\begin{align}\label{gcp_renf_learning}
&\min \sum_{i=1}^{n}f_i(\boldsymbol{\theta}_i)\\\nonumber
& \text{s.t.} \hspace{0.2cm} \boldsymbol{\theta}_1 = \ldots =\boldsymbol{\theta}_n \in\mathbb{R}^p,
\end{align}
where the local objectives are given as:
\begin{equation}\label{local_objective_12}
f_i(\boldsymbol{\theta}_i) = \sum_{j=1}^{m_i}\left[\mathcal{R}(\boldsymbol{\tau}_j)\left(\sum_{t=1}^{T}\left[a^{(j)}_t - \boldsymbol{\theta}^{\mathsf{T}}_i\boldsymbol{b}^{(j)}_t\right]^{2}\right)\right] + \mu_im_i||\boldsymbol{\theta}_i||^2_2,
\end{equation}
with $\mu_i$ being a regularization parameter, and $\sum_{i=1}^n m_i = m$. Easy to see that (\ref{local_objective_12}) can be simplified as:
\begin{align*}
&f_i(\boldsymbol{\theta}_i) = \boldsymbol{\theta}^{\mathsf{T}}_i\underbrace{\left[\sum_{j=1}^{m_i}\mathcal{R}(\boldsymbol{\tau}_j)\left[\sum_{t=1}^{T}\boldsymbol{b}^{(j)}_t\boldsymbol{b}^{(j)\mathsf{T}}_t \right] + \mu_im_i\boldsymbol{I}_{p\times p}  \right]}_{\boldsymbol{F}_i}\boldsymbol{\theta}_i - 2\left(\underbrace{\sum_{j=1}^{m_i}\left[\mathcal{R}(\boldsymbol{\tau}_j)\left(\sum_{t=1}^{T}a^{(j)}_t\boldsymbol{b}^{(j)}_{t}\right)\right]}_{\boldsymbol{g}_i}\right)^{\mathsf{T}}\boldsymbol{\theta}_i + \\\nonumber 
&\underbrace{\sum_{j=1}^{m_i}\left(\mathcal{R}(\boldsymbol{\tau}_j)\sum_{t = 1}^{T}(a^{(j)}_t)^2\right)}_{u_i} = \boldsymbol{\theta}^{\mathsf{T}}_i\boldsymbol{F}_i\boldsymbol{\theta}_i - 2\boldsymbol{g}^{\mathsf{T}}_i\boldsymbol{\theta}_i  + u_i,
\end{align*}
where 
\begin{align}\label{params_expressions}
&\boldsymbol{F}_i = \sum_{j=1}^{m_i}\mathcal{R}(\boldsymbol{\tau}_j)\boldsymbol{B}_j\boldsymbol{B}^{\mathsf{T}}_j + \mu_im_i\boldsymbol{I}_{p\times p}\\\nonumber
&\boldsymbol{g}_i = \sum_{j=1}^{m_i}\mathcal{R}(\boldsymbol{\tau}_j)\boldsymbol{B}_j\boldsymbol{a}_j\\\nonumber
&u_i = \sum_{j=1}^{m_i}\mathcal{R}(\boldsymbol{\tau}_j)\boldsymbol{a}^{\mathsf{T}}_j\boldsymbol{a}_j,
\end{align}
with  
\begin{equation*}
\boldsymbol{B}_j=
\begin{blockarray}{cccc}
\begin{block}{[cccc]}
| & | &  & | \\
\boldsymbol{b}^{(j)}_1 & \boldsymbol{b}^{(j)}_2 & \ldots & \boldsymbol{b}^{(j)}_T \\
| & | &  & | \\
\end{block}
\end{blockarray}\quad \in \mathbb{R}^{p\times T},
\end{equation*}
and
\begin{equation*}
\boldsymbol{a}_j =  \left[
\begin{array}{c}
a^{(j)}_1\\
a^{(j)}_2\\
\vdots\\
a^{(j)}_T
\end{array}
\right]\in\mathbb{R}^{T}.
\end{equation*}
To further simplify expressions (\ref{params_expressions}) let us introduce:
\begin{align*}
&\boldsymbol{\mathcal{B}}^{(i)} = \left[\boldsymbol{B}_1, \ldots \boldsymbol{B}_{m_i}\right]\in\mathbb{R}^{p\times Tm_i}\\\nonumber
&\boldsymbol{\mathcal{R}}^{(i)} = \left[\begin{array}{cccc}
\mathcal{R}(\boldsymbol{\tau}_1)\boldsymbol{I}_{T\times T}&\boldsymbol{0}&\cdots &\boldsymbol{0}\\
\boldsymbol{0}&\mathcal{R}(\boldsymbol{\tau}_2)\boldsymbol{I}_{T\times T}&\cdots &\boldsymbol{0}\\
\vdots & &\ddots &\vdots \\
\boldsymbol{0}&\boldsymbol{0}&\cdots &\mathcal{R}(\boldsymbol{\tau}_{m_i})\boldsymbol{I}_{T\times T}\\
\end{array}\right] \in \mathbb{R}^{Tm_i\times Tm_i}\\\nonumber
&\boldsymbol{\alpha}^{(i)} = \left[
\begin{array}{c}
\boldsymbol{a}_1\\
\boldsymbol{a}_2\\
\vdots\\
\boldsymbol{a}_{m_i}
\end{array}
\right] \in\mathbb{R}^{Tm_i}.
\end{align*}
Then (\ref{params_expressions}) gives:
\begin{align}
&\boldsymbol{F}_i = \boldsymbol{\mathcal{B}}^{(i)}\boldsymbol{\mathcal{R}}^{(i)}\boldsymbol{\mathcal{B}}^{(i) \mathsf{T}} + \mu_im_i\boldsymbol{I}_{p\times p}\\\nonumber
&\boldsymbol{g}_i = \boldsymbol{\mathcal{B}}^{(i)}\boldsymbol{\mathcal{R}}^{(i)}\boldsymbol{\alpha}^{(i)}\\\nonumber
&u_i = \boldsymbol{\alpha}^{(i) \mathsf{T}}\boldsymbol{\mathcal{R}}^{(i)}\boldsymbol{\alpha}^{(i)}.
\end{align}
Similarly to (\ref{y_vectors}) we introduce  vectors $\boldsymbol{y}_1, \ldots, \boldsymbol{y}_p$, then problem (\ref{gcp_renf_learning}) can be written as:
\begin{align}
& \min_{\boldsymbol{y}_1,\ldots \boldsymbol{y}_p}\sum_{i=1}^nf_i(y_1(i), y_2(i), \ldots, y_p(i))\\\nonumber
&s.t\hspace{0.2cm} \boldsymbol{Ly}_1 = \boldsymbol{Ly}_2 = \ldots = \boldsymbol{Ly}_p = \boldsymbol{0} 
\end{align}
where $\boldsymbol{\theta}_i = [y_1(i), y_2(i), \ldots, y_p(i)]^{\mathsf{T}}$, and:
\begin{align*}
&f_i(y_1(i), y_2(i), \ldots, y_p(i)) = \sum_{k=1,l=1}^{p,p}[\boldsymbol{F}_i]_{kl}y_k(i)y_l(i) - 2\sum_{k=1}^p[\boldsymbol{g}_i]_ky_k(i) + u^2_i.
\end{align*}
Hence, the Lagrangian can be written as:
\begin{align*}
&\mathcal{L}(\boldsymbol{y}_1, \ldots, \boldsymbol{y}_p, \boldsymbol{\lambda}_1, \ldots, \boldsymbol{\lambda}_p) = \\\nonumber
&\sum_{i=1}^{n}\left[f_i(y_1(i), y_2(i), \ldots, y_p(i)) + y_1(i)(\boldsymbol{L\lambda}_1)_i + y_2(i)(\boldsymbol{L\lambda}_2)_i + \cdots + y_p(i)(\boldsymbol{L\lambda}_p)_i \right],
\end{align*}
and primal variables can be recovered from the dual by the following equation:
\begin{equation}
\left[
\begin{array}{c}
y_1(i)\\
y_2(i)\\
\vdots\\
y_p(i)
\end{array}
\right]_{[\boldsymbol{\lambda}_1, \ldots, \boldsymbol{\lambda}_p]} = \boldsymbol{F}^{-1}_i\left[\boldsymbol{g}_i - \frac{1}{2}(\boldsymbol{L\Lambda})(i,:)]^{\mathsf{T}}\right],
\end{equation}
where $(\boldsymbol{L\Lambda})(i,:)$ is $i^{th}$ row of matrix $\boldsymbol{L\Lambda}$, and $\boldsymbol{\Lambda}$ is given as (\ref{Lambda_matrix}).\\\newline

\subsubsection{Reinforcement Learning via ADMM}
In this section we will describe distributed ADMM for the reinforcement learning problem in (\ref{gcp_renf_learning}). Recall, that in distributed ADMM each nodes implements the following instructions:
\begin{enumerate}
\item \textbf{{Initialization:}} Chose arbitrary $\boldsymbol{\theta}^{[0]}_{i} \in\mathbb{R}^{p}$ and $\boldsymbol{\lambda}^{[0]}_{ji}\in\mathbb{R}^p$ for $j \in P(i)$ for $i=1,\ldots,n$.
\item \textbf{{For}} $k \ge 0$
\begin{enumerate}
\item Each agent $i$ updates its estimate of $\boldsymbol{\theta}^{[k]}_i$ in a sequential order with
\begin{align}\label{Serega_dalbich_11}
&\boldsymbol{\theta}^{[k+1]}_i = \\\nonumber
&\arg\min_{\boldsymbol{\theta}^{}_i}\left(\underbrace{f_i(\boldsymbol{\theta}^{}_i) + \frac{\beta}{2}\sum_{j\in P(i)}\left|\left|\boldsymbol{\theta}^{[k+1]}_j - \boldsymbol{\theta}^{}_i - \frac{1}{\beta}\boldsymbol{\lambda}^{[k]}_{ji}  \right|\right|^2 + \frac{\beta}{2}\sum_{j\in S(i)}\left|\left|\boldsymbol{\theta}_i - \boldsymbol{\theta}^{[k]}_j - \frac{1}{\beta}\boldsymbol{\lambda}^{[k]}_{ij}\right|\right|^2}_{\xi_i}\right).
\end{align}
\item Each agent updates $\boldsymbol{\lambda}_{ji}$ for $j\in P(i)$ as follows:
\begin{equation*}
\boldsymbol{\lambda}^{[k+1]}_{ji} = \boldsymbol{\lambda}^{[k]}_{ji} - \beta(\boldsymbol{\theta}^{[k+1]}_j - \boldsymbol{\theta}^{[k+1]}_i).
\end{equation*}
\end{enumerate}
\end{enumerate}
We can get the closed form solution for (\ref{Serega_dalbich_1}):
\begin{align*}
&\nabla\zeta_i(\boldsymbol{\theta}_i) = \nabla f_i(\boldsymbol{\theta}_i) + \beta d(i)\boldsymbol{\theta}_i - \beta\left(\sum_{j\in S(i)}\left[\boldsymbol{\theta}^{[k]}_j  + \frac{1}{\beta}\boldsymbol{\lambda}^{[k]}_{ij}\right] + \sum_{j\in P(i)}\left[\boldsymbol{\theta}^{[k+1]}_{j} - \frac{1}{\beta}\boldsymbol{\lambda}^{[k]}_{ji}\right] \right) = \\\nonumber
&2\boldsymbol{F}_i\boldsymbol{\theta}_i - 2\boldsymbol{g}_i + \beta d(i)\boldsymbol{\theta}_i - \beta\left(\sum_{j\in S(i)}\left[\boldsymbol{\theta}^{[k]}_j  + \frac{1}{\beta}\boldsymbol{\lambda}^{[k]}_{ij}\right] + \sum_{j\in P(i)}\left[\boldsymbol{\theta}^{[k+1]}_{j} - \frac{1}{\beta}\boldsymbol{\lambda}^{[k]}_{ji}\right] \right).
\end{align*}
Hence, for the iterative rule (\ref{Serega_dalbich_11}) :
\begin{equation}\label{solution_9}
\boldsymbol{\theta}^{[k+1]}_i = \left[\boldsymbol{F}_i + \frac{\beta d(i)}{2}\boldsymbol{I}_{p\times p}\right]^{-1}\left(\boldsymbol{g}_i + \frac{\beta}{2}\left(\sum_{j\in S(i)}\left[\boldsymbol{\theta}^{[k]}_j  + \frac{1}{\beta}\boldsymbol{\lambda}^{[k]}_{ij}\right] + \sum_{j\in P(i)}\left[\boldsymbol{\theta}^{[k+1]}_{j} - \frac{1}{\beta}\boldsymbol{\lambda}^{[k]}_{ji}\right] \right) \right).
\end{equation} 
\subsubsection{Reinforcement Learning via Distributed Averaging}
In this section, we describe distributed averaging for reinforcement learning (\ref{gcp_renf_learning}). Recall, that:
\begin{enumerate}
\item \textbf{{Initialization:}} For each node $i\in\mathbb{V}$ initialize $\boldsymbol{\theta}_i(1)\in\mathbb{R}^p$ and set
\begin{equation*}
\boldsymbol{z}_i(1) = \boldsymbol{\omega}_i(1) = \boldsymbol{\theta}_i(1).
\end{equation*}
\item Each node $i\in\mathbb{V}$ implements the following instructions:
\begin{align*}
&\boldsymbol{\omega}_i(t+1) = \boldsymbol{\theta}_i(t) + \frac{1}{2}\sum_{j\in\mathbb{N}(i)}\frac{\boldsymbol{\theta}_j(t) - \boldsymbol{\theta}_i(t)}{\max\{d(i), d(j)\}} - \beta \boldsymbol{g}_i(t)\\\nonumber
&\boldsymbol{z}_{i}(t+1) = \boldsymbol{w}_i(t) - \beta \boldsymbol{g}_i(t)\\\nonumber
&\boldsymbol{\theta}_i(t+1) = \boldsymbol{\omega}_i(t+1) + \left(1 - \frac{2}{9n + 1}\right)(\boldsymbol{\omega}_i(t+1) - \boldsymbol{z}_i(t+1)),
\end{align*}
\end{enumerate}
where $\beta$ is a step-size and $\boldsymbol{g}_i(t)$ is the sub-gradient of $f_i$ evaluated at $\boldsymbol{w}_i(t)$, i.e.:
\begin{align*}
&\boldsymbol{g}_i(t) = \nabla f_i(\boldsymbol{w}_i(t)) 
=2\boldsymbol{F}_i\boldsymbol{w}_i(t) - 2\boldsymbol{g}_i.
\end{align*}
After $T$ iterations the node $i\in\mathbb{V}$ computes the average
\begin{equation}
\bar{\boldsymbol{w}}_i = \frac{1}{T}\sum_{t=1}^{T}\boldsymbol{w}_i(t).
\end{equation}
as a solution.

\end{document}